\documentclass[12pt]{article}
\usepackage{psfrag,epsf}
\usepackage{enumerate}
\usepackage{url} 

\usepackage{rotating}
\usepackage{comment}
\usepackage{titling}
\usepackage[title]{appendix}
\usepackage{tikz}
\usetikzlibrary{positioning}

\usepackage{microtype}
\usepackage{booktabs} % for professional tables

\usepackage{amsmath, amssymb, amsfonts, amsthm, bbm}
\usepackage{algorithm, algpseudocode, algorithmicx}
\usepackage{mathtools,commath}
\usepackage{graphicx, multirow}
\usepackage[round]{natbib}
\usepackage{xcolor}
\usepackage{authblk}
\usepackage{hyperref}	
\usepackage[capitalise]{cleveref}

\usepackage{bibunits}

\usepackage{caption,subcaption}

\usepackage{soul,xcolor}
\usepackage[normalem]{ulem}
\setstcolor{red}

\newtheorem{theorem}{Theorem}[section]
\newtheorem{corollary}[theorem]{Corollary}

\newtheorem{assumption}[theorem]{Assumption}
\newtheorem{proposition}[theorem]{Proposition}

\newtheorem{example}{Example}
\newtheorem{remark}[theorem]{Remark}

\newcommand{\zq}[1]{}

\newcommand{\blind}{1}

\begin{document}

\def\spacingset#1{\renewcommand{\baselinestretch}%
{#1}\small\normalsize} \spacingset{1}

%%%%%%%%%%%%%%%%%%%%%%%%%%%%%%%%%%%%%%%%%%%%%%%%%%%%%%%%%%%%%%%%%%%%%%%%%%%%%%

\if1\blind
{
  \title{\bf Distributionally Robust Instrumental Variables Estimation}
   \author{ 
    Zhaonan Qu\thanks{Corresponding author. \url{zq2236@columbia.edu}} \quad Yongchan Kwon\thanks{\url{yk3012@columbia.edu}} \\
    $^*$Columbia University, Data Science Institute \\
    $^\dagger$Columbia University, Department of Statistics
    }
   \date{}
  \maketitle
% \centerline{\href{https://drive.google.com/file/d/1g2ffF4WMaQV9gNRU5m-gEz0FVf75O2Jw/view?usp=sharing}{\textcolor{red}{\large Latest Version Here}}}
} \fi

\if0\blind
{
  \title{\bf Distributionally Robust Instrumental Variables Estimation}
  \author{}
  \date{}
   \maketitle
   \vspace{-3cm}
} \fi

\begin{abstract}
Instrumental variables (IV) estimation is a fundamental method in econometrics and statistics for estimating causal effects in the presence of unobserved confounding. However, challenges such as untestable model assumptions and poor finite sample properties have undermined its reliability in practice. Viewing common issues in IV estimation as distributional uncertainties, we propose DRIVE, a distributionally robust IV estimation method. We show that DRIVE minimizes a square root variant of ridge regularized two stage least squares (TSLS) objective when the ambiguity set is based on a Wasserstein distance. In addition, we develop a novel asymptotic theory for this estimator, showing that it achieves consistency
without requiring the regularization parameter to vanish. %This result relies on a fundamental property of the square root ridge that we call ``delayed shrinkage'', which holds for a class of generalized method of moments (GMM) estimators. 
This novel property ensures that the estimator is robust to distributional uncertainties that persist in large samples. We further derive the asymptotic distribution of Wasserstein DRIVE and propose data-driven procedures to select the regularization parameter based on theoretical results. Simulation studies demonstrate the superior finite sample performance of Wasserstein DRIVE in terms of estimation error and out-of-sample prediction. Due to its regularization and robustness properties, Wasserstein DRIVE presents an appealing option when the practitioner is uncertain about model assumptions or distributional shifts in data. 
\end{abstract}
\noindent%

%\newpage
\spacingset{1.8} % DON'T change the spacing!

% \begin{bibunit}[apalike]

\section{Introduction}
\label{sec:intro}
Instrumental variables (IV) estimation, also known as IV regression, is a fundamental method in econometrics and statistics to infer causal relationships in observational data with unobserved confounding. It leverages access to additional variables (instruments) that affect the outcome exogenously and exclusively through the endogenous regressor to yield consistent causal estimates, even when the standard ordinary least squares (OLS) estimator is biased by unobserved confounding 
\citep{imbens1994identification,angrist1996identification,imbens2015causal}. Over the years, IV estimation has become an indispensable tool for causal inference in empirical works in economics \citep{card1994minimum}, as well as in the study of genetic and epidemiological data \citep{davey2003mendelian}.

Despite the widespread use of IV in empirical and applied works, it has important limitations and challenges, such as invalid instruments \citep{sargan1958estimation,murray2006avoiding}, weak instruments \citep{staiger1997instrumental}, non-compliance \citep{imbens1994identification}, and heteroskedasticity, especially in settings with weak instruments or highly leveraged datasets \citep{andrews2019weak,young2022consistency}. These issues could significantly impact the
validity and quality of estimation and inference using instrumental variables \citep{jiang2017have}. Many works have since been devoted to assessing and addressing these issues, such as statistical tests \citep{hansen1982large,stock2002testing}, sensitivity analysis \citep{rosenbaum1983assessing,bonhomme2022minimizing}, and additional assumptions or structures on the data generating process \citep{kolesar2015identification,kang2016instrumental,guo2018confidence}. 

Recently, an emerging line of works have highlighted interesting connections between causality and the concepts of invariance and robustness \citep{peters2016causal,meinshausen2018causality,rothenhausler2021anchor,buhlmann2020invariance,jakobsen2022distributional,fan2024environment}. Their guiding philosophy is that causal properties can be viewed as \emph{robustness} against changes across heterogeneous environments, represented by a \emph{set} $\mathcal{P}$ of data distributions. The robustness of an estimator against $\mathcal{P}$ is often  represented in a distributionally robust optimization (DRO) framework via the min-max problem
\begin{align}
\label{eq:DRO}
    \min_\beta \sup_{\mathbb{P}\in \mathcal{P}} \mathbb{E}[\ell(W;\beta)],
\end{align}
where $\ell(W;\beta)$ is a loss function of data $W$ and parameter $\beta$ of interest. 
% $\mathcal{P}(\mathbb{P}_0)$ is a set of probability distributions on $W$, parameterized by a robustness parameter $\rho>0$ and a reference distribution $\mathbb{P}_0$, which in practice is often an empirical distribution constructed from data. 

In many estimation and regression settings, one assumes that the true data distribution satisfies certain conditions, e.g., conditional independence or moment equations. Such conditions guarantee that standard statistical procedures based on the empirical distribution $\mathbb{P}_0$ of data, such as M-estimation and generalized method of moments (GMM), enable valid estimation and inference. In practice, however, it is often reasonable to expect that the distribution $\mathbb{P}_0$ of the observed data might deviate from that generated by the {ideal} model that satisfies such conditions, e.g., due to measurement errors or model mis-specifications. DRO addresses such \emph{distributional uncertainties} by explicitly incorporating possible deviations into
an \textbf{ambiguity set} $\mathcal{P}=\mathcal{P}(\mathbb{P}_0,\rho)$ of distributions that are ``close'' to $\mathbb{P}_0$. The parameter $\rho$ quantifies the degree of uncertainty, e.g., as the radius of a ball centered around $\mathbb{P}_0$ and defined by some divergence measure between probability distributions. By minimizing the \emph{worst-case} loss over $\mathcal{P}(\mathbb{P}_0,\rho)$ in the min-max optimization problem \eqref{eq:DRO}, the DRO approach achieves robustness against deviations captured by $\mathcal{P}(\mathbb{P}_0,\rho)$.

DRO provides a useful perspective for understanding the robustness properties of statistical methods. For example, it is well-known that members of the family of $k$-class estimators \citep{anderson1949estimation,nagar1959bias,theil1961economic} are more robust than the standard IV estimator against weak instruments \citep{andrews2007inference}. Recent works by \citet{rothenhausler2021anchor} and \citet{jakobsen2022distributional} show that $k$-class estimators in fact have a DRO representation of the form \eqref{eq:DRO}, where $\ell$ is the square loss, $W=(X,Y)$, and $X,Y$ are endogenous and outcome variables generated from structural equation models parameterized by the natural parameter of $k$-class estimators. See Appendix \ref{sec:anchor-regression} for details.
The general robust optimization problem \eqref{eq:DRO} can trace its roots in the classical robust statistics literature \citep{huber1964robust,huber2011robust} as well as classic works on robustness in economics \citep{hansen2008robustness}. Drawing inspirations from them, recent works in econometrics have also explored the use of robust optimization to account for (local) deviations from model assumptions \citep{kitamura2013robustness,armstrong2021sensitivity,chen2021robust,bonhomme2022minimizing,adjaho2022externally,fan2023quantifying}. These works, together with works on invariance and robustness, highlight the emerging interactions between econometrics, statistics, and robust optimization.

Despite new developments connecting causality and robustness, many questions and opportunities remain. An important challenge in DRO is the choice of the ambiguity set $\mathcal{P}(\mathbb{P}_0,\rho)$ to adequately capture distributional uncertainties. This choice is highly dependent on the structure of the particular problem of interest. While some existing DRO approaches use ambiguity sets $\mathcal{P}(\mathbb{P}_0,\rho)$ based on 
{marginal} or joint distributions of data, such $\mathcal{P}(\mathbb{P}_0,\rho)$ may not effectively capture the structure of IV estimation models. %they may not adequately and efficiently capture uncertainties arising from violations of model assumptions specific to such as the exclusion restriction and heteroskedasticity.
In addition, as the min-max problem \eqref{eq:DRO} minimizes the loss function under the \emph{worst-case} distribution in $\mathcal{P}(\mathbb{P}_0,\rho)$, a common concern is that the resulting estimator is too conservative when $\mathcal{P}(\mathbb{P}_0,\rho)$ is too large. In particular, although DRO estimators enjoy better empirical performance in finite samples, their asymptotic validity typically requires the ambiguity set to vanish to a singleton, i.e., $\rho \rightarrow 0$ \citep{blanchet2019robust, blanchet2022confidence}. %This feature can best be understood using the equivalent formulation of many DRO estimators as regularized regression estimators \citep{blanchet2019robust}, where the size $\rho$ of the ambiguity set $\mathcal{P}(\mathbb{P}_0,\rho)$ becomes the regularization parameter (see \cref{sec:theory}). It is well-known that regularized estimators, such as the ridge and the LASSO \citep{tibshirani1996regression}, enjoy better finite sample properties, but introduce bias that only vanishes in large samples with a vanishing regularization parameter $\rho \rightarrow 0$. 
However, in the context of IV estimation, distributional uncertainties about untestable model assumptions could persist in \emph{large samples}, necessitating the need for an ambiguity set that does not vanish to a singleton. It is therefore important to ask whether and how one can construct an estimator in the IV estimation setting that can sufficiently capture the distributional uncertainties about model assumptions, and at the same time remains asymptotically valid with a non-vanishing robustness parameter. 

In this paper, we propose to view common challenges to IV estimation through the lens of DRO, whereby uncertainties about model assumptions, such as the exclusion restriction and homoskedasticity, are captured by a suitably chosen ambiguity set in \eqref{eq:DRO}. Based on this perspective, we propose DRIVE, a general DRO approach to IV estimation. Instead of constructing the ambiguity set based on marginal or joint distributions as in existing works, we construct $\mathcal{P}(\mathbb{P}_0,\rho)$ from distributions \emph{conditional} on the instrumental variables. More precisely, we construct $\mathbb{P}_0$ as the empirical distribution of outcome and endogenous variables $Y,X$ \emph{projected} onto the space spanned by instrumental variables. When the ambiguity set of DRIVE is based on the 2-Wasserstein metric, we show that the resulting estimator minimizes a square root version of ridge regularized two stage least squares (TSLS) objective, where the radius $\rho$ of the ambiguity set becomes the regularization parameter. This regularized regression formulation relies on the general duality of Wasserstein DRO problems \citep{gao2016distributionally,blanchet2019robust,kuhn2019wasserstein}. 

We next next
reveal a surprising statistical property of the square root ridge by showing that Wasserstein DRIVE is consistent as long as the regularization parameter $\rho$ is bounded above by an estimable constant, which depends on the first stage coefficient of the IV model and can be interpreted as a measure of instrument quality. To our knowledge, this is the first consistency result for regularized regression estimators where the regularization parameter does not vanish as the sample size $n \rightarrow \infty$. %We demonstrate that this property, which we call ``delayed shrinkage'', is a result of the unique geometry of the square root ridge regression, which applies to a class of generalized method of moments (GMM) estimators.  
One implication of our results is that Wasserstein DRIVE, being a regularized regression estimator, enjoys better finite sample properties, but does not introduce bias asymptotically even for non-vanishing $\rho$, unlike standard regularized regression estimators such as the ridge and LASSO. %The consistency of Wasserstein DRIVE with non-vanishing $\rho$ is significant because it ensures the estimator achieves robustness against distributional uncertainties that can persist in large samples, such as those arising from potential violations of standard assumptions in IV estimation. 

We further characterize the asymptotic distribution of Wasserstein DRIVE and propose data-driven procedures to select the regularization parameter. %based on the smallest singular value of the first stage estimate, and the other is based on nonparametric bootstrap of the score quantile. 
We demonstrate with numerical experiments that Wasserstein DRIVE improves over the finite sample performance of IV and $k$-class estimators, thanks to its ridge type regularization, while at the same time retaining asymptotic validity whenever instruments are valid. In particular, Wasserstein DRIVE achieves significant improvements in mean squared errors (MSEs) over IV and OLS when instruments are moderately invalid. These findings suggest that Wasserstein DRIVE can be an attractive option in practice when we are concerned about model assumptions.

The rest of the paper is organized as follows. In \cref{sec:WDRIVE}, we discuss the standard IV estimation framework and common challenges. In \cref{sec:theory}, we propose the Wasserstein DRIVE framework and provide the duality theory. In \cref{sec:bias-analysis}, we develop asymptotic results for the Wasserstein DRIVE, including consistency under a non-vanishing robustness/regularization parameter. %\cref{sec:penalty-selection} discusses data-driven penalty parameter selection procedures, and 
\cref{sec:numerical} conducts numerical studies that compare Wasserstein DRIVE with other estimators including IV, OLS, and $k$-class estimators. Background materials, proofs, and additional results are included in the appendices in the supplementary material.

\textbf{Notation.} Throughout the paper,  $\|v \|_p$ denotes the $p$-norm of a vector $v$, while $\|v \| := \|v \|_2$ denotes the Euclidean norm.
%$\|M\|_F$ denotes the Frobenius norm of a matrix $M$. 
$\text{Tr}(M)$ denotes the trace of a matrix $M$.  $\lambda_{k}(M)$ represents the $k$-th largest eigenvalue of a symmetric matrix $M$. 
 Boldfaced variables, such as $\mathbf{X}$, represents a matrix whose $i$-th row is the variable $X_i$. %$O_p$ denotes the standard boundedness/convergence in probability. 

\vspace{-0.5cm}
\section{Background and Challenges in IV Estimation}
\label{sec:WDRIVE}
 In this section, we first provide a brief review of the standard IV estimation framework. We then motivate the DRO approach to IV estimation by viewing  common challenges from the perspective of distributional uncertainties. In \cref{sec:theory}, we propose the Wasserstein distributionally robust instrumental variables estimation (DRIVE) framework. 
\vspace{-0.5cm}
\subsection{Instrumental Variables Estimation}
 Consider the following standard linear instrumental variables regression model with $X\in\mathbb{R}^{p},Z\in\mathbb{R}^{d}$ where $d\geq p$, and $\beta_{0}\in\mathbb{R}^{p},\gamma\in\mathbb{R}^{d\times p}$: 
% \begin{align}
%     \begin{split}
%         Y &= X^T\beta_0 + Z^T\eta + \epsilon,\\
%     X &= Z^T\gamma + \xi,
%     \end{split}
%     \label{eq:IV-model}
% \end{align}
\begin{align}
\label{eq:IV-model}
    \begin{split}
        Y &= \beta^T_0 X %+ Z\eta + 
        +\epsilon,\\
    X &= \gamma^T Z + \xi.
    \end{split}
\end{align} In \eqref{eq:IV-model}, $X$ are the endogenous variables, $Z$ are the instrumental variables, and $Y$ is the outcome variable. The error terms $\epsilon$ and $\xi$ capture the unobserved (or residual) components of $Y$ and $X$, respectively. We are interested in estimating the causal effects $\beta_0$ of the endogenous variables $X$ on the outcome variable $Y$ given independent and identically distributed (i.i.d.) samples $\{X_i, Y_i, Z_i\}_{i=1}^{n}$. However, $X$ and $Y$ are confounded through some \emph{unobserved} confounders $U$ that are correlated with both $Y$ and $X$, represented graphically in the directed acyclic graph (DAG) below:
\begin{figure}[h!]
  \centering
  \begin{tikzpicture}[node distance=1.5cm]
    % Nodes
    \node (X) {$X$};
    \node (Z) [left=of X] {$Z$};
    \node (Y) [right=of X] {$Y$};
    \node (U) [left=of Y, yshift=1.2cm] {$U$}; % Adjust the distance as needed
    
    % Arrows
    \draw[->] (Z) -- (X) node[midway, above,yshift=-0.1cm] {$\gamma$}; % Greek letter gamma above the arrow
    \draw[->] (U) -- (X);
    \draw[->] (X) -- (Y) node[midway, above,yshift=-0.1cm] {$\beta_0$};
    \draw[->] (U) -- (Y); 
    % Curved dashed arrow from Z to Y
    %\draw[->, red, dashed, bend right] (Z) to (Y) node[midway, below,yshift=-0.27cm] {$\eta$};
    
    % Double-sided dashed red arrow from Z to U
    %\draw[<->, red, dashed] (Z) -- (U);
  \end{tikzpicture}
\end{figure}

Mathematically, the unobserved confounding can be described by the moment condition 
\[\mathbb{E}\left[X\epsilon\right]\neq \mathbf{0}.\]
 As a result of the unobserved confounding, the standard ordinary least squares (OLS) regression estimator of $\beta_0$ that regresses $Y$ on $X$ is biased. To address this problem, the IV estimation approach leverages access to the instrumental variables $Z$, also often called instruments, which satisfy the moment conditions 
\begin{align}
\label{eq:IV-inclusion}
\text{rank}(\mathbb{E}\left[ZX^T\right]) & = p,\\
\label{eq:IV-exclusion}
\mathbb{E}\left[Z\epsilon\right]=\mathbf{0}, \mathbb{E}\left[Z\xi^T\right] &= \mathbf{0}.
\end{align}
 Under these conditions, a popular IV estimator is the two stage least squares (TSLS, sometimes also stylized as 2SLS) estimator \citep{theil1953repeated}. With $\Pi_{Z}:=\mathbf{Z}(\mathbf{Z}^{T}\mathbf{Z})^{-1}\mathbf{Z}^{T}$ and $\mathbf{X,Y,Z}$ matrix representations of $\{X_i, Y_i, Z_i\}_{i=1}^{n}$ whose $i$-th rows correspond to $X_i, Y_i, Z_i$, respectively, the TSLS estimator $\hat \beta^{\text{IV}}:=(\mathbf{X}^T\Pi_{Z}\mathbf{X})^{-1}\mathbf{X}^T\Pi_{Z}\mathbf{Y}$ minimizes the objective 
\begin{align}
\label{eq:IV-objective}
        \min_{\beta}\frac{1}{n}\|\mathbf{Y}-\Pi_{Z}\mathbf{X}\beta\|^{2}.
\end{align}
% \begin{align}
% \label{eq:IV-objective}
%         \min_{\beta}\frac{1}{n} \left(\mathbf{Y}-\Pi_{Z}\mathbf{X}\beta \right)^T \left(\mathbf{Y}-\Pi_{Z}\mathbf{X}\beta \right).
% \end{align}
In contrast, the standard OLS estimator $\hat \beta^{\text{OLS}}$ solves 
the problem
\begin{align}
\label{eq:OLS}\min_{\beta}\frac{1}{n}\|\mathbf{Y}-\mathbf{X}\beta\|^{2}.
\end{align}
When the moment conditions \eqref{eq:IV-inclusion} and \eqref{eq:IV-exclusion} hold, the TSLS estimator is a consistent estimator of the causal effect $\beta_0$ under standard assumptions \citep{wooldridge2020introductory}, and valid inference can be performed by constructing variance estimators based on the asymptotic distribution of $\hat \beta^{\text{IV}}$  \citep{imbens2015causal}. Although not the most common presentation of TSLS, the optimization formulation in \eqref{eq:IV-objective} provides intuition on  how IV estimation works: when the instruments $Z$ are uncorrelated with the unobserved confounders $U$ affecting $X$ and $Y$, the projection operator $\Pi_{Z}$ applied to $\mathbf{X}$ ``removes'' the confounding from $X$, so that $\Pi_{Z}\mathbf{X}$ becomes (asymptotically) uncorrelated with $\epsilon$. Regressing $\mathbf{Y}$ on $\Pi_{Z}\mathbf{X}$ then yields a consistent estimator of $\beta_0$. 

The validity of estimation and inference based on $\hat \beta^{\text{IV}}$ relies critically on the moment conditions \eqref{eq:IV-inclusion} and \eqref{eq:IV-exclusion}. Condition \eqref{eq:IV-inclusion} is often called the \textbf{relevance condition} or rank condition, and requires $\mathbb{E}\left[ZX^T\right]$ to have full rank (recall $p\leq d$). In the special case of one-dimensional instrumental and endogenous variables, i.e., $d=p=1$, it simply reduces to $\mathbb{E}\left[ZX\right]\neq 0$. Intuitively, the relevant condition ensures that the instruments $Z$ can explain sufficient variations in the endogenous variables $X$. In this case, the instruments are said to be relevant and strong. When $\mathbb{E}\left[ZX^T\right]$ is close to being rank deficient, i.e., the smallest eigenvalue
$\lambda_p(\mathbb{E}\left[ZX^T\right])\approx 0$, IV estimation suffers from the so-called weak instrument problem, which results in many issues in estimation and inference, such as small sample bias and non-normal statistics \citep{stock2002survey}. Some $k$-class estimators, such as limited information maximum likelihood (LIML) \citep{anderson1949estimation}, are partially motivated to address these problems. Condition \eqref{eq:IV-exclusion} is often referred to as the \textbf{exclusion restriction} or instrument exogeneity \citep{imbens2015causal}, and instruments that satisfy this condition are called \emph{valid instruments}. When an instrument $Z$ is correlated with the unobserved confounder that confounds $X,Y$, or when $Z$ affects the outcome $Y$ through an unobserved variable other than the endogenous variable $X$, the instrument becomes invalid, resulting in biased estimation and invalid inference of $\beta_0$ \citep{murray2006avoiding}. These issues can often be exacerbated when the instruments are weak, when there is heteroskedasticity \citep{andrews2019weak}, or the data is highly leveraged \citep{young2022consistency}.

Although many works have been devoted to addressing the problems of weak and invalid instruments, there are fundamental limits on the extent to which one can test for these issues. Given the popularity of IV estimation in practice, it is therefore desirable to have estimation and inference procedures that are \emph{robust} to the presence of such issues. Our work is precisely motivated by these considerations. Compared to many existing robust approaches to IV estimation, we take a more agnostic  approach via distributionally robust optimization. More precisely, we argue that many common challenges in IV estimation can be viewed as uncertainties about the data distribution, i.e., deviations from the ideal model that satisfies IV assumptions, which can be explicitly taken into account by choosing an appropriate ambiguity set in a DRO formulation of the standard IV estimation. To demonstrate this perspective more concretely, we now examine some common problems in IV estimation and show that they can be viewed as distributional shifts under a suitable metric, and therefore amenable to a DRO approach. 

\vspace{-0.5cm}
\subsection{Challenges in IV Estimation as Distributional Uncertainties}
\label{subsec:why-dro}
Consider now the following one-dimensional version of the IV model in \eqref{eq:IV-model} 
\begin{align*}
\begin{split}
    Y &= X\beta_0 + \epsilon \\
    X &= Z\gamma + \xi,
    \end{split}
\end{align*}
where we assume that $X,Y$ are confounded by an unobserved confounder $U$ through 
\[\epsilon=Z\eta+U, \quad \xi=U.\]
Note that in addition to $U$, there is also potentially a direct effect $\eta$ from the instrument $Z$ to the outcome variable $Y$. We focus on the resulting model for our subsequent discussions:
\begin{align}
\label{eq:simple-IV}
\begin{split}
    Y &= X\beta_0 + Z\eta + U\\
    X &= Z\gamma + U.
    \end{split}
\end{align}
The standard IV assumptions can be succinctly summarized for \eqref{eq:simple-IV}. The relevance condition \eqref{eq:IV-inclusion} requires that $\gamma\neq0$, while the exclusion restriction \eqref{eq:IV-exclusion} requires that $Z$ is uncorrelated with $U$ and that in addition $\eta=0$. Assume that $U,Z$ are i.i.d. standard normal. $X,Y$ are then determined by $\eqref{eq:simple-IV}$. We are interested in the shifts in data distribution, appropriately defined and measured, when the exogeneity and relevance conditions are violated.

\begin{example}[Invalid Instruments]
\label{example:invalid-instrument}
\normalfont
As $U,Z$ are independent, $Z$ becomes invalid if and only if $\eta \neq 0$, and $|\eta|$ quantifies the degree of instrument invalidity. Let $\mathbb{P}_{\eta}$ denote the joint distribution on $(X,Y,Z)$ in the model \eqref{eq:simple-IV} indexed by $\eta\in \mathbb{R}$. %The distribution $\mathbb{P}_{0}$ corresponds to $\eta=0$, whereby $Z$ is a valid instrument. 
Let $\tilde{\mathbb{P}}_{\eta,Z}$ be the resulting normal distribution on the conditional random variables $(\tilde{X},\tilde{Y})=(X\mid Z,Y\mid Z)$, given $Z$. We are interested in the (expected) distributional shift between $\tilde{\mathbb{P}}_{\eta,Z}$ and $\tilde{\mathbb{P}}_{0,Z}$. We choose the 2-Wasserstein distance $W_2(\cdot,\cdot)$ \citep{kantorovich1942translocation,kantorovich1960mathematical}, also known as the Kantorovich metric, to measure this shift. %In Appendix \ref{subsec:Wasserstein}, we provide more details on the Wasserstein distance. 
Conveniently, the 2-Wasserstein distance between two normal distributions $\mathbb{Q}_{1}=\mathcal{N}(\mu_{1},\Sigma_{1})$
and $\mathbb{Q}_{2}=\mathcal{N}(\mu_{2},\Sigma_{2})$ has an explicit formula due to \citet{olkin1982distance}: 
\begin{align}
\label{eq:normal-Wasserstein-distance}
W_{2}(\mathbb{Q}_{1},\mathbb{Q}_{2})^{2} & =\|\mu_{1}-\mu_{2}\|^{2}+\text{Tr}(\Sigma_{1}+\Sigma_{2}-2(\Sigma_{2}^{1/2}\Sigma_{1}\Sigma_{2}^{1/2})^{1/2}).
\end{align}
Applying \eqref{eq:normal-Wasserstein-distance} to the conditional distributions $\tilde{\mathbb{P}}_{\eta,Z},\tilde{\mathbb{P}}_{0,Z}$, and taking the expectation with respect to $Z$, we obtain the simple formula
\begin{align}
\mathbb{E}W_2(\tilde{\mathbb{P}}_{\eta,Z},\tilde{\mathbb{P}}_{0,Z}) = \sqrt{\frac{2}{\pi}}\cdot |\eta|.
\end{align}
% \begin{align*}
%     W_2(\mathbb{P}_\eta,\mathbb{P}_0) &= \sqrt{\beta^2+\eta^2+\mu^2_Y+1+\beta^2+\mu^2_Y+1-2\sqrt{(\beta^2+\eta^2+\mu^2_Y+1)(\beta^2+\mu^2_Y+1)}} \\
%     &= \sqrt{ \beta^2+\eta^2+\mu^2_Y+1}-\sqrt{ \beta^2+\mu^2_Y+1}
% \end{align*}
% and when $\eta$ is small, the distance is of order 
%\begin{align*}
%\lim_{x\rightarrow0}\frac{\sqrt{\beta^{2}+x+1}-\sqrt{\beta^{2}+1}}{x} & =\frac{1}{2}\lim_{x\rightarrow0}\frac{1}{\sqrt{\beta^{2}+x+1}}=\frac{1}{2}\frac{1}{\sqrt{\beta^{2}+1}}
%\end{align*}
% \begin{align*}
% \sqrt{ \beta^2+\eta^2+\mu^2_Y+1}-\sqrt{ \beta^2+\mu^2_Y+1} \approx\frac{1}{2}\frac{\eta^{2}}{\sqrt{\beta^{2}+\mu^2_Y+1}}
% \end{align*} 
This calculation shows that the degree of instrument invalidity, as measured by the strength of direct effect of instrument on the outcome, is \emph{proportional} to the expected distributional shift of the distribution on $(\tilde{X},\tilde{Y})$ from that under the valid IV assumption. Moreover, the simple form of the expected distributional shift relies on our choice of the Wasserstein distance to measure the distributional shift of the \emph{conditional} random variables $(\tilde{X},\tilde{Y})$. If %instead of the distribution $\tilde{\mathbb{P}}_{\eta,Z}$ on $(\tilde{X},\tilde{Y})$, 
we instead measure shifts in the joint distribution $\mathbb{P}_{\eta}$ on $(X,Y,Z)$, the resulting distributional shift will depend on other model parameters in addition to $\eta$. This example therefore suggests that the Wasserstein metric applied to the conditional distributional shift of $(\tilde{X},\tilde{Y})$ could be an appropriate measure of distributional uncertainty in IV regression models. %A Wasserstein ball of radius $|\eta|$ around the ideal probability distribution $\tilde{\mathbb{P}}_{0,Z}$ on $(\tilde X,\tilde Y)$ will in expectation encompass distributions resulting from invalid instruments that violate the exclusion restriction with direct effects bounded above by $\sqrt{\pi/2}|\eta|$.
\end{example}

 \begin{example}[Weak Instruments]
 \normalfont
 Now consider another common problem with IV estimation, which happens when the first stage coefficient $\gamma$ is close to 0. Let $\tilde{\mathbb{Q}}_{\gamma,Z}$ be the distribution on $(\tilde{X},\tilde{Y})$ indexed by $\gamma\in \mathbb{R}$ and $\eta=0$ in \eqref{eq:simple-IV}. In this case, we can verify that 
 \begin{align*}
\mathbb{E}W_2(\tilde{\mathbb{Q}}_{\gamma_1,Z},\tilde{\mathbb{Q}}_{\gamma_2,Z}) = \sqrt{\frac{2}{\pi}}\sqrt{1+\beta_0^2}|\gamma_1-\gamma_2|.
 \end{align*}
The expected distributional shift between the setting with a ``strong'' instrument with $\gamma=\gamma_0$ and a ``weak'' instrument with $\gamma = \delta \cdot \gamma_0$ where $\delta \rightarrow 0$  in the limit, measured by the 2-Wasserstein metric, is equal to 
\begin{align} \sqrt{\frac{2}{\pi}}\sqrt{1+\beta_0^2}\cdot |\gamma_0|.
\end{align}
Similar to the previous example, the degree of violation of the strong instrument assumption, as measured by the presumed instrument strength $|\gamma_0|$, is proportional to the expected distributional shift on $(\tilde{X},\tilde{Y})$. %which again suggests the 2-Wasserstein distance is a suitable choice to capture distributional uncertainties. 
Note, however, that the distance is also proportional to the magnitude of the causal parameter $\beta_0$. This is reasonable because instrument strength is relative, and should be measured relative to the scale of the true causal parameter. 
 \end{example}
 Next, we consider the distributional shift resulting from heteroskedastic errors, which are known to yield the TSLS estimator inefficient and the standard variance estimator invalid \citep{baum2003instrumental}. Some $k$-class estimators, such as the LIML and the Fuller estimators, also become inconsistent under heteroskedasticity \citep{hausman2012instrumental}.
\begin{example}[Heteroskedasticity]
 \normalfont
In this example, we assume $\eta=0$ in \eqref{eq:simple-IV} and that the conditional distribution of $U$ given $Z$ is centered normal with standard deviation $\alpha\cdot |Z|+1$ where $\alpha \geq 0$. We are interested in the average distributional shift between the heteroskedastic setting ($\alpha >0$) from the homoskedastic setting ($\alpha = 0$). We can verify that the expected 2-Wasserstein distance between the conditional distributions on $(\tilde{X},\tilde{Y})$ is 
\begin{align} \sqrt{\frac{2}{\pi}}\sqrt{1+(\beta_0^2+1)^2}\cdot \alpha,
\end{align}
which is proportional to the degree of heteroskedasticity $\alpha$. 
% \begin{array}{cc}
% 1 & (\beta_{0}+1)^ {}\\
% (\beta_{0}+1)^ {} & (\beta_{0}+1)^{2}
% \end{array}	
% (\alpha|Z|+1)^{2}\cdot\begin{array}{cc}
% 1 & (\beta_{0}+1)^ {}\\
% (\beta_{0}+1) & (\beta_{0}+1)^{2}
% \end{array}	
% 1+(\alpha|Z|+1)^{2}-2(\alpha|Z|+1)	=(1-(\alpha|Z|+1))^{2}(1+(\beta_{0}+1)^{2})
% 	=\alpha|Z|\cdot\sqrt{1+(\beta_{0}+1)^{2}}
\end{example}
The preceding discussions demonstrate that distributional uncertainties resulting from violations of common model assumptions in IV estimation are well captured by the 2-Wasserstein distance on the distributions of the conditional variables $(\tilde{X},\tilde{Y})$. We therefore propose to construct an ambiguity set in \eqref{eq:DRO} using a Wasserstein ball around the empirical distribution on $(\tilde{X},\tilde{Y})$. %We can then obtain a new estimator by solving a DRO problem of the form \eqref{eq:DRO} with the constructed ambiguity set. 
We provide details of this  framework in the next section.

\vspace{-0.5cm}
\section{Wasserstein Distributionally Robust IV Estimation}
\label{sec:theory}
In this section, we propose a distributionally robust IV estimation framework. We propose to use Wasserstein ambiguity sets to account for distributional uncertainties in IV estimation. We develop the dual formulation of Wasserstein DRIVE as regularized regression, %We show that the Wasserstein DRIVE objective is equivalent to a particular convex square root version of ridge regularized IV estimation. 
and discuss its connections and distinctions to other regularized regression estimators. 

\vspace{-0.5cm}
\subsection{DRIVE}
Motivated by the intuition that common challenges to IV estimation in practice, such as violations of model assumptions, can be viewed as distributional uncertainties on the conditional distributions of $(\tilde{X},\tilde{Y})=(X\mid Z, Y\mid Z)$, we propose
the Distributionally Robust IV Estimation (DRIVE) framework, which solves the following DRO problem given a dataset $\{(X_i,Y_i,Z_i)\}_{i=1}^n$ and robustness parameter $\rho$:
\begin{align}
\label{eq:Wasserstein-DRIVE}
\hspace{-1cm}\textbf{(DRIVE Objective)} \quad \min_\beta \sup_{\{\mathbb{Q}:D(\mathbb{Q}, \tilde{\mathbb{P}}_n)\leq\rho\}}\mathbb{E}_{\mathbb{Q}}\left[({Y}-{X}^{T}\beta)^{2}\right],
\end{align}
where $\tilde {\mathbb{P}}_n(\mathcal{X}\times \mathcal{Y})$ is the
empirical distribution on $(X,Y)$ induced by the projected samples 
\[\{\tilde X_i,\tilde Y_i\}_{i=1}^{n}\equiv \{(\Pi_{\mathbf{Z}}\mathbf{X})_i,(\Pi_{\mathbf{Z}}\mathbf{Y})_i\}_{i=1}^n.\]
Here $\mathbf{X}\in \mathbb{R}^{n\times p},\mathbf{Y}\in \mathbb{R}^{n},\mathbf{Z}\in \mathbb{R}^{n\times d}$ are the matrix representations of observations, and $\Pi_{\mathbf{Z}} = \mathbf{Z}(\mathbf{Z}^T\mathbf{Z})^{-1}\mathbf{Z}^T$ is the projection matrix onto the column space of $\mathbf{Z}$. $D(\cdot,\cdot)$ is a metric or divergence measure on the space of probability distributions on $\mathcal{X}\times \mathcal{Y}$. Therefore, in our DRIVE framework, we first regress both the outcome $Y$ and covariate $X$ on the instrument $Z$ to form the $n$ predicted samples $(\Pi_{\mathbf{Z}}\mathbf{X},\Pi_{\mathbf{Z}}\mathbf{Y}$). Then an ambiguity set is constructed using $D$ around the empirical distribution $\tilde {\mathbb{P}}_n$. This choice of the reference distribution $\mathbb{P}_0$ is a key distinction of our work from previous works that leverage DRO in statistical models. In the standard regression/classification setting, the reference distribution is often chosen as the empirical distribution $\hat{\mathbb{P}}_n$ on $\{X_i, Y_i\}_{i=1}^{n}$ \citep{blanchet2019robust}. In the IV estimation setting where we have additional access to instruments $Z$, we have the choice of constructing ambiguity sets around the empirical distribution on the \emph{marginal} quantities $\{(X_i,Y_i,Z_i)\}_{i=1}^n$, which is the approach taken in \citet{bertsimas2022distributionally}. In contrast, we choose to use the empirical distribution on the \emph{conditional} quantities $\{(\tilde X_i,\tilde Y_i)\}_{i=1}^n$. This choice is motivated by the intuition that violations of IV assumptions can be captured by conditional distributional shifts, as illustrated by examples in the previous section. 

The choice of the divergence measure $D(\cdot,\cdot)$ is also important, as it characterizes the potential distributional uncertainties that DRIVE is robust to. In this paper, we propose to use the 2-Wasserstein distance $W_2(\mu,\nu)$ between two probability distributions $\mu,\nu$ \citep{mohajerin2018data,gao2016distributionally}. One advantage of the Wasserstein distance is the tractability of its associated DRO problems \citep{blanchet2019robust}, which can often be formulated as regularized regression problems with unique solutions. See also Appendix \ref{subsec:Wasserstein}. In \cref{subsec:why-dro}, we provided several examples that demonstrate the 2-Wasserstein distance is able to capture common distributional uncertainties in the IV estimation setting. Alternative distance measures of probability distributions, such as the class of $\phi$-divergences \citep{ben2013robust}, can also be used instead of the Wasserstein distance. For example, \citet{kitamura2013robustness} use the Hellinger distance to model local perturbations in robust estimation under moment restrictions, although not in the IV estimation setting. In this paper, we focus on the \textbf{Wasserstein DRIVE} framework based on $D=W_2$, and leave studies of DRIVE with other choices of $D$ to future works. 

 We next begin our formal study of Wasserstein DRIVE. In \cref{subsec:duality}, we will show that the Wasserstein DRIVE objective is dual to a convex regularized regression problem. As a result, the solution to the optimization problem \eqref{eq:Wasserstein-DRIVE} is well-defined, and we denote this estimator by $\hat{\beta}_{\mathrm{DRIVE}}$. In \cref{sec:bias-analysis}, we show $\hat{\beta}_{\mathrm{DRIVE}}$ is consistent with potentially \emph{non-vanishing} choices of the robustness parameter and derive its asymptotic distribution.

\subsection{Dual Representation of Wasserstein DRIVE}
\label{subsec:duality}
It is well-known in the optimization literature that min-max optimization problems such as \eqref{eq:Wasserstein-DRIVE} often have equivalent formulations as regularized regression problems. This correspondence
between regularization and robustness already manifests itself in
the ridge regression, which is equivalent to an $\ell_{2}$-robust
OLS regression \citep{bertsimas2018characterization}. Importantly, the regularized regression formulations are often more tractable in terms of solving the resulting optimization problem, and also facilitate the study of the statistical properties of the estimators. We first show that the Wasserstein DRIVE objective can also be written as a regularized regression problem similar to, but distinct from, the standard TSLS objective with ridge regularization. Proofs can be found in Appendix \ref{sec:proofs}.
\begin{theorem}
\label{thm:duality}
The optimization problem in \eqref{eq:Wasserstein-DRIVE} is  equivalent to the following convex regularized regression problem:
\begin{align}
\label{eq:sqrt-ridge-iv}
\min_{\beta}\sqrt{\frac{1}{n}\|\Pi_{\mathbf{Z}}\mathbf{Y}-\Pi_{\mathbf{Z}}\mathbf{X}\beta\|^{2}}+\sqrt{\rho(\|\beta\|^{2}+1)},
\end{align}
where $\Pi_{\mathbf{Z}} = \mathbf{Z}(\mathbf{Z}^{T}\mathbf{Z})^{-1}\mathbf{Z}^{T}$ is the finite sample projection operator, and $\Pi_{\mathbf{Z}}\mathbf{Y}$ and $\Pi_{\mathbf{Z}}\mathbf{X}$ are the OLS predictions of $\mathbf{X,Y}$ using instruments $\mathbf{Z}$.
\end{theorem}
Note that the robustness parameter $\rho$ of the DRO formulation \eqref{eq:Wasserstein-DRIVE} now has the dual interpretation as the regularization parameter in \eqref{eq:sqrt-ridge-iv}. This convex regularized regression formulation implies that the min-max problem \eqref{eq:Wasserstein-DRIVE} associated with Wassertein DRIVE has a unique solution, thanks to the strict convexity of the regularization term $\sqrt{\rho(\|\beta\|^{2}+1)}$, and is easy to compute despite not having a closed form solution. In particular, \eqref{eq:sqrt-ridge-iv} can be reformulated as a standard second order conic program (SOCP) \citep{el1997robust}, which can be solved efficiently with off-the-shelf convex optimization routines, such as \texttt{CVX}. %See also Appendix \ref{subsec:optimization-considerations}. 
More importantly, we leverage this formulation of Wasserstein DRIVE as a regularized regression problem to study its novel statistical properties in \cref{sec:bias-analysis}. 

The equivalence between Wasserstein DRO problems and regularized regression problem is a familiar general result in recent works. For example,
\citet{blanchet2019robust} and \citet{gao2016distributionally} derive similar duality results for distributionally robust regression with $q$-Wasserstein distances for $q>1$. Compared to previous works, our work is distinct in the following aspects. First, we apply Wasserstein DRO to the IV estimation setting instead of standard regression settings, such as OLS or logistic regression. {Although from an optimization point of view there is no substantial difference, the IV setting motivates a new asymptotic regime that uncovers interesting statistical properties of the resulting estimators.} 
Second, the regularization term in \eqref{eq:sqrt-ridge-iv} is distinct from those in previous works, which often use $\|\beta\|_p$ with $p\geq 1$. This seemingly innocuous difference turns out to be crucial for our novel results on the Wasserstein DRIVE. Lastly, compared to the proof in \citet{blanchet2019robust}, our proof of \cref{thm:duality} is based on a different argument using the Sherman-Morrison formula instead of H\"{o}lder's inequality, which provides an independent proof of the important duality result for Wasserstein distributionally robust optimization. 
\vspace{-0.5cm}
\subsection{Wasserstein DRIVE and Regularized Regression}
 The regularized regression formulation of the Wasserstein DRIVE problem in \eqref{eq:sqrt-ridge-iv} resembles the standard ridge regularized \citep{hoerl1970ridge} TSLS regression:
%\begin{align*}
%\min_{\beta}\frac{1}{n}\sum_{i}(\tilde{Y}_{i}-\beta^{T}\tilde{X}_{i})^{2}+\rho\|\beta\|^{2}
%\end{align*}
\begin{align}
\label{eq:ridge-iv}
\min_{\beta}\frac{1}{n}\sum_{i}({Y}_{i}-\tilde{X}_{i}^T\beta)^{2}+\rho\|\beta\|^{2} \Longleftrightarrow\min_{\beta}\frac{1}{n}\|\mathbf{Y}-\Pi_{\mathbf{Z}}\mathbf{X}\beta\|^{2}+\rho\|\beta\|^{2}.
\end{align}
We therefore refer to \eqref{eq:sqrt-ridge-iv} as the \textbf{square root ridge} regularized TSLS. However, there are three major distinctions between \eqref{eq:sqrt-ridge-iv} and \eqref{eq:ridge-iv} that are essential in guaranteeing the statistical properties of Wasserstein DRIVE not enjoyed by the standard ridge regularized TSLS. First, the presence of square root operations on both the risk term and the penalty term; second, the presence of a constant in the regularization term; third, an additional projection on the outcomes in $\Pi_{\mathbf{Z}}\mathbf{Y}$. We further elaborate on these features in \cref{sec:bias-analysis}. 

In the standard regression setting without instrumental variables, the square root ridge 
\begin{align}
\label{eq:sqrt-ridge-ols}
\min_{\beta}\sqrt{\frac{1}{n}\|\mathbf{Y}-\mathbf{X}\beta\|^{2}}+\sqrt{\rho(1+\|\beta\|^{2})}
\end{align}
also resembles the {``square root LASSO''} of \citet{belloni2011square}:  
% between the standard LASSO regression 
% \begin{align}
% \label{eq:LASSO}
% \min_{\beta}\frac{1}{n}\|\mathbf{Y}-\mathbf{X}\beta\|^{2}+\rho\|\beta\|_1, 
% \end{align}
\begin{align}
\label{eq:sqrt-LASSO}
\min_{\beta}\sqrt{\frac{1}{n}\|\mathbf{Y}-\mathbf{X}\beta\|^{2}}+\lambda\|\beta\|_1. 
\end{align}
In particular, both can be written as dual problems of Wasserstein DRO problems \citep{blanchet2019robust}. %In addition, when $X$ and $\beta$ are one-dimensional, we can verify that the solutions to \eqref{eq:sqrt-ridge-ols} and \eqref{eq:sqrt-LASSO} coincide when $\sqrt{\rho}=\lambda$. 
However, the square root LASSO is motivated by high-dimensional regression settings where the dimension of $X$ is potentially larger than the sample size $n$, but $\beta$ is very sparse. In contrast, our study of the square root ridge is motivated by its robustness properties in the IV estimation setting, where the dimension of the endogenous variable is small (often one-dimensional). In other words, variable selection is not the main focus of this paper. A variant of the square root ridge estimator in \eqref{eq:sqrt-ridge-ols} was also considered in the standard regression setting by \citet{owen2007robust}, who instead uses the penalty term $\|\beta\|_2$. 

As is well-known in the regularized regression literature \citep{fu2000asymptotics}, when the regularization parameter decays to 0 at a rate $O_p(1/\sqrt{n})$, the ridge estimator is consistent. A similar result also holds for the square root ridge 
\eqref{eq:sqrt-ridge-ols} in the standard regression setting as $\rho \rightarrow 0$. However, in the IV estimation setting, our distributional uncertainties about model assumptions, such as the validity of instruments, could persist even in large samples. Recall that $\rho$ is also the robustness parameter in the DRO formulation \eqref{eq:Wasserstein-DRIVE}. Therefore, the usual requirement that $\rho \rightarrow 0$ as $n \rightarrow \infty$ cannot adequately capture distributional uncertainties in the IV estimation setting.
In the next section, we study the asymptotic properties of Wasserstein DRIVE when $\rho$ does not necessarily vanish. In particular, we establish the consistency of Wasserstein DRIVE leveraging the three distinct features of \eqref{eq:sqrt-ridge-iv} that are absent in the standard ridge regularized TSLS regression \eqref{eq:ridge-iv}. 
% Unlike the OLS setting, in the IV regression setting the errors $\|\Pi_{\mathbf{Z}}\mathbf{Y}-(\Pi_{\mathbf{Z}}\mathbf{X})\beta_0\|_\infty \rightarrow 0$, where $\beta_0$ is the \emph{true} parameter in the model. This property, combined with the geometry of \eqref{eq:sqrt-ridge-iv}, yields the interesting
This asymptotic result is in stark contrast to the conventional wisdom on regularized regression that regularized regression achieves lower variance at the cost of non-zero bias.

\vspace{-0.5cm}
\section{Asymptotic Theory of Wasserstein DRIVE}
\label{sec:bias-analysis}
In this section, we leverage distinct geometric features of the square root ridge regression to study the asymptotic properties of the Wasserstein DRIVE. 
In \cref{subsec:theory-consistency}, we show that the Wasserstein DRIVE estimator is consistent for any $\rho \in [0,\overline \rho]$, where $\overline{\rho}$ depends on the first stage coefficient $\gamma$. This property is a consequence of the consistency of the square root ridge estimator in settings where the objective value at the true parameter vanishes, such as the GMM estimation setting. It ensures that Wasserstein DRIVE can achieve better finite sample performance thanks to its ridge type regularization, while at the same time retaining asymptotic validity when instruments are valid. In \cref{subsec:theory-asymptotic-distribution}, we characterize the asymptotic distribution of Wasserstein DRIVE, and discuss several special settings particularly relevant in practice, such as the just-identified setting with one-dimensional instrumental and endogenous variables. %Our theory provides  guidance on how to use Wasserstein DRIVE in practice to assess the validity of model assumptions and obtain robust estimates of causal effects.
\vspace{-0.5cm}
\subsection{Consistency of Wasserstein DRIVE}
\label{subsec:theory-consistency}
Recall the linear IV regression model in \eqref{eq:IV-model}
\begin{align*}
\begin{split}
        Y &= \beta^T_0 X %+ Z\eta + 
        +\epsilon,\\
    X &= \gamma^T Z + \xi,
    \end{split}
\end{align*}
 where $X\in\mathbb{R}^{p}, Z\in\mathbb{R}^{d}$, and $\beta_{0}\in\mathbb{R}^{p},\gamma\in\mathbb{R}^{d\times p}$
with $d\geq p$ to ensure identification. In this section, we make the standard assumptions that the instruments satisfy the relevance and exogeneity conditions in \eqref{eq:IV-inclusion} and \eqref{eq:IV-exclusion}, $\epsilon,\xi$ are homoskedastic, the instruments $Z$ are not perfectly collinear, and that $\mathbb{E}\|Z\|^{2k}<\infty, \mathbb{E}\|\xi\|^{2k}<\infty, \mathbb{E}|\epsilon|^{2k}<\infty$ for some $k>2$. The results can be extended in a straightforward manner when we relax these assumptions, e.g., only requiring that exogeneity holds \emph{asymptotically}. Given i.i.d. samples from the linear IV model, recall the regularized regression formulation of the Wasserstein DRIVE
objective 
\begin{align}
\label{eq:sqrt-IV-sample}
\min_{\beta}\sqrt{\frac{1}{n}\sum_{i}(\Pi_{\mathbf{Z}}\mathbf{Y}-\Pi_{\mathbf{Z}}\mathbf{X}\beta)_{i}^{2}}+\sqrt{\rho_n(\|\beta\|^{2}+1)},
\end{align}
 where $\Pi_{\mathbf{Z}}=\mathbf{Z}(\mathbf{Z}^{T}\mathbf{Z})^{-1}\mathbf{Z}^{T}\in \mathbb{R}^{n\times n}$,
and $\Pi_{\mathbf{Z}}\mathbf{Y}\in\mathbb{R}^{n}$ and $\Pi_{\mathbf{Z}}\mathbf{X}\in \mathbb{R}^{n\times p}$
are $\mathbf{Y}\in\mathbb{R}^{n},\mathbf{X}\in\mathbb{R}^{n\times p}$
projected onto the instrument space spanned by $\mathbf{Z}\in\mathbb{R}^{n\times d}$. 
\begin{theorem}[Consistency of Wasserstein DRIVE] 
\label{thm:DRIVE-consistency}
Let $\hat \beta_n^{\text{DRIVE}}$ be the unique minimizer of the objective in \eqref{eq:sqrt-IV-sample}. Let $\rho_n \rightarrow \rho \geq 0$ and $\frac{1}{n}\mathbf Z^T\mathbf Z\rightarrow_p=\mathbb{E}[ZZ^T]=\Sigma_Z$. Under the relevance and exogeneity conditions \eqref{eq:IV-inclusion} and \eqref{eq:IV-exclusion},
the Wasserstein DRIVE estimator  
$\hat \beta_n^{\text{DRIVE}}$ converges to $\beta^{\text{DRIVE}}$ in probability as $n\rightarrow \infty$, where $\beta^{\text{DRIVE}}$ is the unique minimizer of
\begin{align}
\label{eq:limiting-objective}
\min_\beta\sqrt{(\beta-\beta_{0})^{T}\gamma^T \Sigma_Z \gamma(\beta-\beta_{0})}+\sqrt{\rho(\|\beta\|^{2}+1)}.
\end{align}
Moreover, whenever $\rho\in[0,\overline{\rho}]$ where $\overline{\rho}=\lambda_{p}(\gamma^T \Sigma_Z  \gamma)$ is the smallest eigenvalue of $\gamma^T \Sigma_Z \gamma \in \mathbb{R}^{p\times p}$,
the unique minimizer of the objective \eqref{eq:limiting-objective} is the true causal effect, i.e.,  $\beta^{\text{DRIVE}}\equiv \beta_{0}$.
\end{theorem}
Therefore, Wasserstein DRIVE is consistent as long as the limit of the regularization parameter is bounded above by $\overline{\rho
}=\lambda_{p}(\gamma^T \Sigma_Z \gamma)$. In the case when $\Sigma_Z = \sigma^2_Z I_d$ for $\sigma_Z^2>0$, the upper bound $\overline{\rho
}$ is proportional to the square of the smallest singular value of the first stage coefficient $\gamma$, which is positive under the relevance condition \eqref{eq:IV-inclusion}. Recall that $\rho_n$ is the radius of the Wasserstein ball in the min-max formulation of DRIVE in \eqref{eq:Wasserstein-DRIVE}. \cref{thm:DRIVE-consistency} therefore guarantees that even when the robustness parameter $\rho_n \equiv \rho \neq 0$, which implies the solution to the min-max problem is different from the TSLS estimator ($\rho =0$), the resulting estimator always has the same limit as long as  $\rho \leq \lambda_{p}(\gamma^T \Sigma_Z \gamma)$. %Previous works have also leveraged spectral information in the IV estimation setting. For example, \citet{cragg1993testing,stock2002testing} use the smallest singular value of $\gamma$ to construct tests for weak instruments. The $k$-class parameter $\kappa$ for the LIML estimator is also chosen as the solution to a generalized minimum eigenvalue problem that minimizes the impact of weak instruments \citep{amemiya1985advanced}.

The significance of \cref{thm:DRIVE-consistency} is twofold. First, it provides a meaningful interpretation of the robustness parameter $\rho$ in Wasserstein DRIVE in terms of problem parameters, more precisely the variance covariance matrix $\Sigma_Z$ of $Z$ and the first stage coefficient $\gamma$ in the IV regression model. The maximum amount of robustness that can be afforded by Wasserstein DRIVE without sacrificing consistency is $\lambda_{p}(\gamma^T \Sigma_Z \gamma)$, which directly depends on the strength and variance of the instrument. This relation can be described more precisely when $\Sigma_Z = \sigma^2_Z I_d$, in which case the bound is proportional to $\sigma^2_Z$ and $\lambda_p(\gamma^T\gamma)$. Both quantities improve the quality of the instruments: $\sigma^2_Z$ improves the proportion of variance of $X$ and $Y$ explained by the instrument vs. noise, while a $\gamma$ far from rank deficiency avoids the weak instrument problem. Therefore, the robustness of Wasserstein DRIVE depends intrinsically on the strength of the instruments. The quantity $\gamma^T \Sigma_Z \gamma$ is not unfamiliar in the IV setting, as it is proportional to the \emph{inverse} of the asymptotic variance of the standard TSLS estimator when errors are homoskedastic. This observation suggests an intrinsic connection between robustness and efficiency in the IV setting. See the discussions after \cref{thm:asymptotic-distribution} for more on this point.

More importantly, \cref{thm:DRIVE-consistency} is the first consistency result for regularized regression estimators where the regularization parameter does not vanish with sample size. Although regularized regression such as ridge and LASSO is often associated with better finite sample performance at the cost of introducing some bias, our work demonstrates that, in the IV estimation setting, we can get the best of both worlds. On one hand, the ridge type regularization in Wasserstein DRIVE improves upon the  finite sample properties of the standard IV estimators, which aligns with conventional wisdom on regularized regression. On the other hand, with a bounded level of the regularization parameter $\rho$,
Wasserstein DRIVE can still achieve consistency. This is in stark contrast to existing asymptotic results on regularized regression \citep{fu2000asymptotics}. Therefore, in the context of IV estimation, with Wasserstein DRIVE we can achieve consistency and a certain amount of robustness at the same time, by leveraging additional information in the form of valid instruments. The maximum degree of robustness that can be achieved also has a natural interpretation in terms of the strength and variance of the instruments.
 
\cref{thm:DRIVE-consistency} also suggests the following procedure to construct a feasible and valid robustness/regularization parameter $\hat \rho$ given data $\{X_i, Y_i, Z_i\}_{i=1}^{n}$. Let $\hat{\gamma}=(\mathbf{Z}^{T}\mathbf{Z})^{-1}\mathbf{Z}^{T}\mathbf{X}$ be the OLS regression estimator of the first stage coefficient $\gamma$ and $\hat \Sigma_Z$ an estimator of $\Sigma_Z$, such as the heteroskedasticity-robust estimator \citep{white1980heteroskedasticity}. We can use $\hat \rho \leq \lambda_p(\hat{\gamma}^T \hat \Sigma_Z\hat{\gamma})$ to construct the Wasserstein DRIVE objective, i.e., any value bounded above by the 
smallest eigenvalue of $\hat{\gamma}^T \hat \Sigma_Z\hat{\gamma}$. Under the assumptions in \cref{thm:DRIVE-consistency}, $\lambda_p(\hat{\gamma}^T\hat \Sigma_Z\hat{\gamma})\rightarrow \lambda_p({\gamma}^T\Sigma_Z{\gamma})$, which guarantees that the Wasserstein DRIVE estimator with parameter $\hat \rho$ is consistent. In Appendix \ref{sec:penalty-selection}, we discuss the construction of feasible regularization parameters in more detail. We demonstrate the validity and superior finite sample performance of DRIVE based on these proposals in simulation studies in \cref{sec:numerical}.
 
\begin{figure}
\begin{centering}
\includegraphics[scale=0.5]{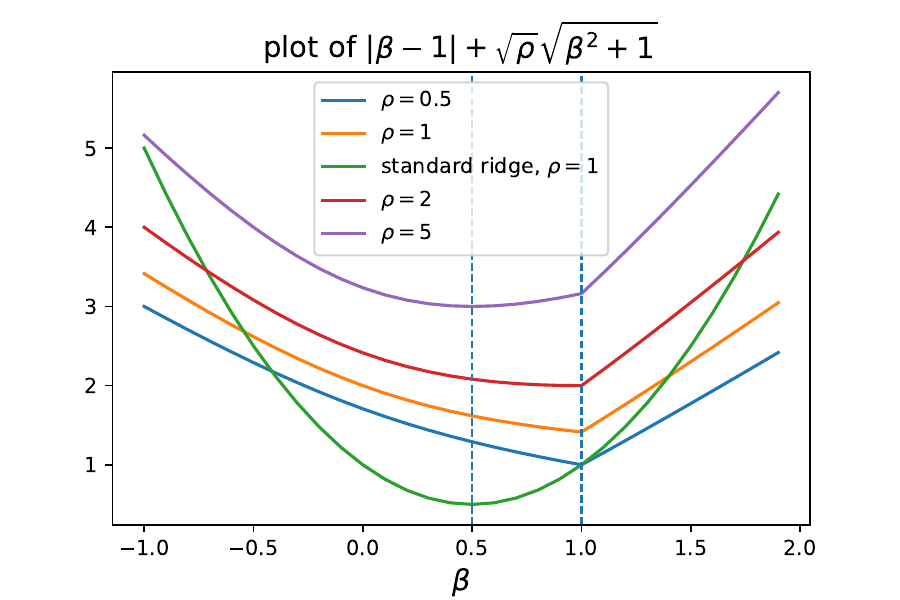}
\par\end{centering}
\caption{Plot of $|\beta-1|+\sqrt{\rho}\sqrt{(\beta^2+1)}$, which is the dual limit objective function in the one-dimensional case with $\sigma^2_Z=\gamma=\beta_0=1$. We also plot limit of standard ridge loss $(\beta-1)^2+\beta^2$. For $\rho\leq 2$, the minimum is achieved at $\beta=1$, while for $\rho=5$ and for the standard ridge, the minimum is achieved at $\beta=0.5$.}
\label{fig:loss}
\end{figure}

One may wonder why Wasserstein DRIVE can achieve consistency with a non-zero regularization $\rho$. Here we briefly discuss the phenomenon that the limiting objective \eqref{eq:limiting-objective}
\begin{align*}
\min_\beta\sqrt{(\beta-\beta_{0})^{T}\gamma^T \Sigma_Z \gamma(\beta-\beta_{0})}+\sqrt{\rho(\|\beta\|^{2}+1)}
\end{align*}
has a unique minimizer at $\beta_0$ for bounded $\rho>0$. The first term $\sqrt{(\beta-\beta_{0})^{T}\gamma^T \Sigma_Z \gamma(\beta-\beta_{0})}$ achieves its minimum value of 0 at $\beta=\beta_0$. When $\rho$ is small, the effect of adding the regularization term $\sqrt{\rho(\|\beta\|^{2}+1)}$ does not overwhelm the first term, especially when its curvature at $\beta_0$ is large. As a result, we may expect the minimizer to not deviate much from $\beta_0$. While this intuition is reasonable qualitatively, it does not fully explain the fact that the minimizer does not \emph{change} for small $\rho$. In the standard regression setting, the same intuition can be applied to the standard ridge regularization, but we know shrinkage occurs as soon as $\rho>0$. The key distinction of \eqref{eq:sqrt-IV-sample} turns out to be the square root operations we apply to the loss and regularization terms, which endows the objective with a geometric interpretation, and ensures that the minimizer does not deviate from $\beta_0$ unless $\rho$ is above some positive threshold. We call this phenomenon the ``delayed shrinkage'' of the square root ridge, as shrinkage does not happen until the regularization is large enough. %In Appendix \ref{sec:sqrt-ridge}, we provide more details on this property of the square root ridge, as well as the unique features of the Wasserstein DRIVE that guarantee it. 
We illustrate it with a simple example in \cref{fig:loss}, where the minimizer of the limiting square root objective does not change for a bounded range of $\rho$. 

% Consider the case when $d=p=1$, and $\sigma^2_Z=\gamma=\beta_0=1$. We can verify that the limiting objective \eqref{eq:limiting-objective} reduces to the following expression:
% \begin{align*}
% f(\beta)=|\beta-1|+\sqrt{\rho}\sqrt{(\beta^{2}+1)}.
% \end{align*}
%  Note that the function is strictly convex and differentiable except
% at $\beta=1$. When $\beta<1$, the derivative is $-1+\beta\sqrt{\rho}\frac{1}{\sqrt{(1+\beta^{2})}}$, which is negative if and only $\rho\leq 2$. For $\beta>1$, the derivative becomes $1+\beta\sqrt{\rho}\frac{1}{\sqrt{(1+\beta^{2})}}>0$. As a result, $f(\beta)$
% has a unique minimizer at $\beta=1$ as long as
% $\rho\leq 2$. See \cref{fig:loss} for plots of $f(\beta)$ at different values of $\rho$.

Lastly, we comment on the importance of projection operations in Wasserstein DRIVE. A crucial feature of the Wasserstein DRIVE objective is that both the outcome and the covariates are regressed on the instrument to compute their predicted values. In other words, the objective \eqref{eq:Wasserstein-DRIVE} uses $\Pi_{\mathbf{Z}}\mathbf{Y}-\Pi_{\mathbf{Z}}\mathbf{X}\beta$ instead of $\mathbf{Y}-\Pi_{\mathbf{Z}}\mathbf{X}\beta$. For standard IV estimation ($\rho=0)$, there is no substantial difference between the two objectives, since their minimizers are exactly the same, due to the idempotent property $\Pi^2_\mathbf{Z}=\Pi_\mathbf{Z}$. In fact, in applications of TSLS, the outcome variable is often not regressed on the instrument. However, Wasserstein DRIVE is consistent for positive $\rho$ \emph{only} if the outcome $Y$ is also projected onto the instrument space. In other words, the following problem does not yield a consistent estimator when $\rho > 0$:
    \begin{align*}
\min_{\beta}\sqrt{\frac{1}{n}\|\mathbf{Y}-\Pi_{\mathbf{Z}}\mathbf{X}\beta\|^{2}}+\sqrt{\rho(\|\beta\|^{2}+1)}.
\end{align*}
The reason behind this phenomenon is that ${\frac{1}{n}\|\Pi_{\mathbf{Z}}\mathbf{Y}-\Pi_{\mathbf{Z}}\mathbf{X}\beta\|^{2}}$ is a GMM objective 
\begin{align*}
    (\frac{1}{n}\sum_{i}Z_{i}(Y_{i}-\beta^{T}X_{i}))^{T}(\frac{1}{n}\mathbf{Z}^{T}\mathbf{Z})^{-1}(\frac{1}{n}\sum_{i}Z_{i}(Y_{i}-\beta^{T}X_{i})),
\end{align*}
which when $n \rightarrow \infty$ achieves a minimal value of 0 at $\beta_0$, while the limit of ${\frac{1}{n}\|\mathbf{Y}-\Pi_{\mathbf{Z}}\mathbf{X}\beta\|^{2}}$ does not vanish even at $\beta_0$. In the former case, the geometric properties of the square root ridge ensure that the minimizer of the regularized objective is $\beta_0$. 

\subsection{Asymptotic Distribution of Wasserstein DRIVE}
\label{subsec:theory-asymptotic-distribution}
 Having established the consistency of Wasserstein DRIVE with bounded $\rho$, we now turn to the characterization of its asymptotic distribution. In general, the asymptotic distribution of Wasserstein DRIVE is different from that of the standard IV estimator. However, we will also examine several special cases relevant in practice where they coincide.
\begin{theorem}
 [Asymptotic Distribution]
 \label{thm:asymptotic-distribution}
When $\lim_{n\rightarrow \infty}\rho_n=\rho\leq{\lambda_{p}(\gamma^T \Sigma_Z \gamma)}$, the Wasserstein DRIVE estimator $\hat \beta_n^{\text{DRIVE}}$ has asymptotic distribution characterized by the following optimization problem:
\begin{align}
\label{eq:asymptotic-distribution}
\sqrt{n}(\hat \beta^{\text{DRIVE}}-\beta_{0}) & \rightarrow_{d}\arg\min_{\delta}\sqrt{(\mathcal{Z}+\Sigma_Z\gamma\delta)^{T}\Sigma_Z^{-1}(\mathcal{Z}+\Sigma_Z\gamma\delta)}+\frac{\sqrt{\rho}\beta_{0}^{T}}{\sqrt{(1+\|\beta_{0}\|^{2})}}\cdot\delta,
\end{align}
where $\mathcal{Z}=\mathcal{N}(0,\sigma^2\Sigma_Z)$ and $\sigma^2=\mathbb{E}\epsilon^2$. 

In particular, when $\rho_n\rightarrow0$ at any rate, we have 
\begin{align*}
\sqrt{n}(\hat \beta^{\text{DRIVE}}-\beta_{0}) & \rightarrow_{d}\mathcal{N}(0,\sigma^2(\gamma^{T}\Sigma_Z\gamma)^{-1}),
\end{align*}
which is the asymptotic distribution for TSLS estimators with homoskedastic errors $\epsilon$.
\end{theorem}
Recall that the maximal robustness parameter $\rho$ of Wasserstein DRIVE while still being consistent is equal to the smallest eigenvalue of $\gamma^{T}\Sigma_Z\gamma$, which is proportional to the inverse of the asymptotic variance of the TSLS estimator. Therefore, as the efficiency of TSLS increases, so does the robustness of the associated Wasserstein DRIVE estimator. The ``price'' to pay for robustness when $\rho>0$ is an interesting question. It is clear from \cref{fig:loss} that the curvature of the population objective decreases as $\rho$ increases. Since the objective is not continuous at $\beta_0$, however, a generalized notion of curvature is needed to precisely characterize this behavior. Note that the asymptotic distribution of the TSLS estimator minimizes the objective 
\[(\mathcal{Z}+\Sigma_Z\gamma\delta)^{T}\Sigma_Z^{-1}(\mathcal{Z}+\Sigma_Z\gamma\delta),\] \cref{thm:asymptotic-distribution} implies that in general the asymptotic distributions of Wasserstein DRIVE and TSLS are different when $\rho>0$. However, there are still several cases relevant in practice where their asymptotic distributions do coincide, which we discuss next.
\begin{corollary}\label{cor:asymptotic-distribution-coincide}
In the following cases, the asymptotic distribution of Wasserstein DRIVE
is the same as that of the standard TSLS estimator:
\begin{enumerate}
    \item When $\rho=0$;
\item When $\rho\leq{\lambda_{p}(\gamma^T \Sigma_Z\gamma)}$ and $\beta_{0}$ is identically $\mathbf{0}$;
\item When  $\rho\leq{\lambda_{p}(\gamma^T \Sigma_Z\gamma)}$ and both $\beta_{0}$ and $\gamma$ are one-dimensional, i.e., $d=p=1$.
\end{enumerate}
\end{corollary}
In particular, the just-identified case with $d=p=1$ covers many empirical applications of IV estimation, since in practice we are often interested in the causal effect of a single endogenous variable, for which we have a single instrument. The case when $\beta_{0}\equiv \mathbf{0}$ is also very relevant, since an important question in practice is whether the causal effect of a variable is zero. Our theory suggests that the asymptotic 
distribution of Wasserstein DRIVE should be the same as that of the
TSLS when the causal effect is zero and $d>1$, even for $\rho>0$. Based on this observation, intuitively, we should expect that Wasserstein DRIVE and TSLS estimators to be ``close'' to each other. If the estimators or their asymptotic variance estimators differ significantly, then $\beta_0$ may not be identically 0. We can design statistical tests by leveraging this intuition. For example, we can construct test statistics using the TSLS estimator and the DRIVE estimator with $\rho>0$, such as the difference $\hat \beta^{\text{DRIVE}}-\hat \beta^{\text{TSLS}}$. Then we can use bootstrap-based tests, such as a bootstrapped permutation test, to assess the null hypothesis that $\hat \beta^{\text{DRIVE}}-\hat \beta^{\text{TSLS}}=0$. If we fail to reject the null hypothesis, then there is evidence that the true causal effect $\beta_{0}=\mathbf{0}$. %Similarly, the Wasserstein DRIVE estimator can be used as a robustness check for the standard IV estimation. For example, in the one-dimensional just-identified case, we can compare the estimators $\hat \beta^{\text{DRIVE}}$ and $\hat \beta^{\text{IV}}$ or test statistics based on them. Since their asymptotic distributions coincide, the statistics should be close to each other. If they differ substantially but we are certain that the true causal effect $\beta_0$ is not null, then there is evidence that we should be concerned about the validity of model assumptions in IV estimation. This reasoning could also be applied in the general setting, and it would be interesting to study the connections to sensitivity analysis. 

%\yc{I am less sure about this paragraph. When assumptions are not valid, there is nothing we can guarantee. For instance, if $\hat \beta^{\text{DRIVE}}$ and $\hat \beta^{\text{IV}}$ are changed in the same direction, then we may not be able to detect invalidity. } \zq{Good point. I think we can still emphasize that if we do see they are different, then we have evidence of invalidity. If they are the same, then we may not conclude anything.}

\cref{cor:asymptotic-distribution-coincide} can be seemingly pessimistic because it demonstrates that the asymptotic distribution of Wasserstein DRIVE could be the same as that of the TSLS in special cases. However, recall that Wasserstein DRIVE is formulated to minimize the \emph{worst-case} risk over a set of distributions that are designed to capture deviations from model assumptions. Therefore, there is not actually any \emph{a priori} reason that it should coincide with the TSLS when $\rho>0$. In this sense, the fact that the Wasserstein DRIVE is consistent with $\rho>0$ and may even coincide with TSLS is rather surprising. In the latter case, the worst-case distribution for Wasserstein DRIVE in the large sample limit must coincide with that of the standard population distribution, which may be worth further investigation. 

The asymptotic results we develop in this section provide the basis on which one can perform estimation and inference with the Wasserstein DRIVE estimator. In the next section, we study the finite sample properties of DRIVE in simulation studies and demonstrate that it is superior in terms of estimation error and out of sample prediction compared to other popular estimators. %when the instruments are mildly invalid. 

\section{Numerical Studies}
\label{sec:numerical}
In this section, we study the empirical performance of Wasserstein DRIVE. Our results deliver three main messages. First, we demonstrate with simulations that Wasserstein DRIVE, with non-zero robustness parameter $\rho$ based on \cref{thm:DRIVE-consistency}, has comparable performance as the standard IV estimator whenever instruments are valid. Second, when instruments become invalid, Wasserstein DRIVE %with $\rho$ selected using either first stage coefficient or nonparametric bootstrap estimate of the score quantile (Appendix \ref{sec:penalty-selection}) 
outperforms other methods in terms of RMSE. Third, on the education dataset of \citet{card1993using}, Wasserstein DRIVE also has superior performance at prediction for a heterogeneous target population. %These results demonstrate that Wasserstein DRIVE is easy to implement and enjoys superior empirical performance compared to the standard IV and $k$-class estimators. 
\subsection{MSE of Wasserstein DRIVE}
We use the data generating process 
\begin{align*}
\begin{split}
    Y &= X\beta_0 + Z\eta + U%\mu_Y+ \epsilon_{XY}
    \\
    X &= \gamma Z + U%\mu_X + \epsilon_{ZX}
    \\
    %Z \mid U & \sim\mathcal{N}(\mu_{Z}+\beta_{UZ}U,\Sigma_{Z})
    Z &= U\beta_{UZ} + \epsilon_Z,\\ 
    \end{split}
\end{align*}
% \begin{align*}
% U & \sim\mathcal{N}(\mu_{U},\Sigma_{U})\\
% Z & \sim\mathcal{N}(\mu_{Z},\Sigma_{Z})+\beta_{UZ}U\\
% X & =\beta_{ZX}Z+\beta_{UX}U\\
% Y & =\beta_{ZY}Z+\beta_{XY}X+\beta_{UY}U
% \end{align*}
where $U,\epsilon_Z \sim \mathcal{N}(0,\sigma^2)$ and we allow a direct effect $\eta$ from the instruments $Z$ to the outcome $Y$. Moreover, the instruments $Z$ can also be correlated with the unobserved confounder $U$ ($\beta_{UZ}\neq 0)$. We fix the true parameters and generate independent datasets from the model, varying the degree of instrument invalidity. In \cref{tab:strong-instrument}, we report the MSE of estimators averaged over 500 repeated experiments. We control the degree of instrument invalidity by varying $\eta$, the direct effect of instruments on the outcome, and $\beta_{UZ}$, the correlation between unobserved confounder and instruments. Results in \cref{tab:strong-instrument} are based on data where $\|\gamma\| \gg 0$ is large. We see that when instruments are strong, Wasserstein DRIVE performs as well as TSLS when instruments are valid, but performs significantly better than OLS, TSLS, anchor, and TSLS ridge when instruments become invalid. This suggests that DRIVE could be preferable in practice when we are concerned about instrument validity. 

\begin{table}[t]
\begin{centering}
\resizebox{0.5\textwidth}{!}{
\begin{tabular}{c|c||c|c|c|c|c}
$\eta$ & $\beta_{UZ}$ & OLS & TSLS & anchor & TSLS ridge & DRIVE\tabularnewline
\hline 
\hline 
0 & 0 & 0.21 & \textbf{0.03} & 0.19 & \textbf{0.03} & \textbf{0.03}\tabularnewline
\hline 
0.4 & 0 & 0.20 & 0.07 & 0.16 & 0.06 & \textbf{0.03}\tabularnewline
\hline 
0.4 & 0.4 & 0.26 & 0.25 & 0.24 & 0.21 & \textbf{0.07}\tabularnewline
\hline 
0.4 & 0.8 & 0.29 & 0.62 & 0.29 & 0.56 & \textbf{0.09}\tabularnewline
\hline 
0.8 & 0 & 0.26 & 0.23 & 0.23 & 0.22 & \textbf{0.06}\tabularnewline
\hline 
0.8 & 0.4 & 0.32 & 0.51 & 0.31 & 0.46 & \textbf{0.10}\tabularnewline
\hline 
0.8 & 0.8 & 0.37 & 0.82 & 0.38 & 0.81 & \textbf{0.14}\tabularnewline
\end{tabular}
}
\par\end{centering}
\caption{MSE of estimators when instruments are potentially invalid. $\beta_0=1, n=2000$, $\sigma=0.5$. %Wasserstein DRIVE consistently outperforms the other estimators. 
For TSLS ridge the regularization parameter is selected using cross validation based
on out-of-sample prediction errors. For anchor regression the regularization
parameter is selected based on the proposal in \citet{rothenhausler2021anchor}. For DRIVE the regularization parameter is selected using nonparametric bootstrap of the score quantile (Appendix \ref{sec:penalty-selection}). \zq{std information}
}
\label{tab:strong-instrument}
\end{table}

\begin{figure}[ht]
%\hspace*{-1.8cm} 
\begin{centering}
\includegraphics[scale=0.4]{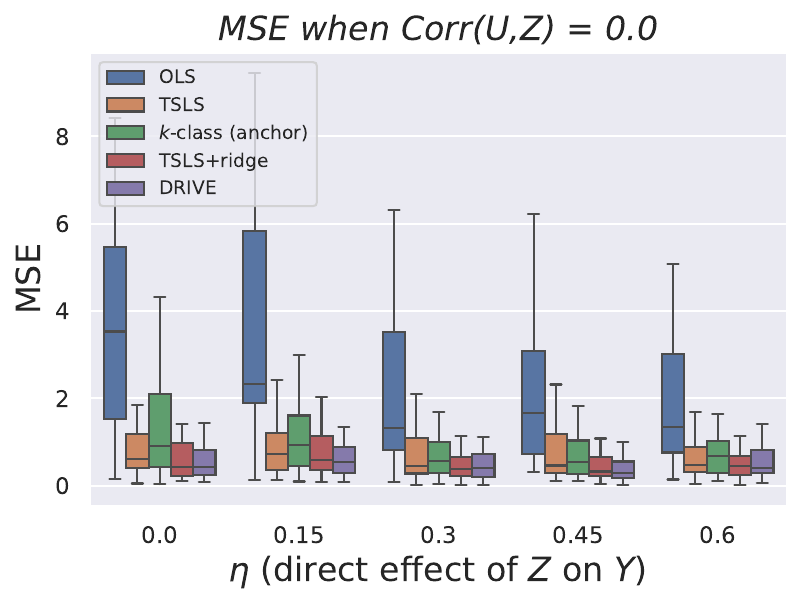}\includegraphics[scale=0.4]{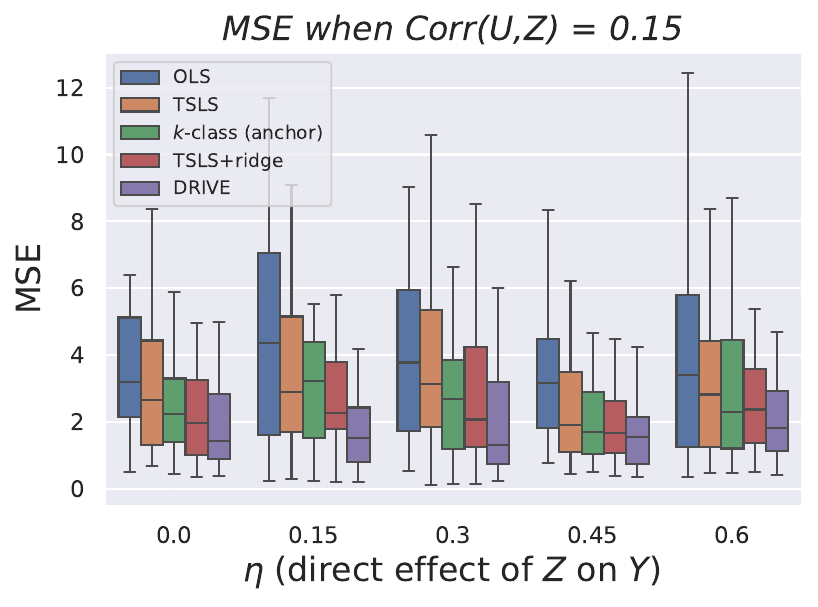}\includegraphics[scale=0.4]{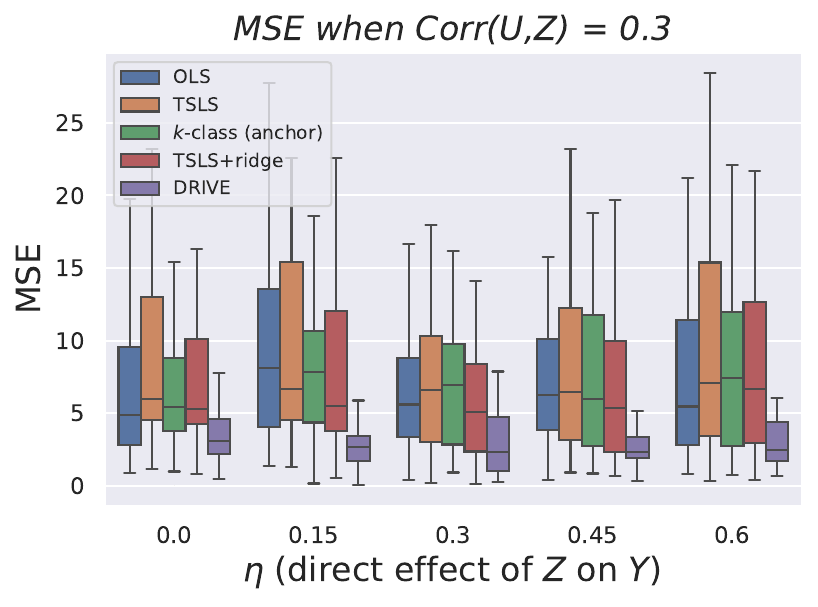}
\par\end{centering}
\caption{MSEs of estimators when instruments are potentially invalid. Instrument $Z$ can have direct effects $\eta$ on the outcome $Y$, or be correlated with the unobserved confounder $U$.
Wasserstein DRIVE consistently outperforms the other estimators.
}
\label{fig:strongly-invalid-estimation-error}
\end{figure}

We further investigate the empirical performance of Wasserstein DRIVE when instruments are potentially invalid or weak. We present box plots (omitting outliers) of MSEs in \cref{fig:strongly-invalid-estimation-error}. %and \cref{fig:mildly-invalid-estimation-error}. 
The Wasserstein DRIVE estimator with regularization parameter $\rho$ based on bootstrapped
quantiles of the score function consistently outperforms OLS, TSLS, anchor ($k$-class), and TSLS with ridge regularization. Moreover, the selected
penalties increase as the direct effect of $Z$ on $Y$ or the correlation
between the unobserved confounder $U$ and the instrument $Z$ increases,
i.e., as the model assumption of valid instruments becomes increasingly
invalid. See \cref{fig:quantile-increase-invalid-iv} in Appendix \cref{sec:penalty-selection} for more details. This property is highly desirable, because based on the DRO formulation of DRIVE, $\rho$ represents the amount of robustness against distributional shifts associated with the estimator, which
should increase as the instruments become more invalid (larger distributional shift). Box plots of estimation errors in \cref{fig:strongly-invalid-estimation-error} also verify that even when instruments are valid, the finite sample performance of Wasserstein DRIVE is still better compared to the standard IV estimator, suggesting that there is no additional cost in terms of MSE when applying Wasserstein DRIVE, even when instruments are valid.

\subsection{Prediction under Distributional Shifts on Education Data}
\zq{Add mathematical formulations of the train-test framework. }
We now turn our attention to a different task that has received more attention in recent years, especially in the context of policy learning and estimating causal effects across heterogeneous populations \citep{dehejia2021local,adjaho2022externally,menzel2023transfer}. We study the prediction performance of estimators when they are estimated on a dataset (training set) that has a potentially different distribution from the dataset for which they are used to make predictions (test set). We demonstrate that whenever the distributions between training and
test datasets have significant differences, the prediction error of Wasserstein DRIVE is significantly smaller than that of OLS, IV, and anchor ($k$-class) estimators.

We conduct our numerical study using the classic dataset on the return of education to wage compiled by David Card \citep{card1993using}. Here, the causal inference problem is estimating the effect of additional school years on the increase in wage later in life. The dataset contains demographic information about interviewed subjects. Importantly, each sample comes from one of nine regions in the United States, which differ in the average number of years of schooling and other characteristics, i.e., there are \emph{covariate shifts} in data collected from different regions. Our strategy is to divide the dataset into a training set and a test set based on the relative ranks of their average years of schools, which is the endogenous variable. We expect that if there are distributional shifts between different regions, then predicting wages using education and other information using conventional models trained on the training data may not yield a good performance on the test data.

\begin{sidewaystable}
\begin{centering}
\resizebox{\columnwidth}{!}{
\begin{tabular}{c|c|c|c|c|c|c|c}
training set (size) & test set (size) & OLS & TSLS & DRIVE & anchor regression & ridge & TSLS ridge\tabularnewline
\hline 
\hline 
\multirow{4}{*}{bottom 3 educated states (1247)} & \multirow{2}{*}{top 3 educated regions (841)} & 0.444 & 0.537 & \textbf{0.364} & 0.444 & 0.444 & 0.421\tabularnewline
 &  & (0.009) & (0.031) & \textbf{(0.002)} & (0.009) & (0.009) & (0.019)\tabularnewline
\cline{2-8} \cline{3-8} \cline{4-8} \cline{5-8} \cline{6-8} \cline{7-8} \cline{8-8} 
 & \multirow{2}{*}{top 6 educated regions (1763)} & 0.451 & 1.064 & \textbf{0.371} & 0.451 & 0.451 & 0.430\tabularnewline
 &  & (0.011) & (0.274) & \textbf{(0.003)} & (0.011) & (0.011) & (0.027)\tabularnewline 
\multirow{2}{*}{top 6 educated regions (1763)} & \multirow{2}{*}{bottom 3 educated regions (1247)} & 0.390 & 0.584 & \textbf{0.356} & 0.390 & 0.390 & 0.377\tabularnewline
 &  & (0.007) & (0.120) & \textbf{(0.002)} & (0.007) & (0.007) & (0.015)\tabularnewline
\hline 
\multirow{4}{*}{top 3 educated regions (841)} & \multirow{2}{*}{bottom 3 educated regions (1247)} & 0.389 & 1.99 & 0.355 & 0.388 & 0.359 & \textbf{0.344}\tabularnewline
 &  & (0.013) & (0.775) & (0.005) & (0.013) & (0.014) & \textbf{(0.004)}\tabularnewline
\cline{2-8} \cline{3-8} \cline{4-8} \cline{5-8} \cline{6-8} \cline{7-8} \cline{8-8} 
 & \multirow{2}{*}{middle 3 educated regions (922)} & 0.328 & 3.18 & 0.364 & 0.328 & \textbf{0.326} & 0.361\tabularnewline
 &  & (0.001) & (1.213) & (0.005) & (0.001) & \textbf{(0.001)} & (0.005)\tabularnewline
\hline 
\multirow{8}{*}{middle 3 educated regions (922)} & \multirow{2}{*}{top 3 educated regions (841)} & \textbf{0.332} & 0.410 & \textbf{0.332} & \textbf{0.332} & 0.333 & \textbf{0.332}\tabularnewline
 &  & \textbf{(0.001)} & (0.025) & \textbf{(0.001)} & \textbf{(0.001)} & (0.001) & \textbf{(0.001)}\tabularnewline
\cline{2-8} \cline{3-8} \cline{4-8} \cline{5-8} \cline{6-8} \cline{7-8} \cline{8-8} 
 & \multirow{2}{*}{bottom 3 educated regions (1247)} & 0.416 & 0.538 & \textbf{0.363} & 0.416 & 0.409 & 0.386\tabularnewline
 &  & (0.014) & (0.063) & \textbf{(0.005)} & (0.014) & (0.016) & (0.019)\tabularnewline
\cline{2-8} \cline{3-8} \cline{4-8} \cline{5-8} \cline{6-8} \cline{7-8} \cline{8-8} 
 & \multirow{2}{*}{most+least educated regions (374)} & 0.395 & 2.47 & 0.362 & 0.395 & 0.392 & \textbf{0.355}\tabularnewline
 &  & (0.001) & (1.81) & (0.004) & (0.011) & (0.012) & \textbf{(0.004)}\tabularnewline
\cline{2-8} \cline{3-8} \cline{4-8} \cline{5-8} \cline{6-8} \cline{7-8} \cline{8-8} 
 & \multirow{2}{*}{top 3+bottom 3 educated regions (2088)} & 0.396 & 0.451 & \textbf{0.358} & 0.396 & 0.382 & 0.366\tabularnewline
 &  & (0.005) & (0.032) & \textbf{(0.003)} & (0.009) & (0.006) & (0.009)\tabularnewline
\end{tabular}%
}
\par\end{centering}
\caption{Comparison of estimation methods in terms of MSE on test data. Here the training and test datasets are split according to the 9 regions
in the Card college proximity dataset based on their average education
levels. In this specification, we did not include experience squared.
Standard errors are obtained using 10 bootstrapped datasets.}
\label{tab:card_education}
\end{sidewaystable}

Since each sample is labeled
as working in 1976 in one of nine regions in the U.S., we split the samples based on these labels, using number of years
of education as the splitting variable. For example, we can construct the training set by including samples from the top 6 regions with the highest average years of schooling, and the test set to consist of samples coming from the bottom 3 regions with the lowest average years of schooling. In this case, we would expect the training and test sets to have come from different distributions. Indeed, the average years of schooling differs by more than 1 year, and is statistically significant.

In splitting the samples based on the distribution of the endogenous variable, we are also motivated by the long-standing debates revolving around the use of instrumental variables in classic economic studies \citep{card1994minimum}. A leading concern is the validity of instruments. In the case of the study on educational returns, the validity of estimation and inference require that the instruments (proximity to college and quarter of birth) are not correlated with unobserved characteristics that may also affect their earnings. The following quote from \citet{card1999causal} illustrates this concern: 
\begin{quote}
    ``In the case of quasi or natural experiments, however, inferences
are based on difference between groups of individuals who attended schools at different times, or in different locations, or had differences in other characteristics such as month of birth. The use of these differences to draw causal inferences about the effect of schooling
requires careful consideration of the maintained assumption that the groups are otherwise
identical.''
\end{quote}
When this assumption is violated, the estimates based on a particular subpopulation becomes unreliable for the wider population, and we evaluate the performance based on how well they generalize to other groups of the population with potential distributional or covariate shifts.  In Table \ref{tab:card_education}, we compare the test set MSE of OLS, IV, Wasserstein DRIVE,
anchor regression, ridge, and ridge regularized IV estimators. We see that Wasserstein DRIVE consistently outperforms other estimators commonly used in practice.

\section{Concluding Remarks}
In this paper, we propose a distributionally robust instrumental variables estimation framework. Our approach is motivated by two main considerations in practice. The first is the concern about model mis-specification in IV estimation, most notably the validity of instruments. Second, going beyond estimating the causal effect for the endogenous variable, practitioners may also be interested in making good predictions with the help of instruments when there is heterogeneity between training and test datasets, e.g., generalizing from findings using samples from a particular population/geographical group to other groups. We argue that both challenges can be naturally unified as problems of distributional shifts, and then addressed using frameworks from distributionally robust optimization. 

We provide a dual representation of our Wasserstein DRIVE framework as a regularized TSLS problem, and reveal a distinct property of the resulting estimator: it is consistent with \emph{non-vanishing} penalty parameter. %This property suggests that, unlike standard regression settings, in the IV estimation setting we can achieve robustness and asymptotic validity at the same time with a non-vanishing robustness parameter $\rho$. 
%Leveraging this result, we propose data-drive procedures to select the penalty parameter, which is demonstrated to adapt to increasing degrees of instrument invalidity. 
We further characterize the asymptotic distribution of the Wasserstein DRIVE, and establish a few special cases when it coincides with that of the standard TSLS estimator. Numerical studies suggest that Wasserstein DRIVE has superior finite sample performance in two regards. First, it has lower estimation error when instruments are potentially invalid, but performs as well as the TSLS when instruments are valid. Second, it outperforms existing methods at the task of predicting outcomes under distributional shifts between training and test data. These findings provide support for the appeal of our DRO approach to IV estimation, and suggest that Wasserstein DRIVE could be preferable in practice to standard IV methods. Finally, there are many future research directions of interest, such as further results on inference and testing,  as well as connections to sensitivity analysis. %Currently, ambiguity sets are built as a ball around the empirical distribution on the predicted outcomes and regressors. However, this construction may be too conservative, and creating more specific uncertainty sets could be more efficient. For example, if we have extra information that only a subset of the available instruments may violate IV assumptions, we can construct less conservative distributional uncertainty sets. 
Extensions to nonlinear models would also be useful in practice.

\section*{Acknowledgements}
We are indebted to Han Hong, Guido Imbens, and Yinyu Ye for invaluable advice and guidance throughout this project, and to Agostino Capponi, Timothy Cogley, Rajeev Dehejia, Yanqin Fan, Alfred Galichon, Rui Gao, Wenzhi Gao, Vishal Kamat, Samir Khan, Frederic Koehler, Michal Koles{\'a}r, Simon Sokbae Lee, Greg Lewis, Elena Manresa, Konrad Menzel, Axel Peytavin, Debraj Ray, Martin Rotemberg, Vasilis Syrgkanis, Johan Ugander, and Ruoxuan Xiong for helpful discussions and suggestions. This work was supported in part by a Stanford Interdisciplinary Graduate Fellowship (SIGF).

\bibliography{ref_DRIVE}

\begin{thebibliography}{}

\bibitem[Adjaho and Christensen, 2022]{adjaho2022externally}
Adjaho, C. and Christensen, T. (2022).
\newblock Externally valid treatment choice.
\newblock {\em arXiv preprint arXiv:2205.05561}, 1.

\bibitem[Anderson and Rubin, 1949]{anderson1949estimation}
Anderson, T.~W. and Rubin, H. (1949).
\newblock Estimation of the parameters of a single equation in a complete system of stochastic equations.
\newblock {\em The Annals of mathematical statistics}, 20(1):46--63.

\bibitem[Andrews, 1994]{andrews1994empirical}
Andrews, D.~W. (1994).
\newblock Empirical process methods in econometrics.
\newblock {\em Handbook of econometrics}, 4:2247--2294.

\bibitem[Andrews, 1999]{andrews1999consistent}
Andrews, D.~W. (1999).
\newblock Consistent moment selection procedures for generalized method of moments estimation.
\newblock {\em Econometrica}, 67(3):543--563.

\bibitem[Andrews, 2007]{andrews2007inference}
Andrews, D.~W. (2007).
\newblock Inference with weak instruments.
\newblock {\em Advances in Economics and Econometrics}, 3:122--173.

\bibitem[Andrews et~al., 2019]{andrews2019weak}
Andrews, I., Stock, J.~H., and Sun, L. (2019).
\newblock Weak instruments in instrumental variables regression: Theory and practice.
\newblock {\em Annual Review of Economics}, 11(1).

\bibitem[Angrist et~al., 1996]{angrist1996identification}
Angrist, J.~D., Imbens, G.~W., and Rubin, D.~B. (1996).
\newblock Identification of causal effects using instrumental variables.
\newblock {\em Journal of the American statistical Association}, 91(434):444--455.

\bibitem[Angrist and Pischke, 2009]{angrist2009mostly}
Angrist, J.~D. and Pischke, J.-S. (2009).
\newblock {\em Mostly harmless econometrics: An empiricist's companion}.
\newblock Princeton university press.

\bibitem[Armstrong and Koles{\'a}r, 2021]{armstrong2021sensitivity}
Armstrong, T.~B. and Koles{\'a}r, M. (2021).
\newblock Sensitivity analysis using approximate moment condition models.
\newblock {\em Quantitative Economics}, 12(1):77--108.

\bibitem[Basmann, 1960a]{basmann1960finite}
Basmann, R.~L. (1960a).
\newblock On finite sample distributions of generalized classical linear identifiability test statistics.
\newblock {\em Journal of the American Statistical Association}, 55(292):650--659.

\bibitem[Basmann, 1960b]{basmann1960asymptotic}
Basmann, R.~L. (1960b).
\newblock On the asymptotic distribution of generalized linear estimators.
\newblock {\em Econometrica, Journal of the Econometric Society}, pages 97--107.

\bibitem[Baum et~al., 2003]{baum2003instrumental}
Baum, C.~F., Schaffer, M.~E., and Stillman, S. (2003).
\newblock Instrumental variables and gmm: Estimation and testing.
\newblock {\em The Stata Journal}, 3(1):1--31.

\bibitem[Bekker, 1994]{bekker1994alternative}
Bekker, P.~A. (1994).
\newblock Alternative approximations to the distributions of instrumental variable estimators.
\newblock {\em Econometrica: Journal of the Econometric Society}, pages 657--681.

\bibitem[Belloni et~al., 2012]{belloni2012sparse}
Belloni, A., Chen, D., Chernozhukov, V., and Hansen, C. (2012).
\newblock Sparse models and methods for optimal instruments with an application to eminent domain.
\newblock {\em Econometrica}, 80(6):2369--2429.

\bibitem[Belloni et~al., 2018]{belloni2018high}
Belloni, A., Chernozhukov, V., Chetverikov, D., Hansen, C., and Kato, K. (2018).
\newblock High-dimensional econometrics and regularized gmm.
\newblock {\em arXiv preprint arXiv:1806.01888}.

\bibitem[Belloni et~al., 2011]{belloni2011square}
Belloni, A., Chernozhukov, V., and Wang, L. (2011).
\newblock Square-root lasso: pivotal recovery of sparse signals via conic programming.
\newblock {\em Biometrika}, 98(4):791--806.

\bibitem[Ben-Tal et~al., 2013]{ben2013robust}
Ben-Tal, A., Den~Hertog, D., De~Waegenaere, A., Melenberg, B., and Rennen, G. (2013).
\newblock Robust solutions of optimization problems affected by uncertain probabilities.
\newblock {\em Management Science}, 59(2):341--357.

\bibitem[Bennett and Kallus, 2023]{bennett2023variational}
Bennett, A. and Kallus, N. (2023).
\newblock The variational method of moments.
\newblock {\em Journal of the Royal Statistical Society Series B: Statistical Methodology}, 85(3):810--841.

\bibitem[Berkowitz et~al., 2008]{berkowitz2008nearly}
Berkowitz, D., Caner, M., and Fang, Y. (2008).
\newblock Are “nearly exogenous instruments” reliable?
\newblock {\em Economics Letters}, 101(1):20--23.

\bibitem[Bertsimas and Copenhaver, 2018]{bertsimas2018characterization}
Bertsimas, D. and Copenhaver, M.~S. (2018).
\newblock Characterization of the equivalence of robustification and regularization in linear and matrix regression.
\newblock {\em European Journal of Operational Research}, 270(3):931--942.

\bibitem[Bertsimas et~al., 2022]{bertsimas2022distributionally}
Bertsimas, D., Imai, K., and Li, M.~L. (2022).
\newblock Distributionally robust causal inference with observational data.
\newblock {\em arXiv preprint arXiv:2210.08326}.

\bibitem[Bertsimas and Popescu, 2005]{bertsimas2005optimal}
Bertsimas, D. and Popescu, I. (2005).
\newblock Optimal inequalities in probability theory: A convex optimization approach.
\newblock {\em SIAM Journal on Optimization}, 15(3):780--804.

\bibitem[Bickel et~al., 2009]{bickel2009simultaneous}
Bickel, P.~J., Ritov, Y., and Tsybakov, A.~B. (2009).
\newblock {Simultaneous analysis of Lasso and Dantzig selector}.
\newblock {\em The Annals of Statistics}, 37(4):1705 -- 1732.

\bibitem[Blanchet et~al., 2019]{blanchet2019robust}
Blanchet, J., Kang, Y., and Murthy, K. (2019).
\newblock Robust wasserstein profile inference and applications to machine learning.
\newblock {\em Journal of Applied Probability}, 56(3):830--857.

\bibitem[Blanchet and Murthy, 2019]{blanchet2019quantifying}
Blanchet, J. and Murthy, K. (2019).
\newblock Quantifying distributional model risk via optimal transport.
\newblock {\em Mathematics of Operations Research}, 44(2):565--600.

\bibitem[Blanchet et~al., 2022]{blanchet2022confidence}
Blanchet, J., Murthy, K., and Si, N. (2022).
\newblock Confidence regions in wasserstein distributionally robust estimation.
\newblock {\em Biometrika}, 109(2):295--315.

\bibitem[Bonhomme and Weidner, 2022]{bonhomme2022minimizing}
Bonhomme, S. and Weidner, M. (2022).
\newblock Minimizing sensitivity to model misspecification.
\newblock {\em Quantitative Economics}, 13(3):907--954.

\bibitem[Bound et~al., 1995]{bound1995problems}
Bound, J., Jaeger, D.~A., and Baker, R.~M. (1995).
\newblock Problems with instrumental variables estimation when the correlation between the instruments and the endogenous explanatory variable is weak.
\newblock {\em Journal of the American statistical association}, 90(430):443--450.

\bibitem[Bowden et~al., 2015]{bowden2015mendelian}
Bowden, J., Davey~Smith, G., and Burgess, S. (2015).
\newblock Mendelian randomization with invalid instruments: effect estimation and bias detection through egger regression.
\newblock {\em International journal of epidemiology}, 44(2):512--525.

\bibitem[Bowden et~al., 2016]{bowden2016consistent}
Bowden, J., Davey~Smith, G., Haycock, P.~C., and Burgess, S. (2016).
\newblock Consistent estimation in mendelian randomization with some invalid instruments using a weighted median estimator.
\newblock {\em Genetic epidemiology}, 40(4):304--314.

\bibitem[B{\"u}hlmann, 2020]{buhlmann2020invariance}
B{\"u}hlmann, P. (2020).
\newblock Invariance, causality and robustness.
\newblock {\em Statistical Science}, 35(3):404--426.

\bibitem[Burgess et~al., 2016]{burgess2016robust}
Burgess, S., Bowden, J., Dudbridge, F., and Thompson, S.~G. (2016).
\newblock Robust instrumental variable methods using multiple candidate instruments with application to mendelian randomization.
\newblock {\em arXiv preprint arXiv:1606.03729}.

\bibitem[Burgess et~al., 2020]{burgess2020robust}
Burgess, S., Foley, C.~N., Allara, E., Staley, J.~R., and Howson, J.~M. (2020).
\newblock A robust and efficient method for mendelian randomization with hundreds of genetic variants.
\newblock {\em Nature communications}, 11(1):376.

\bibitem[Burgess et~al., 2011]{burgess2011avoiding}
Burgess, S., Thompson, S.~G., and Collaboration, C. C.~G. (2011).
\newblock Avoiding bias from weak instruments in mendelian randomization studies.
\newblock {\em International journal of epidemiology}, 40(3):755--764.

\bibitem[Calafiore and Ghaoui, 2006]{calafiore2006distributionally}
Calafiore, G.~C. and Ghaoui, L.~E. (2006).
\newblock On distributionally robust chance-constrained linear programs.
\newblock {\em Journal of Optimization Theory and Applications}, 130:1--22.

\bibitem[Caner, 2009]{caner2009lasso}
Caner, M. (2009).
\newblock Lasso-type gmm estimator.
\newblock {\em Econometric Theory}, 25(1):270--290.

\bibitem[Caner and Kock, 2018]{caner2018high}
Caner, M. and Kock, A.~B. (2018).
\newblock High dimensional linear gmm.
\newblock {\em arXiv preprint arXiv:1811.08779}.

\bibitem[Card, 1993]{card1993using}
Card, D. (1993).
\newblock Using geographic variation in college proximity to estimate the return to schooling.
\newblock {\em NBER Working Paper}, (w4483).

\bibitem[Card, 1999]{card1999causal}
Card, D. (1999).
\newblock The causal effect of education on earnings.
\newblock {\em Handbook of labor economics}, 3:1801--1863.

\bibitem[Card and Krueger, 1994]{card1994minimum}
Card, D. and Krueger, A.~B. (1994).
\newblock Minimum wages and employment: A case study of the fast-food industry in new jersey and pennsylvania.
\newblock {\em American Economic Review}, 84(4).

\bibitem[Chamberlain and Imbens, 2004]{chamberlain2004random}
Chamberlain, G. and Imbens, G. (2004).
\newblock Random effects estimators with many instrumental variables.
\newblock {\em Econometrica}, 72(1):295--306.

\bibitem[Chao and Swanson, 2005]{chao2005consistent}
Chao, J.~C. and Swanson, N.~R. (2005).
\newblock Consistent estimation with a large number of weak instruments.
\newblock {\em Econometrica}, 73(5):1673--1692.

\bibitem[Chen et~al., 2021]{chen2021robust}
Chen, X., Hansen, L.~P., and Hansen, P.~G. (2021).
\newblock Robust inference for moment condition models without rational expectations.
\newblock {\em Journal of Econometrics, forthcoming}.

\bibitem[Chernozhukov et~al., 2015]{chernozhukov2015post}
Chernozhukov, V., Hansen, C., and Spindler, M. (2015).
\newblock Post-selection and post-regularization inference in linear models with many controls and instruments.
\newblock {\em American Economic Review}, 105(5):486--490.

\bibitem[Cigliutti and Manresa, 2022]{cigliutti2022adversarial}
Cigliutti, I. and Manresa, E. (2022).
\newblock Adversarial method of moments.

\bibitem[Conley et~al., 2012]{conley2012plausibly}
Conley, T.~G., Hansen, C.~B., and Rossi, P.~E. (2012).
\newblock Plausibly exogenous.
\newblock {\em Review of Economics and Statistics}, 94(1):260--272.

\bibitem[Cragg and Donald, 1993]{cragg1993testing}
Cragg, J.~G. and Donald, S.~G. (1993).
\newblock Testing identifiability and specification in instrumental variable models.
\newblock {\em Econometric Theory}, 9(2):222--240.

\bibitem[Davey~Smith and Ebrahim, 2003]{davey2003mendelian}
Davey~Smith, G. and Ebrahim, S. (2003).
\newblock ‘mendelian randomization’: can genetic epidemiology contribute to understanding environmental determinants of disease?
\newblock {\em International journal of epidemiology}, 32(1):1--22.

\bibitem[Dehejia et~al., 2021]{dehejia2021local}
Dehejia, R., Pop-Eleches, C., and Samii, C. (2021).
\newblock From local to global: External validity in a fertility natural experiment.
\newblock {\em Journal of Business \& Economic Statistics}, 39(1):217--243.

\bibitem[Delage and Ye, 2010]{delage2010distributionally}
Delage, E. and Ye, Y. (2010).
\newblock Distributionally robust optimization under moment uncertainty with application to data-driven problems.
\newblock {\em Operations research}, 58(3):595--612.

\bibitem[Duchi et~al., 2021]{duchi2021statistics}
Duchi, J.~C., Glynn, P.~W., and Namkoong, H. (2021).
\newblock Statistics of robust optimization: A generalized empirical likelihood approach.
\newblock {\em Mathematics of Operations Research}, 46(3):946--969.

\bibitem[Dupa{\v{c}}ov{\'a}, 1987]{dupavcova1987minimax}
Dupa{\v{c}}ov{\'a}, J. (1987).
\newblock The minimax approach to stochastic programming and an illustrative application.
\newblock {\em Stochastics: An International Journal of Probability and Stochastic Processes}, 20(1):73--88.

\bibitem[El~Ghaoui and Lebret, 1997]{el1997robust}
El~Ghaoui, L. and Lebret, H. (1997).
\newblock Robust solutions to least-squares problems with uncertain data.
\newblock {\em SIAM Journal on matrix analysis and applications}, 18(4):1035--1064.

\bibitem[Emdin et~al., 2017]{emdin2017mendelian}
Emdin, C.~A., Khera, A.~V., and Kathiresan, S. (2017).
\newblock Mendelian randomization.
\newblock {\em Jama}, 318(19):1925--1926.

\bibitem[Fan et~al., 2024]{fan2024environment}
Fan, J., Fang, C., Gu, Y., and Zhang, T. (2024).
\newblock Environment invariant linear least squares.
\newblock {\em The Annals of Statistics}, 52(5):2268--2292.

\bibitem[Fan et~al., 2023]{fan2023quantifying}
Fan, Y., Park, H., and Xu, G. (2023).
\newblock Quantifying distributional model risk in marginal problems via optimal transport.
\newblock {\em arXiv preprint arXiv:2307.00779}.

\bibitem[Fisher, 1961]{fisher1961cost}
Fisher, F.~M. (1961).
\newblock On the cost of approximate specification in simultaneous equation estimation.
\newblock {\em Econometrica: journal of the Econometric Society}, pages 139--170.

\bibitem[Fu and Knight, 2000]{fu2000asymptotics}
Fu, W. and Knight, K. (2000).
\newblock Asymptotics for lasso-type estimators.
\newblock {\em The Annals of statistics}, 28(5):1356--1378.

\bibitem[Fuller, 1977]{fuller1977some}
Fuller, W.~A. (1977).
\newblock Some properties of a modification of the limited information estimator.
\newblock {\em Econometrica: Journal of the Econometric Society}, pages 939--953.

\bibitem[Galichon, 2018]{galichon2018optimal}
Galichon, A. (2018).
\newblock {\em Optimal transport methods in economics}.
\newblock Princeton University Press.

\bibitem[Galichon, 2021]{galichon2021unreasonable}
Galichon, A. (2021).
\newblock The unreasonable effectiveness of optimal transport in economics.
\newblock {\em arXiv preprint arXiv:2107.04700}.

\bibitem[Gao and Kleywegt, 2023]{gao2016distributionally}
Gao, R. and Kleywegt, A. (2023).
\newblock Distributionally robust stochastic optimization with wasserstein distance.
\newblock {\em Mathematics of Operations Research}, 48(2):603--655.

\bibitem[Goodfellow et~al., 2014]{goodfellow2014generative}
Goodfellow, I., Pouget-Abadie, J., Mirza, M., Xu, B., Warde-Farley, D., Ozair, S., Courville, A., and Bengio, Y. (2014).
\newblock Generative adversarial nets.
\newblock {\em Advances in neural information processing systems}, 27.

\bibitem[Guo et~al., 2018a]{guo2018testing}
Guo, Z., Kang, H., Cai, T.~T., and Small, D.~S. (2018a).
\newblock Testing endogeneity with high dimensional covariates.
\newblock {\em Journal of Econometrics}, 207(1):175--187.

\bibitem[Guo et~al., 2018b]{guo2018confidence}
Guo, Z., Kang, H., Tony~Cai, T., and Small, D.~S. (2018b).
\newblock Confidence intervals for causal effects with invalid instruments by using two-stage hard thresholding with voting.
\newblock {\em Journal of the Royal Statistical Society Series B: Statistical Methodology}, 80(4):793--815.

\bibitem[Hahn and Hausman, 2005]{hahn2005estimation}
Hahn, J. and Hausman, J. (2005).
\newblock Estimation with valid and invalid instruments.
\newblock {\em Annales d'Economie et de Statistique}, pages 25--57.

\bibitem[Hahn et~al., 2004]{hahn2004estimation}
Hahn, J., Hausman, J., and Kuersteiner, G. (2004).
\newblock Estimation with weak instruments: Accuracy of higher-order bias and mse approximations.
\newblock {\em The Econometrics Journal}, 7(1):272--306.

\bibitem[Hall, 2003]{hall2003generalized}
Hall, A.~R. (2003).
\newblock Generalized method of moments.
\newblock {\em A companion to theoretical econometrics}, pages 230--255.

\bibitem[Hansen, 1982]{hansen1982large}
Hansen, L.~P. (1982).
\newblock Large sample properties of generalized method of moments estimators.
\newblock {\em Econometrica: Journal of the econometric society}, pages 1029--1054.

\bibitem[Hansen and Sargent, 2008]{hansen2008robustness}
Hansen, L.~P. and Sargent, T.~J. (2008).
\newblock {\em Robustness}.
\newblock Princeton university press.

\bibitem[Hansen and Sargent, 2010]{hansen2010wanting}
Hansen, L.~P. and Sargent, T.~J. (2010).
\newblock Wanting robustness in macroeconomics.
\newblock In {\em Handbook of monetary economics}, volume~3, pages 1097--1157. Elsevier.

\bibitem[Hausman et~al., 2012]{hausman2012instrumental}
Hausman, J.~A., Newey, W.~K., Woutersen, T., Chao, J.~C., and Swanson, N.~R. (2012).
\newblock Instrumental variable estimation with heteroskedasticity and many instruments.
\newblock {\em Quantitative Economics}, 3(2):211--255.

\bibitem[Hoerl and Kennard, 1970]{hoerl1970ridge}
Hoerl, A.~E. and Kennard, R.~W. (1970).
\newblock Ridge regression: applications to nonorthogonal problems.
\newblock {\em Technometrics}, 12(1):69--82.

\bibitem[Hu and Hong, 2013]{hu2013kullback}
Hu, Z. and Hong, L.~J. (2013).
\newblock Kullback-leibler divergence constrained distributionally robust optimization.
\newblock {\em Available at Optimization Online}, pages 1695--1724.

\bibitem[Huber, 1964]{huber1964robust}
Huber, P.~J. (1964).
\newblock Robust estimation of a location parameter.
\newblock {\em The Annals of Mathematical Statistics}, 35(1):73--101.

\bibitem[Huber and Ronchetti, 2011]{huber2011robust}
Huber, P.~J. and Ronchetti, E.~M. (2011).
\newblock {\em Robust statistics}.
\newblock John Wiley \& Sons.

\bibitem[Imbens and Angrist, 1994]{imbens1994identification}
Imbens, G.~W. and Angrist, J.~D. (1994).
\newblock Identification and estimation of local average treatment effects.
\newblock {\em Econometrica}, 62(2):467--475.

\bibitem[Imbens and Rubin, 2015]{imbens2015causal}
Imbens, G.~W. and Rubin, D.~B. (2015).
\newblock {\em Causal inference in statistics, social, and biomedical sciences}.
\newblock Cambridge university press.

\bibitem[Iyengar, 2005]{iyengar2005robust}
Iyengar, G.~N. (2005).
\newblock Robust dynamic programming.
\newblock {\em Mathematics of Operations Research}, 30(2):257--280.

\bibitem[Jakobsen and Peters, 2022]{jakobsen2022distributional}
Jakobsen, M.~E. and Peters, J. (2022).
\newblock Distributional robustness of k-class estimators and the pulse.
\newblock {\em The Econometrics Journal}, 25(2):404--432.

\bibitem[Jiang, 2017]{jiang2017have}
Jiang, W. (2017).
\newblock Have instrumental variables brought us closer to the truth.
\newblock {\em Review of Corporate Finance Studies}, 6(2):127--140.

\bibitem[Kadane and Anderson, 1977]{kadane1977comment}
Kadane, J.~B. and Anderson, T. (1977).
\newblock A comment on the test of overidentifying restrictions.
\newblock {\em Econometrica: Journal of the Econometric Society}, pages 1027--1031.

\bibitem[Kaji et~al., 2020]{kaji2020adversarial}
Kaji, T., Manresa, E., and Pouliot, G. (2020).
\newblock An adversarial approach to structural estimation.
\newblock {\em arXiv preprint arXiv:2007.06169}.

\bibitem[Kallus and Zhou, 2021]{kallus2021minimax}
Kallus, N. and Zhou, A. (2021).
\newblock Minimax-optimal policy learning under unobserved confounding.
\newblock {\em Management Science}, 67(5):2870--2890.

\bibitem[Kang et~al., 2016]{kang2016instrumental}
Kang, H., Zhang, A., Cai, T.~T., and Small, D.~S. (2016).
\newblock Instrumental variables estimation with some invalid instruments and its application to mendelian randomization.
\newblock {\em Journal of the American statistical Association}, 111(513):132--144.

\bibitem[Kantorovich, 1942]{kantorovich1942translocation}
Kantorovich, L.~V. (1942).
\newblock On the translocation of masses.
\newblock In {\em Dokl. Akad. Nauk. USSR (NS)}, volume~37, pages 199--201.

\bibitem[Kantorovich, 1960]{kantorovich1960mathematical}
Kantorovich, L.~V. (1960).
\newblock Mathematical methods of organizing and planning production.
\newblock {\em Management science}, 6(4):366--422.

\bibitem[Kitamura et~al., 2013]{kitamura2013robustness}
Kitamura, Y., Otsu, T., and Evdokimov, K. (2013).
\newblock Robustness, infinitesimal neighborhoods, and moment restrictions.
\newblock {\em Econometrica}, 81(3):1185--1201.

\bibitem[Koles{\'a}r, 2018]{kolesar2018minimum}
Koles{\'a}r, M. (2018).
\newblock Minimum distance approach to inference with many instruments.
\newblock {\em Journal of Econometrics}, 204(1):86--100.

\bibitem[Koles{\'a}r et~al., 2015]{kolesar2015identification}
Koles{\'a}r, M., Chetty, R., Friedman, J., Glaeser, E., and Imbens, G.~W. (2015).
\newblock Identification and inference with many invalid instruments.
\newblock {\em Journal of Business \& Economic Statistics}, 33(4):474--484.

\bibitem[Koopmans, 1949]{koopmans1949optimum}
Koopmans, T.~C. (1949).
\newblock Optimum utilization of the transportation system.
\newblock {\em Econometrica: Journal of the Econometric Society}, pages 136--146.

\bibitem[Kuhn et~al., 2019]{kuhn2019wasserstein}
Kuhn, D., Esfahani, P.~M., Nguyen, V.~A., and Shafieezadeh-Abadeh, S. (2019).
\newblock Wasserstein distributionally robust optimization: Theory and applications in machine learning.
\newblock In {\em Operations research \& management science in the age of analytics}, pages 130--166. Informs.

\bibitem[Kunitomo, 1980]{kunitomo1980asymptotic}
Kunitomo, N. (1980).
\newblock Asymptotic expansions of the distributions of estimators in a linear functional relationship and simultaneous equations.
\newblock {\em Journal of the American Statistical Association}, 75(371):693--700.

\bibitem[Lei et~al., 2023]{lei2023policy}
Lei, L., Sahoo, R., and Wager, S. (2023).
\newblock Policy learning under biased sample selection.
\newblock {\em arXiv preprint arXiv:2304.11735}.

\bibitem[Lewis and Syrgkanis, 2018]{lewis2018adversarial}
Lewis, G. and Syrgkanis, V. (2018).
\newblock Adversarial generalized method of moments.
\newblock {\em arXiv preprint arXiv:1803.07164}.

\bibitem[McDonald, 1977]{mcdonald1977k}
McDonald, J.~B. (1977).
\newblock The k-class estimators as least variance difference estimators.
\newblock {\em Econometrica: Journal of the Econometric Society}, pages 759--763.

\bibitem[Meinshausen, 2018]{meinshausen2018causality}
Meinshausen, N. (2018).
\newblock Causality from a distributional robustness point of view.
\newblock In {\em 2018 IEEE Data Science Workshop (DSW)}, pages 6--10. IEEE.

\bibitem[Menzel, 2023]{menzel2023transfer}
Menzel, K. (2023).
\newblock Transfer estimates for causal effects across heterogeneous sites.
\newblock {\em arXiv preprint arXiv:2305.01435}.

\bibitem[Metzger, 2022]{metzger2022adversarial}
Metzger, J. (2022).
\newblock Adversarial estimators.
\newblock {\em arXiv preprint arXiv:2204.10495}.

\bibitem[Mohajerin~Esfahani and Kuhn, 2018]{mohajerin2018data}
Mohajerin~Esfahani, P. and Kuhn, D. (2018).
\newblock Data-driven distributionally robust optimization using the wasserstein metric: performance guarantees and tractable reformulations.
\newblock {\em Mathematical Programming}, 171(1-2):115--166.

\bibitem[Murray, 2006]{murray2006avoiding}
Murray, M.~P. (2006).
\newblock Avoiding invalid instruments and coping with weak instruments.
\newblock {\em Journal of economic Perspectives}, 20(4):111--132.

\bibitem[Nagar, 1959]{nagar1959bias}
Nagar, A.~L. (1959).
\newblock The bias and moment matrix of the general k-class estimators of the parameters in simultaneous equations.
\newblock {\em Econometrica: Journal of the Econometric Society}, pages 575--595.

\bibitem[Nelson and Startz, 1990a]{nelson1990distribution}
Nelson, C.~R. and Startz, R. (1990a).
\newblock The distribution of the instrumental variables estimator and its t-ratio when the instrument is a poor one.
\newblock {\em Journal of Business}, pages S125--S140.

\bibitem[Nelson and Startz, 1990b]{nelson1990some}
Nelson, C.~R. and Startz, R. (1990b).
\newblock Some further results on the exact small sample properties of the instrumental variable estimator.
\newblock {\em Econometrica: Journal of the Econometric Society}, pages 967--976.

\bibitem[Olkin and Pukelsheim, 1982]{olkin1982distance}
Olkin, I. and Pukelsheim, F. (1982).
\newblock The distance between two random vectors with given dispersion matrices.
\newblock {\em Linear Algebra and its Applications}, 48:257--263.

\bibitem[Owen, 2007]{owen2007robust}
Owen, A.~B. (2007).
\newblock A robust hybrid of lasso and ridge regression.
\newblock {\em Contemporary Mathematics}, 443(7):59--72.

\bibitem[Peters et~al., 2016]{peters2016causal}
Peters, J., Bühlmann, P., and Meinshausen, N. (2016).
\newblock {Causal Inference by using Invariant Prediction: Identification and Confidence Intervals}.
\newblock {\em Journal of the Royal Statistical Society Series B: Statistical Methodology}, 78(5):947--1012.

\bibitem[Pollard, 1991]{pollard1991asymptotics}
Pollard, D. (1991).
\newblock Asymptotics for least absolute deviation regression estimators.
\newblock {\em Econometric Theory}, 7(2):186--199.

\bibitem[Pr{\'e}kopa, 2013]{prekopa2013stochastic}
Pr{\'e}kopa, A. (2013).
\newblock {\em Stochastic programming}, volume 324.
\newblock Springer Science \& Business Media.

\bibitem[Richardson, 1968]{richardson1968exact}
Richardson, D.~H. (1968).
\newblock The exact distribution of a structural coefficient estimator.
\newblock {\em Journal of the American Statistical Association}, 63(324):1214--1226.

\bibitem[Rosenbaum and Rubin, 1983]{rosenbaum1983assessing}
Rosenbaum, P.~R. and Rubin, D.~B. (1983).
\newblock Assessing sensitivity to an unobserved binary covariate in an observational study with binary outcome.
\newblock {\em Journal of the Royal Statistical Society: Series B (Methodological)}, 45(2):212--218.

\bibitem[Rothenh{\"a}usler et~al., 2021]{rothenhausler2021anchor}
Rothenh{\"a}usler, D., Meinshausen, N., B{\"u}hlmann, P., Peters, J., et~al. (2021).
\newblock Anchor regression: Heterogeneous data meet causality.
\newblock {\em Journal of the Royal Statistical Society Series B}, 83(2):215--246.

\bibitem[Ruszczy{\'n}ski, 2010]{ruszczynski2010risk}
Ruszczy{\'n}ski, A. (2010).
\newblock Risk-averse dynamic programming for markov decision processes.
\newblock {\em Mathematical programming}, 125:235--261.

\bibitem[Sahoo et~al., 2022]{sahoo2022learning}
Sahoo, R., Lei, L., and Wager, S. (2022).
\newblock Learning from a biased sample.
\newblock {\em arXiv preprint arXiv:2209.01754}.

\bibitem[Sanderson and Windmeijer, 2016]{sanderson2016weak}
Sanderson, E. and Windmeijer, F. (2016).
\newblock A weak instrument f-test in linear iv models with multiple endogenous variables.
\newblock {\em Journal of econometrics}, 190(2):212--221.

\bibitem[Sargan, 1958]{sargan1958estimation}
Sargan, J.~D. (1958).
\newblock The estimation of economic relationships using instrumental variables.
\newblock {\em Econometrica: Journal of the econometric society}, pages 393--415.

\bibitem[Scarf, 1958]{scarf1958min}
Scarf, H. (1958).
\newblock A min-max solution of an inventory problem.
\newblock {\em Studies in the mathematical theory of inventory and production}.

\bibitem[Shapiro and Kleywegt, 2002]{shapiro2002minimax}
Shapiro, A. and Kleywegt, A. (2002).
\newblock Minimax analysis of stochastic problems.
\newblock {\em Optimization Methods and Software}, 17(3):523--542.

\bibitem[Sherman and Morrison, 1950]{sherman1950adjustment}
Sherman, J. and Morrison, W.~J. (1950).
\newblock Adjustment of an inverse matrix corresponding to a change in one element of a given matrix.
\newblock {\em The Annals of Mathematical Statistics}, 21(1):124--127.

\bibitem[Sinha et~al., 2017]{sinha2017certifying}
Sinha, A., Namkoong, H., Volpi, R., and Duchi, J. (2017).
\newblock Certifying some distributional robustness with principled adversarial training.
\newblock {\em arXiv preprint arXiv:1710.10571}.

\bibitem[Small, 2007]{small2007sensitivity}
Small, D.~S. (2007).
\newblock Sensitivity analysis for instrumental variables regression with overidentifying restrictions.
\newblock {\em Journal of the American Statistical Association}, 102(479):1049--1058.

\bibitem[Staiger and Stock, 1997]{staiger1997instrumental}
Staiger, D. and Stock, J.~H. (1997).
\newblock Instrumental variables regression with weak instruments.
\newblock {\em Econometrica: Journal of the Econometric Society}, pages 557--586.

\bibitem[Stock and Wright, 2000]{stock2000gmm}
Stock, J.~H. and Wright, J.~H. (2000).
\newblock Gmm with weak identification.
\newblock {\em Econometrica}, 68(5):1055--1096.

\bibitem[Stock et~al., 2002]{stock2002survey}
Stock, J.~H., Wright, J.~H., and Yogo, M. (2002).
\newblock A survey of weak instruments and weak identification in generalized method of moments.
\newblock {\em Journal of Business \& Economic Statistics}, 20(4):518--529.

\bibitem[Stock and Yogo, 2002]{stock2002testing}
Stock, J.~H. and Yogo, M. (2002).
\newblock Testing for weak instruments in linear iv regression.

\bibitem[Theil, 1953]{theil1953repeated}
Theil, H. (1953).
\newblock Repeated least squares applied to complete equation systems.
\newblock {\em The Hague: central planning bureau}.

\bibitem[Theil, 1961]{theil1961economic}
Theil, H. (1961).
\newblock Economic forecasts and policy.

\bibitem[Tibshirani, 1996]{tibshirani1996regression}
Tibshirani, R. (1996).
\newblock Regression shrinkage and selection via the lasso.
\newblock {\em Journal of the Royal Statistical Society Series B: Statistical Methodology}, 58(1):267--288.

\bibitem[van~der Vaart and Wellner, 1996]{van_der_vaart1996}
van~der Vaart, A.~W. and Wellner, J.~A. (1996).
\newblock {\em Weak Convergence and Empirical Processes: With Applications to Statistics}.
\newblock Springer.

\bibitem[VanderWeele et~al., 2014]{vanderweele2014methodological}
VanderWeele, T.~J., Tchetgen, E. J.~T., Cornelis, M., and Kraft, P. (2014).
\newblock Methodological challenges in mendelian randomization.
\newblock {\em Epidemiology (Cambridge, Mass.)}, 25(3):427.

\bibitem[Vaserstein, 1969]{vaserstein1969markov}
Vaserstein, L.~N. (1969).
\newblock Markov processes over denumerable products of spaces, describing large systems of automata.
\newblock {\em Problemy Peredachi Informatsii}, 5(3):64--72.

\bibitem[Vershik, 2013]{vershik2013long}
Vershik, A.~M. (2013).
\newblock Long history of the monge-kantorovich transportation problem: (marking the centennial of lv kantorovich’s birth!).
\newblock {\em The Mathematical Intelligencer}, 35:1--9.

\bibitem[Villani, 2009]{villani2009optimal}
Villani, C. (2009).
\newblock {\em Optimal transport: old and new}, volume 338.
\newblock Springer.

\bibitem[Von~Neumann and Morgenstern, 1947]{von1947theory}
Von~Neumann, J. and Morgenstern, O. (1947).
\newblock Theory of games and economic behavior, 2nd rev.

\bibitem[Wang et~al., 2016]{wang2016likelihood}
Wang, Z., Glynn, P.~W., and Ye, Y. (2016).
\newblock Likelihood robust optimization for data-driven problems.
\newblock {\em Computational Management Science}, 13:241--261.

\bibitem[White, 1980]{white1980heteroskedasticity}
White, H. (1980).
\newblock A heteroskedasticity-consistent covariance matrix estimator and a direct test for heteroskedasticity.
\newblock {\em Econometrica}, pages 817--838.

\bibitem[Windmeijer et~al., 2018]{windmeijer2018use}
Windmeijer, F., Farbmacher, H., Davies, N., and Smith, G.~D. (2018).
\newblock On the use of the lasso for instrumental variables estimation with some invalid instruments.
\newblock {\em Journal of the American Statistical Association}.

\bibitem[Windmeijer et~al., 2021]{windmeijer2021confidence}
Windmeijer, F., Liang, X., Hartwig, F.~P., and Bowden, J. (2021).
\newblock The confidence interval method for selecting valid instrumental variables.
\newblock {\em Journal of the Royal Statistical Society Series B: Statistical Methodology}, 83(4):752--776.

\bibitem[Wooldridge, 2020]{wooldridge2020introductory}
Wooldridge, J.~M. (2020).
\newblock Introductory econometrics: a modern approach.

\bibitem[Young, 2022]{young2022consistency}
Young, A. (2022).
\newblock Consistency without inference: Instrumental variables in practical application.
\newblock {\em European Economic Review}, page 104112.

\end{thebibliography}
\bibliographystyle{apalike}	
% {\spacingset{1.45}
% \putbib[ref_DRIVE]}

% \end{bibunit}

% \begin{bibunit}[apalike]
\begin{appendices}
    \newpage
\section{Background and Preliminaries}
\subsection{Wasserstein Distributionally Robust Optimization}
\label{subsec:Wasserstein}
We first formally define the Wasserstein distance and discuss relevant results useful in this paper. The Wasserstein distance is a metric on the space of probability distributions defined based on the optimal transport problem. More specifically, given any Polish space $\mathcal{X}$ with metric $d$, let $\mathcal{P}(\mathcal{X})$ be the set of Borel probability measures on $\mathcal{X}$ and $\mathbb{P},\mathbb{Q}\in \mathcal{P(\mathcal{X})}$. For exposition, we assume they have densities $f_1$ and $f_2$, respectively, although the Wasserstein distance is well-defined for more general probability measures using the concept of push-forwards \citep{villani2009optimal}. The optimal transport problem, whose studied was pioneered by \citet{kantorovich1942translocation,kantorovich1960mathematical}, aims to find the joint probability distribution between $\mathbb{P}$ and $\mathbb{Q}$ with the smallest \emph{cost}, as specified by the metric $d$: 
\begin{align}
    \label{eq:Wasserstein}
    \min_{\pi \in \mathcal{P}(\mathcal{X}\times\mathcal{X}): \int_{x_1} \pi(x_1,x_2) dx_1 = f_2(x_2); \int_{x_2} \pi(x_1,x_2) dx_2 = f_1(x_1)} \int_{\mathcal{X}\times\mathcal{X}}d^p(x_1,x_2)\pi(x_1,x_2) dx_1 dx_2,
\end{align}
where $p\geq 1$. The $p$-\textbf{Wasserstein distance} $W_p(\mathbb{P},\mathbb{Q})$ is defined to be the $p$-th root of the optimal value of the optimal transport problem above. %\yc{I am less sure the following part is necessary. I think the duality result of the Wasserstein distance is not critically important in our problem.}
The Wasserstein distance is a metric on the space $\mathcal{P(\mathcal{X})}$ of probability measures, and the dual problem of \eqref{eq:Wasserstein} is derived the following important duality result due to Kantorovich \citep{villani2009optimal}:
\begin{align*}
    W^p_p(\mathbb{P},\mathbb{Q})=\sup_{u\in L^1(\mathbb{P}),v\in L^1(\mathbb{Q}): u(x_1)+v(x_2)\leq d^p(x_1,x_2)} \{ \int_{x_1} u(x_1) f_1(x_1) dx_1 + \int_{x_2} v(x_2) f_2(x_2) dx_2 \}
\end{align*}
It should be noted that the term ``Wasserstein metric'' for the optimal transport distance defined above is an unfortunate mistake, as \citet{kantorovich1942translocation} should be credited with pioneering the theory of optimal transport and proposing the metrics. However, due to a work of Wasserstein \citep{vaserstein1969markov}, which briefly discussed the optimal transport distance, being more well-known in the West initially, the terminology of Wasserstein metric persisted until today \citep{vershik2013long}. The optimal transport problem has also been studied in the seminal work of \citet{koopmans1949optimum}.

One of the appeals of the Wasserstein distance when formulating distributionally robust optimization problems lies in the tractability of the dual DRO problem. Specifically, in \eqref{eq:Wasserstein-DRIVE}, the inner maximization problem requires solving an infinite-dimensional optimization problem for every $\beta$, and is in generally not tractable. However, if $D$ is the Wasserstein distance, the inner problem has a tractable dual minimization problem, which when combined with the outer minimization problem over $\beta$, yields a simple and tractable minimization problem. This will allow us to efficiently compute the WDRO estimator. Moreover, it establishes connections with the popular statistical approach of ridge regression. 

Let $c\in L^1(\mathcal{X})$ be a general loss function and $\mathbb{P}\in \mathcal{P}(\mathcal{X})$ with density $f_1$. The following general duality result \citep{gao2016distributionally,blanchet2019quantifying} provides a tractable reformulation of the Wasserstein DRIVE objective introduced in Section \ref{sec:WDRIVE}:
\begin{align}
    \label{eq:WDRO-duality}
    \sup_{\mathbb{Q}\in\mathcal{P}(\mathcal{X}): W_p(\mathbb{Q},\mathbb{P})\leq \rho} \int f_2(x) c(x)dx = \inf_{\lambda \geq 0} \{ \lambda \theta^p  - \int \inf_{x_2 \in \mathcal{X}} [\lambda d^p(x_1,x_2)-c(x_2)] f_1(x_1) dx_1\},
\end{align}
where $f_2$ is the density of $\mathbb{Q}$.
\subsection{Anchor Regression of \citet{rothenhausler2021anchor}}
\label{sec:anchor-regression}
In the anchor regression framework of \citet{rothenhausler2021anchor}, the baseline distribution $\mathbb{P}_0$ on $(X,Y,U,A)$ is prescribed by the following linear structural equation model (SEM), given well-defined $\mathbf{B},\mathbf{M}$ and distributions of $A,\epsilon$:
\begin{align*}
\begin{pmatrix}X\\
Y\\
U
\end{pmatrix} =\mathbf{B}\begin{pmatrix}X\\
Y\\
U
\end{pmatrix}+\mathbf{M}A+\epsilon \Longleftrightarrow \begin{pmatrix}X\\
Y\\
U
\end{pmatrix} = (I-\mathbf{B})^{-1}(\mathbf{M}A+\epsilon). 
\end{align*}
Here $U$ represents unobserved confounders, $Y$ is the outcome variable, $X$ are observed regressors, and $A$ are ``anchors'' that can be understood as potentially invalid instrumental variables that may violate the exclusion restriction. Under this SEM, \citet{rothenhausler2021anchor} posit that the potential deviations from the reference distribution $\mathbb{P}_0$ are driven by \emph{bounded} uncertainties in the anchors $A$. Their main result provides a DRO interpretation of a modified population version of the IV regression that interpolates between the IV and OLS
objectives for $\gamma>1$:
\begin{align}
\label{eq:anchor-duality}
      \min_\beta    \mathbb{E}[(Y-X^T\beta)^2]  + (\gamma-1) \mathbb{E}[(P_A(Y-X^T\beta))^2] = \min_\beta \sup_{v \in C^\gamma} \mathbb{E}_v [(Y-X^T\beta)^2]. 
\end{align}
The set of distributions $\mathbb{E}_v$ induced by $v \in C^\gamma$ are defined via the following SEM with a bounded set $C^\gamma$:
\begin{align}
\label{eq:anchor-ambiguity-set}
    \begin{pmatrix}X\\
Y\\
U
\end{pmatrix} = (I-\mathbf{B})^{-1}v, \quad C^\gamma := \{v: vv^T \preceq\gamma \mathbf{M}\mathbb{E}(AA^T)\mathbf{M}^T\}.
\end{align}
In the interpolated objective in \cref{eq:anchor-duality}, $P_A(\cdot)=\mathbb{E}(\cdot \mid A)$ and $\mathbb{E}[(P_A(Y-X^T\beta))^2]$ is the population version of the IV (TSLS) regression objective with  $A$ as the instrument. Letting $\kappa = 1-{1}/{\gamma}$, we can rewrite the interpolated objective on the left hand side in \eqref{eq:anchor-duality} equivalently as  
\begin{align}
\label{eq:k-class}
   \min_\beta  \mathbb{E}[(P_A(Y-X^T\beta))^2] + \frac{1-\kappa}{\kappa} \mathbb{E}[(Y-X^T\beta)^2],
\end{align}
which can be interpreted as ``regularizing'' the IV objective with the OLS objective, with penalty parameter $(1-\kappa)/\kappa$. \citet{jakobsen2022distributional} observe that the finite sample version of the objective in \eqref{eq:k-class} is precisely that of a $k$-class estimator with parameter $\kappa$ \citep{theil1961economic,nagar1959bias}. This observation together with \eqref{eq:anchor-duality} therefore provides a DRO interpretation of $k$-class estimators, which is also extended by \citet{jakobsen2022distributional} to more general instrumental variables estimation settings. Moreover, when $\kappa=1$, or equivalently $\gamma \rightarrow \infty$, we recover the standard IV objective in \eqref{eq:k-class}. Therefore, the IV estimator has a distributionally robust interpretation via \eqref{eq:anchor-duality} when distributional shifts $v$ are \emph{unbounded}.

The DRO interpretation \eqref{eq:anchor-duality} of $k$-class estimators sheds new light on some old wisdom on IV estimation. As has already been observed and studied by a large literature \citep{richardson1968exact,nelson1990distribution,nelson1990some,bound1995problems,staiger1997instrumental,hahn2004estimation,burgess2011avoiding,andrews2019weak,young2022consistency}, when instruments
are weak, the usual normal approximation to the distribution of the IV estimator may be very poor, and the IV estimator is biased in small samples and in the weak instruments asymptotics. Moreover, a small violation of the exclusion restriction, i.e., direct effect of instruments on the outcome, can result in large bias when instruments are weak \citep{angrist2009mostly}. Consequently, IV may not perform as well as the OLS
estimator in such a setting. Regularizing the IV objective by the OLS
objective in \eqref{eq:k-class} can therefore alleviate the weak instrument problem. This improvement has also been observed for $k$-class estimators with particular choices of $\kappa$ \citep{fuller1977some,hahn2004estimation}. The DRO interpretation complements the intuition above based on regularizing the IV objective with the OLS objective. In so far as weak instruments can be understood as a form of distributional shift from standard modeling assumptions (strong first stage effects), a distributionally robust regression approach is a natural solution to address the challenge of weak instruments. In the case of the anchor regression, the distribution uncertainty set indexed by $v\in C^\gamma$ always contains distributions on $(X,Y,U,A)$ where the association between $A$ and $X$ is weak, by selecting appropriate $\|v\|\approx 0$. Therefore, the DRO formulation \eqref{eq:anchor-duality} of $k$-class estimators demonstrates that they are robust against the weak instrument problem by design. An additional insight of the DRO formulation is that $k$-class estimators and anchor regression are also optimal
in terms of \emph{predicting} $Y$ with $X$ when the distribution of $(X,Y)$ could change between the training and test datasets in a bounded manner induced by the anchors $A$. 

On the other hand, the DRO interpretation of $k$-class estimators also exposes its potential limitations. First of all, the ambiguity set in \eqref{eq:anchor-ambiguity-set} does not in fact contain the reference distribution $\mathbb{P}_0$ itself for any finite robustness parameter, which is unsatisfactory. Moreover, the SEM in \eqref{eq:anchor-ambiguity-set} also implies that the instrument (anchors) $A$ cannot be influenced by the unobserved confounder $U$, which is a major source of uncertainty regarding the validity of instruments in applications of IV estimation. In this sense, we may understand $k$-class estimators as being robust against weak instruments \citep{young2022consistency}, since they minimize an objective that interpolates between OLS and IV. On the other hand, the DRIVE approach we propose in this paper is by design robust against invalid instruments, as the ambiguity set captures distributional shifts arising from conditional correlations between the instrument and the outcome variable, conditional on the endogenous variable.

\section{Related Works}
\label{sec:related-works} 
Our work is related to several literatures, including distributionally robust optimization, instrumental variables estimation, and regularized (penalized) regression. Although historically they developed largely independent of each other, recent works have started to explore their interesting connections, and our work can be viewed as an effort in this direction. 

\subsection{Distributionally Robust Optimization and Min-max Optimization}
DRO has an important research area in operations research, and traces its origin to game theory \citep{von1947theory}. \citet{scarf1958min} first studied DRO in the context of inventory control under uncertainties about future demand distributions. 
% DRO has long been an important research area in operations research, and traces its origins to the seminal work of \citet{scarf1958min}, which considered maximizing worst-case expected profit in a newsvendor problem with respect to a set of demand distributions with known mean and variance.
This work was followed by a line of research in min-max stochastic optimization models, notably the works of \citet{shapiro2002minimax}, \citet{calafiore2006distributionally}, and \citet{ruszczynski2010risk}. Distributional uncertainty sets based on moment conditions are considered by \citet{dupavcova1987minimax,prekopa2013stochastic,bertsimas2005optimal,delage2010distributionally}. Distributional uncertainty sets based on distance or divergence measures are considered by \citet{iyengar2005robust,wang2016likelihood}. In recent years, distributional uncertainty sets based on the Wasserstein metric have gained traction, appearing in \citet{mohajerin2018data,blanchet2019robust,blanchet2019quantifying,duchi2021statistics}, partly due to their close connections to regularized regression, such as the LASSO \citep{tibshirani1996regression,belloni2011square} and regularized logistic regression. Other works employ alternative divergence measures, such as the KL divergence \citep{hu2013kullback} and more generally $\phi$-divergence \citep{ben2013robust}. In this work, we focus on DRO based on the Wasserstein metric, originally proposed by \citet{kantorovich1942translocation} in the context of optimal transport, which has also become a popular tool in economics in recent years \citep{galichon2018optimal,galichon2021unreasonable}.

DRO has gained traction in causal inference problems in econometrics and statistics very recently. For example, \citet{kallus2021minimax,adjaho2022externally,lei2023policy} apply DRO in policy learning to handle distributional uncertainties. \citet{chen2021robust} apply DRO to address the possibility of mis-specification of rational expectation in estimation of structural models. \citet{sahoo2022learning} use distributional shifts to model sampling bias. \citet{bertsimas2022distributionally} study DRO versions of classic causal inference frameworks. \citet{fan2023quantifying} studies distributional model risk when data comes from multiple sources and only marginal reference measures are identified. DRO is also connected to the literature in macroeconomics on robust control \citep{hansen2010wanting}. A related recent line of works in econometrics also employs a min-max approach to estimation \citep{lewis2018adversarial,kaji2020adversarial,metzger2022adversarial,cigliutti2022adversarial,bennett2023variational}, inspired by adversarial networks from machine learning \citep{goodfellow2014generative}. These works leverage \emph{adversarial} learning to enforce a large, possibly infinite, number of (conditional) moment constraints, in order to achieve efficiency gains. In contrast, the emphasize of the min-max formulation in our paper is to capture the potential \emph{violations} of model assumptions using a distributional uncertainty set. 

The DRO approach that we propose in this paper is motivated by a recent line of works that reveal interesting connections between causality and notions of invariance and distributional robustness \citep{peters2016causal,meinshausen2018causality,rothenhausler2021anchor,buhlmann2020invariance,jakobsen2022distributional}.

Another important literature has studied the connections between causal inference and concepts of invariance and robustness \citep{peters2016causal,meinshausen2018causality,rothenhausler2021anchor,buhlmann2020invariance,jakobsen2022distributional}. Our work is closely related to this line of works, whereby causality is interpreted as an invariance or robustness property under distributional shifts. In particular,  \citet{rothenhausler2021anchor,jakobsen2022distributional} provide a distributionally robust interpretation of the classic $k$-class estimators. In our work, instead of constructing the distribution set based on marginal or joint distributions as is commonly done in previous works, we propose a Wasserstein DRO version of the IV estimation problem based on distributional shifts in \emph{conditional} quantities, which is then reformulated as a ridge type regularized IV estimation problem. In this regard, our estimator is fundamentally different from the $k$-class estimators, which minimize an IV regression objective regularized by an OLS objective.

\subsection{Instrumental Variables Estimation}
Our work is also closely related to the classic  literatures in econometrics and statistics on instrumental variables estimation (regression), which is originally proposed and developed by \citet{theil1953repeated} and \citet{nagar1959bias}, and became widely used in applied fields in economics. Since then, many works have investigated potential challenges to instrumental variables estimation and their solutions, including invalid instruments \citep{fisher1961cost,hahn2005estimation,berkowitz2008nearly,kolesar2015identification} and weak instruments \citep{nelson1990distribution,nelson1990some,staiger1997instrumental,murray2006avoiding,andrews2019weak}. Tests of weak instrument have been proposed by  \citet{stock2002testing} and \citet{sanderson2016weak}. Notably, the test of \citet{stock2002testing} for multi-dimensional instruments is based on the minimum eigenvalue rank test statistic of \citet{cragg1993testing}. In our Wasserstein DRIVE framework, the penalty/robustness parameter can also be selected using the minimum eigenvalue of the first stage coefficient. It remains to further study the connections between our work and the weak instrument literature in this regard. The related econometric literature on many (weak) instruments studies the regime where the number of instruments is allowed to diverge proportionally with the sample size \citep{kunitomo1980asymptotic,bekker1994alternative,chamberlain2004random,chao2005consistent,kolesar2018minimum}. In this work, we will assume a fixed number of instruments to best illustrate the Wasserstein DRIVE approach. However, it would be interesting to extend the framework and analysis in the current work to the many instruments setting. 

Testing for invalid instruments is possible in the over-identified regime, where there are more instruments than endogenous variables \citep{sargan1958estimation,kadane1977comment,hansen1982large,andrews1999consistent}. These tests have been used in combination with variable selection methods, such as LASSO and thresholding, to select valid instruments under certain assumptions \citep{kang2016instrumental,windmeijer2018use,guo2018testing,windmeijer2021confidence}. In our paper, we propose a regularization selection procedure for Wasserstein DRIVE based on bootstrapped score quantile. In simulations, we find that the selected $\rho$ increases with the degree of instrument invalidity. It remains to further study the relation of this score quantile and test statistics for instrument invalidity in the over-identified setting. Lastly, our framework can be viewed as complementary to the post-hoc sensitivity analysis of invalid instruments \citep{angrist1996identification,small2007sensitivity,conley2012plausibly}, where instead of bounding the potential bias of IV estimators arising from violations of model assumptions \emph{after} estimation, we incorporate such potential deviations directly into the estimation procedure.

Instrumental variables estimation has also gained wide adoption in epidemiology and genetics, where it is known as Mendelian randomization (MR) \citep{vanderweele2014methodological,bowden2015mendelian,sanderson2016weak,emdin2017mendelian}. An important consideration in MR is invalid instruments, because many genetic variants, which are candidate instruments in Mendelian randomization, could be correlated with the outcome variable through unknown mechanisms that are either direct effects (horizontal pleitropy) or correlations with unobserved confounders. Methods have been proposed to address these challenges, based on robust regression and regularization ideas \citep{bowden2015mendelian,bowden2016consistent,burgess2016robust,burgess2020robust}. Our proposed DRIVE framework contributes to this area by providing a novel regularization method robust against potentially invalid instruments. 

\subsection{Regularized Regression}

Our Wasserstein DRIVE framework can be viewed as an instance of data-driven regularized IV method. In this regard, it complements the classic $k$-class estimators, which regularize the IV objective with OLS \citep{rothenhausler2021anchor}. Data-driven $k$-class estimators have been shown to enjoy better finite sample properties. These include the LIML \citep{anderson1949estimation} and the Fuller estimator \citep{fuller1977some}, which is a modification of LIML that works well when instrument are weak \citep{stock2002survey}. More recently, \citet{jakobsen2022distributional} proposed another data-driven $k$-class estimator called the PULSE, which minimizes the OLS objective but with a constraint set defined by statistical tests of independence between instrument and residuals. \citet{kolesar2015identification} propose a modification of the $k$-class estimator that is consistent with invalid instruments whose direct effects on the outcome are independent of the first stage effect on the endogenous regressor. 

There is a rich literature that explores the interactions and connections between regularized regression and instrumental variables methods. One line of works seeks to improve the finite-sample performance and asymptotic properties of IV type estimators using methods from regularized regression. For example, \citet{windmeijer2018use} applies LASSO regression to the first stage, motivated by applications in genetics where one may have access to many weak or invalid instruments. \citet{belloni2012sparse,chernozhukov2015post} also apply LASSO, but the task is to select optimal instruments in the many instruments setting or when covariates are high-dimensional. \citep{caner2009lasso,caner2018high,belloni2018high} apply LASSO to GMM estimators, generalizing regularized regression results from the M-estimation setting to the moment estimation setting, which also includes IV estimation.

Another line of works on regularized regressions, which is more closely related to our work, have investigated the connections and equivalences between regularized regression and causal effect estimators in econometrics based on instrumental variables. \citet{basmann1960finite,basmann1960asymptotic,mcdonald1977k} are the first to connect $k$-class estimators to regularized regressions. \citet{rothenhausler2021anchor} and \citet{jakobsen2022distributional} further study the distributional robustness of $k$-class estimators as minimizers of the TSLS objective \emph{regularized} by the OLS objective. The Wasserstein DRIVE estimator that we propose in this work applies a different type of regularization, namely a square root ridge regularization, to the second stage coefficient in the TSLS objective. As such it has different behaviors compared to the anchor and $k$-class estimators, which regularize using the OLS objective. It is also different from works that apply regularization to the first stage.

\section{Square Root Ridge Regression}
\label{sec:sqrt-ridge}

In this section, we turn our attention to the square root ridge estimator in the standard regression setting. We first establish the $\sqrt{n}$-consistency of the square root ridge when the regularization parameter vanishes at the appropriate rate. We then consider a novel regime with non-vanishing regularization parameter and vanishing noise, revealing
properties that are strikingly different from the standard ridge regression. As we will see, these observations in the standard setting help motivate and provide the essential intuitions for our results in the IV estimation setting. In short, the interesting behaviors of the square root ridge arise from its unique geometry in the regime of vanishing noise, where $\sum_{i=1}^n\epsilon^2_i = {o}_p(n)$ as $n\rightarrow \infty$. This regime is rarely studied in conventional regression settings in statistics, but it precisely captures features of the instrumental variables estimation setting, where projected residuals $(\mathbf{Y}-\mathbf{X}\beta_0)^T\mathbf{\Pi}_\mathbf{Z}(\mathbf{Y}-\mathbf{X}\beta_0)=o_p(n)$ when instruments are valid and $\beta_0$ is the true effect coefficient. In addition to providing intuitions for the IV estimation setting, the regularization parameter selection procedure proposed for the square root LASSO in the standard regression setting by \citet{belloni2011square} also inspires us to propose a novel procedure for the IV setting in \cref{sec:penalty-selection}, which is shown to perform well in simulations. 

\subsection{$\sqrt{n}$-Consistency of the Square Root Ridge}

We now consider the square root ridge estimator in the standard regression setting, and prove its $\sqrt{n}$-consistency. We will build on the results of \citet{belloni2011square} on the non-asymptotic estimation error of the square root LASSO estimator. Conditional on a fixed design $X_{i}\in\mathbb{R}^{p}$, and with $\Phi$ the CDF of $\epsilon_i$, we consider the data generating process, 
\begin{align*}
Y_{i} & =X_{i}^{T}\beta_{0}+\sigma\epsilon_{i}.
\end{align*}
In this section, we rewrite the objective of the square root ridge estimation \eqref{eq:sqrt-ridge-ols} as 
\begin{align}
\label{eq:sqrt-ridge-belloni}
\min_{\beta}\sqrt{\hat{Q}(\beta)}+\frac{\lambda}{n}\sqrt{\|\beta\|^{2}+1}\\
\hat{Q}(\beta)=\frac{1}{n}\sum_{i=1}^n(Y_{i}-X_{i}^{T}\beta)^{2},
\end{align}
and denote $\hat{\beta}$ as the minimizer of the objective.
Without loss of generality, we assume for all $j$, 
\begin{align*}
\frac{1}{n}\sum_{i=1}^nX_{ij}^{2} & =1
\end{align*}
 In other words, each covariate (feature) is normalized to have unit
norm. Similar to the square root LASSO case, we will show that by
selecting $\lambda=\mathcal{O}(\sqrt{n})$ properly, or equivalently $\rho=\mathcal{O}(n^{-1})$ in \eqref{eq:sqrt-ridge-ols}, we can achieve, with probability
$1-\alpha$, a $\sqrt{n}$-consistency result:
\begin{align*}
\|\hat{\beta}-\beta\|_{2} & \lesssim\sigma\sqrt{p\log(2p/\alpha)/n}.
\end{align*}
 Compare this with the bound of the square root LASSO, which is 
\begin{align*}
\|\hat{\beta}-\beta\|_{2} & \lesssim\sigma\sqrt{s\log(2p/\alpha)/n},
\end{align*}
 where $s$ is the number of non-zero entries of $\beta_{0}$, and
is allowed to diverge as $n\rightarrow\infty$. Since we do not impose assumptions on the
size of the support of the $p$-dimensional vector $\beta_{0}$, if $s=p$ is finite in the square root LASSO framework, we achieve the same bound on the estimation error. Our bound for the
square root ridge is therefore sharp in this sense. 

An important quantity in the analysis is the score $\tilde{S}$, which
is the gradient of $\sqrt{\hat{Q}(\beta)}$ evaluated at the true
parameter value $\beta=\beta_{0}$:
\begin{align*}
\tilde{S} & =\nabla\sqrt{\hat{Q}(\beta)}(\beta_{0})=\frac{\nabla\hat{Q}(\beta_{0})}{2\sqrt{\hat{Q}(\beta_{0})}}=\frac{E_{n}(X\sigma\epsilon)}{\sqrt{E_{n}(\sigma^{2}\epsilon^{2})}}=\frac{E_{n}(X\epsilon)}{\sqrt{E_{n}(\epsilon^{2})}},
\end{align*}
 where $E_{n}$ denotes the empirical average of the quantities. Similar
to the lower bound on the regularization parameter in terms of the score
function $\lambda\geq cn\|\tilde{S}\|_{\infty}$ in \citet{belloni2011square}, we will aim to impose
the condition that $\lambda\geq cn\|\tilde{S}\|_{2}$ for some $c>1$.
Conveniently, this condition is already implied by $\lambda=\sqrt{p}\lambda^{*}$,
where $\lambda^{\ast}$ follows the selection procedures proposed
in that paper. To see this point, note that $\|\tilde{S}\|_{2}\leq\sqrt{p}\|\tilde{S}\|_{\infty}$,
so that with high probability, $\sqrt{p}\lambda^{*}\geq\sqrt{p}cn\|\tilde{S}\|_{\infty}\geq cn\|\tilde{S}\|_{2}$.
Thus we may use the exact same selection procedure to
achieve the desired bound, although there are other selection procedures
for $\lambda$ that would guarantee $\lambda\geq cn\|\tilde{S}\|_{2}$
with high probability. For example, choose the $(1-\alpha)$-quantile
of $n\|\tilde{S}\|_{2}$ given $X_{i}$'s. We will for now adopt the selection procedure and the model assumptions in \citet{belloni2011square}.
\begin{assumption}
\label{ass:belloni-asymptotics}
We have $\log^{2}(p/\alpha)\log(1/\alpha)=o(n)$ and $p/\alpha\rightarrow\infty$
as $n\rightarrow\infty$.
\end{assumption}
Under this assumption, and assuming that $\epsilon$ is normal, the
selected regularization $\lambda=\sqrt{p}\lambda^{*}$ satisfies 
\begin{align*}
\lambda & \lesssim\sqrt{pn\log(2p/\alpha)}
\end{align*}
with probability $1-\alpha$ for all large $n$, using the same
argument as Lemma 1 of \citet{belloni2011square}. 

An important quantity in deriving the bound on the estimation error
is the ``prediction'' norm 
\begin{align*}
\|\hat{\beta}-\beta_{0}\|_{2,n}^{2} & :=\frac{1}{n}\sum_{i}(X_{i}^{T}(\hat{\beta}-\beta_{0}))^{2}\\
 & =(\hat{\beta}-\beta_{0})^{T}\frac{1}{n}\sum_{i}X_{i}X_{i}^{T}(\hat{\beta}-\beta_{0}),
\end{align*}
 which is related to the Euclidean norm $\|\hat{\beta}-\beta_{0}\|_{2}$
through the Gram matrix $\frac{1}{n}\sum_{i}X_{i}X_{i}^{T}$. We need
to make an assumption on the modulus of continuity. 
\begin{assumption}
\label{ass:belloni-modulus}
There exists a constant $\kappa$ and $n_{0}$ such that for all $n\ge n_{0}$,
$\kappa\|\delta\|_{2}\leq\|\delta\|_{2,n}$ for all $\delta\in\mathbb{R}^{p}$.
\end{assumption}
When $p\leq n$, the Gram matrix $\frac{1}{n}\sum_{i}X_{i}X_{i}^{T}$
will be full rank (with high probability with random design), and
concentrate around the population covariance matrix. This setting
of $p\leq n$ is different from the high-dimensional setting in the
square root LASSO paper, as LASSO-type penalties are able to achieve
selection consistency when $p>n$ under sparsity, whereas ridge-type
penalties generally cannot. %(However, it is possible square root ridge can achieve high-dimensional consistency in the noiseless setting, need to look into this further!). 
Note also that when $p>n$, the
restricted eigenvalues are necessary when defining $\kappa$, and
it is necessary to prove that $\hat{\beta}-\beta_{0}$ belongs to
a restricted subset of $\mathbb{R}^{p}$ on which the bound with $\kappa$
holds. When $p\leq n$, the restricted subset and eigenvalues are
not necessary, and $\kappa$ can be understood as the minimum eigenvalue
of the Gram matrix, which would be bonded away from $0$ (with high
probability). The exact value of $\kappa$ is a function of the data
generating process. For example, if we assume covariates are generated
independent of each other, then $\kappa\approx1$. 
\begin{theorem}
\label{thm:sqrt-ridge-root-n-consistency}
Assume that $p\leq n$ but $p$ is allowed to grow with $n$. Let
the regularization $\lambda=\sqrt{p}\lambda^{*}$ where $\lambda^\ast = c\sqrt{n}\Phi^{-1}(1-\alpha/2p)$, and under \cref{ass:belloni-asymptotics} and \cref{ass:belloni-modulus},
the solution $\hat{\beta}$ to the square root ridge problem 
\begin{align*}
\min_{\beta}\sqrt{\frac{1}{n}\sum_{i}(Y_{i}-X_{i}^{T}\beta)^{2}}+\frac{\lambda}{n}\sqrt{\|\beta\|^{2}+1}
\end{align*}
 satisfies 
\begin{align*}
\|\hat{\beta}-\beta_{0}\|_{2} & \leq\frac{2(\frac{1}{c}+1)}{1-(\frac{\lambda}{n})^{2}\kappa^{2}}\frac{\lambda}{n}\cdot\sigma\sqrt{E_{n}(\epsilon^{2})}\lesssim\sigma\sqrt{p\log(2p/\alpha)/n}
\end{align*}
 with probability at least $1-\alpha$ for all $n$ large enough.%
\begin{comment}
The quantity $\sqrt{E_{n}(\epsilon^{2})}$ can be bounded by some
constant as in Belloni.
\end{comment}
\end{theorem}
We remark that the quantile of the score function $\frac{E_{n}(X\epsilon)}{\sqrt{E_{n}(\epsilon^{2})}}$ is not only critical for establishing the $\sqrt{n}$-consistency of the square root ridge. It is also important in practice as the basis for regularization parameter selection. In \cref{sec:penalty-selection}, we propose a data-driven regularization selection procedure that uses nonparametric bootstrap to estimate the quantile of the score, and demonstrate in \cref{sec:numerical} that it has very good empirical performance. The nonparametric bootstrap procedure may be of independent as well. Before we discuss regularization parameter selection in detail, we first focus on the statistical properties of the square root ridge under the novel vanishing noise regime. 

\subsection{Delayed Shrinkage of Square Root Ridge}
Conventional wisdom on regressions with ridge type penalties is that they induce \emph{shrinkage} on parameter estimates, and this shrinkage happens for any non-zero regularization. Asymptotically, if the regularization parameter does not vanish as the sample size increases, the limit of the estimator, when it exists, is not equal to the true parameter. The same behavior may be expected of the square root ridge regression. Indeed, this is the case in the standard linear regression setting with constant variance, i.e., $\text{Var}(\epsilon_i)=\sigma^2>0$ and
\begin{align*}
%\label{eq:ols}
Y_i=X_i^T\beta_0 + \epsilon_i.
\end{align*} 
However, as we will see, when $\text{Var}(\epsilon_i)$ depends on the sample size, and vanishes as $n\rightarrow \infty$, the square root ridge estimator can be consistent
for \emph{non-vanishing} penalties.

To best illustrate the intuition behind this property of the square root ridge, we start with the following simple example. Consider the data generating process written in matrix vector form: 
\begin{align}
\label{eq:ols-matrix}
\mathbf{Y}=\mathbf{X}\beta_0 + \mathbf{\epsilon},
\end{align}
where the rows of $\mathbf{X}\in \mathbb{R}^{n\times p}$ are i.i.d. $\mathcal{N}(0,I_p)$ and independent of $\mathbf{\epsilon}\sim \mathcal{N}(0,\sigma^2_n I_p)$. Suppose that the variance of the noises vanishes: $\sigma^2_n \rightarrow 0$ as $n\rightarrow \infty$. This is not a standard regression setup, but captures the essence of the IV estimation setting, as we show in \cref{sec:bias-analysis}.

\begin{figure}
\begin{centering}
\includegraphics[scale=0.5]{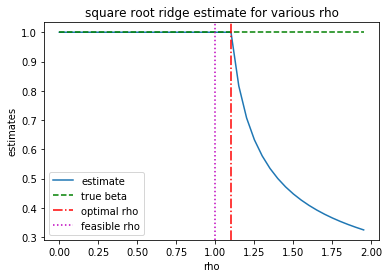}
\par\end{centering}
\caption{Limit of the square root ridge estimator in a one-dimensional example with vanishing noise, as a function of the regularization parameter $\rho$. Optimal $\rho=1+\frac{1}{\|\beta_0\|^2}$ is the largest regularization level that guarantees consistency of square root ridge. %\textcolor{red}{The figure quality should be improved. Capitalization + font size + line width is too thin.}
}
\label{fig:noiseless-sqrt}
\end{figure}

Recall the square root ridge regression problem in \eqref{eq:sqrt-ridge-ols}, which is strictly convex:
\begin{align*}
\min_{\beta}\sqrt{\frac{1}{n}\|\mathbf{Y}-\mathbf{X}\beta\|^{2}}+\sqrt{\rho(1+\|\beta\|^{2})}. 
\end{align*}
Let $\hat{\beta}^{(n)}_{\mathrm{sqrt}}$
be its unique minimizer. As the sample size $n\rightarrow \infty$, we will fix the regularization parameter $\rho \equiv 1$, instead of letting $\rho \rightarrow 0$. Standard asymptotic theory implies that $\hat{\beta}^{(n)}_{\mathrm{sqrt}} \rightarrow_p {\beta}_{\mathrm{sqrt}}$, where ${\beta}_{\mathrm{sqrt}}$ is the minimizer of the limit of the square root ridge objective. For the simple model \eqref{eq:ols-matrix}, we can verify that
\begin{align*}
\sqrt{\frac{1}{n}\|\mathbf{Y}-\mathbf{X}\beta\|^{2}}+\sqrt{(1+\|\beta\|^{2})} \rightarrow_p
\|\beta_0-\beta\|+\sqrt{(1+\|\beta\|^{2})},
\end{align*}
where we have used the crucial property $\sigma_n^2 \rightarrow 0$. 
% \begin{align*}
% \sqrt{\frac{1}{n}\|\mathbf{Y}-\mathbf{X}\hat{\beta}_{\mathrm{sqrt}}\|^{2}}+\sqrt{(1+\|\hat{\beta}_{\mathrm{sqrt}}\|^{2})} & =\sqrt{\frac{1}{n}(\mathbf{X}\beta-\mathbf{X}\hat{\beta}_{\mathrm{sqrt}})^{T}(\mathbf{X}\beta-\mathbf{X}\hat{\beta}_{\mathrm{sqrt}})}+\sqrt{\rho(1+\|\hat{\beta}_{\mathrm{sqrt}}\|^{2})}\\
%  & =\sqrt{\frac{1}{n}(\beta-\hat{\beta}_{\mathrm{sqrt}})^{T}\mathbf{X}^{T}\mathbf{X}(\beta-\hat{\beta}_{\mathrm{sqrt}})}+\sqrt{\rho(1+\|\hat{\beta}_{\mathrm{sqrt}}\|^{2})}
% \end{align*}
Therefore, under standard conditions, we have 
\begin{align}
\label{eq:example-sqrt-ridge}
\hat{\beta}^{(n)}_{\mathrm{sqrt}} \rightarrow_p {\beta}_{\mathrm{sqrt}}:=\arg \min_\beta \|\beta_0-\beta\|+\sqrt{(1+\|\beta\|^{2})}.
\end{align}
Note that the limiting objective above is strictly convex and hence has a unique minimizer. Moreover, 
\begin{align*}
\|\beta_0-\beta\|+\sqrt{(1+\|\beta\|^{2})}=\|(\beta_0,-1)-(\beta,-1)\|+\|(\beta,-1)\|\geq\|(\beta_0,-1)\|,
\end{align*}
using the triangle inequality.
% \begin{align*}
% \|(\beta_0,-1)-(\beta,-1)\|+\|(\beta,-1)\| & \geq\|(\beta_0,-1)\|.
% \end{align*}
On the other hand, setting $\beta=\beta_0$ in \eqref{eq:example-sqrt-ridge} achieves
the lower bound $\|(\beta_0,-1)\|$. Therefore, ${\beta}_{\mathrm{sqrt}}=\beta_0$ is the unique minimizer of the limiting objective, and so 
\[\hat{\beta}^{(n)}_{\mathrm{sqrt}} \rightarrow_p {\beta}_0\] 
with $\rho = 1$ non-vanishing. We have therefore demonstrated that with a non-vanishing regularization parameter, the square root ridge regression can still produce a consistent estimator. This phenomenon holds more generally: the square root ridge estimator is consistent for any (limiting) regularization parameter $\rho \in [0, 1+\frac{1}{\|\beta_0\|^2}]$, as long as the noise vanishes, in the sense that $\sum_{i=1}^n\epsilon^2_i = {o}_p(n)$. This condition is achieved for a wide variety of empirical risk minimization objectives, including the IV estimation objective.
\begin{theorem}
\label{lem:sqrt-ridge-noiseless}
In the linear model \eqref{eq:ols-matrix} where the rows of $\mathbf{X}$ are distributed i.i.d. $\mathcal{N}(0,I_p)$, if $\sum_{i=1}^n\epsilon^2_i = {o}_p(n)$ as sample size $n\rightarrow \infty$, then for any $\rho \in [0, 1+\frac{1}{\|\beta_0\|^2}]$, the unique solution $\hat{\beta}_{\mathrm{sqrt}}$ to \eqref{eq:sqrt-ridge-ols} is consistent:
\begin{align*}   \hat{\beta}^{(n)}_{\mathrm{sqrt}} \rightarrow_p \beta_0.
\end{align*}
\end{theorem}
%Proofs can be found in \cref{sec:proofs}.
In \cref{fig:noiseless-sqrt}, we plot the solution of the limiting square root ridge objective in a one-dimensional example. As we can see, (asymptotic) shrinkage is \emph{delayed} until regularization $\rho$ exceeds the limit $1+\frac{1}{\|\beta_0\|^2}$ in the vanishing noise regime. This behavior is in stark contrast with the regular ridge regression estimator, for which shrinkage starts from the origin, even in the vanishing noise setting.

\begin{remark}[Necessary Requirements of Delayed Shrinkage]
   \normalfont 
 Although the delayed shrinkage property of the square root ridge is essentially a simple consequence of the triangle inequality, it relies crucially on three features of the square root ridge estimation procedure. 

First, even though the square root ridge shares similarities with the standard ridge regression 
\begin{align*}
\min_{\beta}{\frac{1}{n}\|\mathbf{Y}-\mathbf{X}\beta\|^{2}}+{\rho\|\beta\|^{2}},
\end{align*}
only the former has delayed shrinkage: the square root operations applied to the mean squared loss and the squared norm of the parameter above are essential. To see this, note that with vanishing noise, the limit of the standard ridge estimator for the model in \eqref{eq:ols-matrix} is the solution to the following problem:
\begin{align*}
    \min_\beta \|\beta_0-\beta\|^2+\rho \|\beta\|^{2},
\end{align*}
which results in the optimal solution $\beta = \beta_0/(1+\rho)$. Therefore, with any non-zero $\rho$, the ridge estimator exhibits shrinkage, even with vanishing noise. 

Second, the inclusion of the extra constant 1 in the regularization term $\sqrt{(1+\|\beta\|^{2})}$ is crucial in guaranteeing that $\beta_{\mathrm{sqrt}}=\beta_0$ is the \emph{unique} limit of the square root ridge estimator. To see this, consider instead the following modified square root ridge problem, which appears in \citet{owen2007robust,blanchet2019robust}: 
\begin{align*}
\min_{\beta}\sqrt{\frac{1}{n}\|\mathbf{Y}-\mathbf{X}\beta\|^{2}}+\sqrt{\rho}\|\beta\|_2,  
\end{align*}
where the regularization term does not include an additive constant in the square root, so simplifies to $\|\beta\|$. Under model \eqref{eq:ols-matrix} with vanishing noise and $\rho=1$, this objective has limit $\|\beta_0-\beta_{\mathrm{sqrt}}\|+\|\beta_{\mathrm{sqrt}}\|$. 
Without the ``curvature'' guaranteed by the additional constant in the regularization term, the limiting objective is no longer strictly convex, and there is actually an infinite number of solutions that achieves the lower bound in the triangle inequality
\begin{align*}
\|\beta_0-\beta_{\mathrm{sqrt}}\|+\|\beta_{\mathrm{sqrt}}\| & \geq\|\beta_0\|,
\end{align*}
including $\beta_{\mathrm{sqrt}}=0$. This implies that the solution to the modified objective is no longer guaranteed to be a consistent estimator of $\beta_0$. Indeed, the inconsistency of this curvature-less version of the square root ridge estimator has also been corroborated by simulations.

Third, given that small penalties in the square root ridge objective could achieve \emph{regularization} in finite samples without sacrificing consistency, one may wonder why it is not widely used. This is partly due to the standard ridge being easier to implement computationally, but the main reason is that the delayed shrinkage of the square root ridge estimator is \emph{only} present in the
vanishing noise regime. To see this, assume now  $\mathbf{\epsilon} \sim \mathcal{N}(0,\sigma^2 I)$ with non-vanishing $\sigma^2>0$ in \eqref{eq:ols-matrix}. As sample size $n\rightarrow \infty$,
\begin{align*}
\sqrt{\frac{1}{n}\|\mathbf{Y}-\mathbf{X}{\beta}\|^{2}}+\sqrt{\rho(1+\|{\beta}\|^{2})} & =\sqrt{\frac{1}{n}(\mathbf{X}\beta_0+\mathbb{\mathbf{\epsilon}}-\mathbf{X}\beta)^{T}(\mathbf{X}\beta_0+\mathbb{\mathbf{\epsilon}}-\mathbf{X}\beta)}+\sqrt{\rho(1+\|\beta\|^{2})}\\
 & \rightarrow_p\sqrt{\|\beta_0-\beta\|^{2}+\sigma^{2}}+\sqrt{\rho(1+\|\beta\|^{2})},
\end{align*}
and as before $\hat{\beta}^{(n)}_{\mathrm{sqrt}} \rightarrow_p \beta_{\mathrm{sqrt}}$, the unique minimizer of the limiting objective above. The optimal condition is given by
\begin{align*}
\frac{(\beta-\beta_0)}{\sqrt{\|\beta-\beta_0\|^{2}+\sigma^{2}}}+\sqrt{\rho}\frac{\beta}{\sqrt{(1+\|\beta\|^{2})}} & =0,
\end{align*}
 and now only when $\rho \rightarrow 0$ is $\hat{\beta}^{(n)}_{\mathrm{sqrt}}$ a consistent
estimator of $\beta_0$, \emph{unless} $\beta_0 \equiv 0$. For this reason, the fact that square
root ridge can be consistent with non-vanishing regularization may not be particularly useful in standard regression settings. In the presence of non-vanishing noise, shrinkage happens for any non-zero regularization, which has also been confirmed in simulations. 
\end{remark}

 Although the consistency of the square root ridge estimator with non-vanishing regularization does not have immediate practical implications for conventional regression problems, it is actually very well suited for the instrumental variables estimation setting. The reason is that IV and TSLS regressions involve projecting the endogenous (and outcome) variables onto the space spanned by the instrumental variables in the first stage. When instruments are valid, this projection cancels out the noise terms asymptotically, resulting in
 $\frac{1}{n}(\mathbf{Y}-\mathbf{X}\beta_0)^T\mathbf{\Pi}_\mathbf{Z}(\mathbf{Y}-\mathbf{X}\beta_0)\rightarrow_p 0$. The subsequent second-stage regression involving the projected variables therefore  precisely corresponds to the vanishing noise regime, and we may expect a similar delayed shrinkage effect. This  is indeed the case, and with non-zero regularization and (asymptotically) valid instruments, we show in \cref{sec:bias-analysis} that the Wasserstein DRIVE estimator is consistent. This result suggests that we can introduce robustness and regularization to the standard IV estimation through the square root ridge objective without sacrificing asymptotic validity, and has important implications in practice. 
\subsection{Square Root Ridge vs. Ridge for GMM and M-Estimators}

 We also remark on the distinction between the square root
ridge and the standard ridge in the case when $\rho_n\rightarrow 0$.
From \citet{fu2000asymptotics}, we know that if $\rho$ approaches 0 at
a rate of or slower than $O(1/\sqrt{n})$, then the ridge estimator
 has asymptotic bias, i.e., it is not centered at $\beta_0$. However, for square root ridge (and DRIVE),
 as long as $\rho_n\rightarrow0$ at any rate, the estimator will not have any bias. This feature is a result of the self-normalization property of the square root ridge. In \eqref{eq:asymptotic-distribution}, the second term results from
\begin{align*}
\sqrt{n\rho(1+\|\beta_{0}+\delta/\sqrt{n}\|^{2})}-\sqrt{n\rho(1+\|\beta_{0}\|^{2})} & =\frac{n\rho\beta_{0}^{T}}{\sqrt{n\rho(1+\|\beta_{0}\|^{2})}}\cdot\delta/\sqrt{n}+o(\delta/\sqrt{n})\\
 & \rightarrow \frac{\sqrt{\rho}\beta_{0}^{T}\delta}{\sqrt{(1+\|\beta_{0}\|^{2})}},
\end{align*}
which does not depend on $n$. In this sense, the parameter $\rho$ in square root ridge is scale-free, unlike the regularization parameter in the standard ridge case, whose natural scale is $O(1/\sqrt{n})$.
In the same spirit, when $\rho$ does not vanish, the resulting square
root ridge estimator will have similar behaviors as that of a standard
ridge estimator with a vanishing regularization parameter with rate
$\Theta(1/\sqrt{n}$). Moreover, the amount of shrinkage essentially does
not depend on the magnitude of $\beta_{0}$ due to the normalization of $\beta_0$ by $\sqrt{(1+\|\beta_{0}\|^{2})}$, which is also different from
the standard ridge setting.

Lastly, we discuss the distinction between our work and that of \citet{blanchet2022confidence}, which analyzes the asymptotic properties of a
general class of DRO estimators. In that work, the original estimators
are based on minimizing a sample loss of the form 
\begin{align*}
\frac{1}{n}\sum_{i=1}^n\ell(X_i,Y_i,\beta),
\end{align*}
 which encompasses most M-estimators, including the maximum likelihood estimator, and they focus on the case when $\rho_n \rightarrow 0$. However,
the IV (TSLS) estimator is different in that it is a moment-based
estimator, more precisely a GMM estimator \citep{hansen1982large}. The key distinction
between these estimators is that the objective function
of GMM estimators (and Z-estimators based on estimating equations) usually converges to a weighted distance function that
evaluates to 0 at the true parameter $\beta_{0}$, whereas the objectives of M-estimators tend to converge to a limit that does not vanish even at
the true parameter. To see this distinction more precisely, consider the limit of the OLS objective
under the linear model $Y_{i}=X_{i}^{T}\beta+\epsilon_{i}$ with $\mathbb{E}(X_{i}\epsilon_{i})=0$
and $\frac{1}{n}\mathbf{X}^{T}\mathbf{X}\rightarrow_p I_p$:
\begin{align*}
\frac{1}{n}(\mathbf{Y}-\mathbf{X}\beta)^{T}(\mathbf{Y}-\mathbf{X}\beta) & =\frac{1}{n}(\mathbf{X}\beta_{0}+\epsilon-\mathbf{X}\beta)^{T}(\mathbf{X}\beta_{0}+\epsilon-\mathbf{X}\beta)\\
 & \rightarrow(\beta_{0}-\beta)^{T}(\beta_{0}-\beta)+\sigma^{2}(\epsilon),
\end{align*}
 which is minimized at $\beta_{0}$, achieving a minimum of $\sigma^{2}(\epsilon)$.
On the other hand, consider the following GMM version of the OLS estimator,
based on the moment condition that $\mathbb{E}(X_{i}\epsilon_{i})=0$:
\begin{align*}
\min_{\beta}\frac{1}{n}\left\{ (\mathbf{Y}-\mathbf{X}\beta)^{T}\mathbf{X}\right\} W\left\{ \mathbf{X}^{T}(\mathbf{Y}-\mathbf{X}\beta)\right\} ,
\end{align*}
 where $W$ is a weighting matrix, with the optimal choice being $(\mathbf{X}^{T}\mathbf{X})^{-1}$
in this setting. We have, assuming again $\frac{1}{n}\mathbf{X}^{T}\mathbf{X} \rightarrow_p I_p$,  
\begin{align*}
\frac{1}{n}\left\{ (\mathbf{Y}-\mathbf{X}\beta)^{T}\mathbf{X}\right\} (\mathbf{X}^{T}\mathbf{X})^{-1}\left\{ \mathbf{X}^{T}(\mathbf{Y}-\mathbf{X}\beta)\right\}  & =\left\{ \frac{1}{n}(\mathbf{Y}-\mathbf{X}\beta)^{T}\mathbf{X}\right\} (\frac{1}{n}\mathbf{X}^{T}\mathbf{X})^{-1}\left\{ \frac{1}{n}\mathbf{X}^{T}(\mathbf{Y}-\mathbf{X}\beta)\right\} \\
 & \rightarrow(\beta_{0}-\beta)^{T}I(\beta_{0}-\beta)=\|\beta_{0}-\beta\|^2,
\end{align*}
 which is also minimized at $\beta_{0}$ but achieves a minimum value
of $0$. This distinction between M-estimators and Z-estimators (and GMM estimators)
is negligible in the standard setting without the distributionally robust optimization component, and in fact the
standard OLS estimator is preferable to the GMM version for being
more stable \citep{hall2003generalized}. However, when we apply square root ridge regularization to these estimators, they start behaving differently. Only regularized regression based on GMM and Z-estimators enjoys consistency with a non-zero $\rho>0$. In Appendix \ref{sec:WDRO-GMM}, we exploit this property to generalize our results and develop asymptotic results for a general class of GMM estimators.

\section{Extensions to GMM Estimation and $q$-Wasserstein Distances}
\label{sec:extensions}
In this section, we consider generalizations of the framework and results in the main paper. We first formulate a Wasserstein Distributionally Robust GMM Estimation Framework, and generalize the asymptotic results on Wasserstein DRIVE in this setting. We then consider Wasserstein DRIVE with $q$-Wasserstein distance where $q\neq 2$, and demonstrate that the resulting estimator enjoys a similar consistency property with non-vanishing robustness/regularization parameter.
\subsection{Wasserstein Distributionally Robust GMM}
\label{sec:WDRO-GMM}
In this section, we consider general GMM estimation and propose a
distributionally robust GMM estimation framework. Let $\theta_{0}\in\mathbb{R}^{p}$
be the true parameter vector in the interior of some compact space
$\Theta\subseteq\mathbb{R}^{p}$. Let $\psi(W,\theta)$ be a vector
of moments that satisfy 
\begin{align*}
\mathbb{E}[\psi(W_{i},\theta_{0})] & =0,
\end{align*}
for all $i$, where $\{W_{1},\dots,W_{n}\}$ are independent but not
necessarily identically distributed variables. Let $\psi_{i}(\theta_{0})=\psi(W_{i},\theta)$.
We consider the GMM estimators that minimize the objective 
\begin{align*}
\min_{\theta}\left(\frac{1}{n}\sum_{i}\psi_{i}(\theta)\right)^{T}W_{n}(\theta)\left(\frac{1}{n}\sum_{i}\psi_{i}(\theta)\right)
\end{align*}
 where $W_{n}$ is a positive definite weight matrix, e.g., the weight
matrix corresponding to the two-step or continuous updating estimator,
and $\frac{1}{n}\sum_{i}\psi_{i}(\theta)$ are the sample moments
under the empirical distribution $\mathbb{P}_{n}$ on $\psi(\theta)$.
Both the IV estimation and GMM formulation of OLS regression fall
under this framework. When we are uncertain about the validity of
the moment conditions, similarly to the Wasserstein DRIVE, we consider
a regularized regression objective given by 
\begin{align}\min_{\theta}\sqrt{\left(\frac{1}{n}\sum_{i}\psi_{i}(\theta)\right)^{T}W_{n}(\theta)\left(\frac{1}{n}\sum_{i}\psi_{i}(\theta)\right)}+\sqrt{\rho(1+\|\theta\|^{2})}.\end{align}
We will study the asymptotic properties of this regularized GMM objective. 
We make use of the following sufficient technical conditions in \citet{caner2009lasso} on GMM estimation to simplify the proof.

\begin{assumption}
\label{ass:gmm}
The following conditions are satisfied:
\begin{enumerate}
\item For all $i$ and $\theta_{1},\theta_{2}\in\Theta$, we have $|\psi_{i}(\theta_{1})-\psi_{i}(\theta_{2})|\le B_{t}|\theta_{1},\theta_{2}|$,
with $\lim_{n\rightarrow\infty}\frac{1}{n}\sum_{i=1}^{n}\mathbb{E}B_{t}^{d}<\infty$
for some $d>2$; $\sup_{\theta\in\Theta}\mathbb{E}|\psi_{i}(\theta)|^{d}<\infty$
for some $d>2$.
\item Let $m_{n}(\theta):=\frac{1}{n}\mathbb{E}\sum_{i}\psi(\theta)$ and
assume that $m_{n}(\theta)\rightarrow m(\theta)$ uniformly over $\Theta$,
$m_{n}(\theta)$ is continuously differentiable in $\theta$, $m_{1}(\theta_{0})=0$
if and only if $\theta=\theta_{0}$, and $m(\theta)$ is continuous
in $\theta$; The Jacobian matrix $\partial m_{n}(\theta)/\partial\theta\rightarrow_{p}J(\theta)$
in a neighborhood of $\theta$, and $J(\theta_{0})$ has full rank.
\item $W_{n}(\theta)$ is positive definite and continuous on $\Theta$,
and $W_{n}(\theta)\rightarrow_{p}W(\theta)$ uniformly in $\theta$.
$W(\theta)$ is continuous in $\theta$ and positive definite for
all $\theta\in\Theta$. 
\item The population objective $m(\theta)^{T}W(\theta)m(\theta)$ is lower
bounded by the squared distance $\|\theta-\theta_{0}\|^{2}$, i.e.,
$m(\theta)^{T}W(\theta)m(\theta)\geq\overline{\rho}\|\theta-\theta_{0}\|^{2}$
for all $\theta\in\Theta$ and some $\overline{\rho}>0$.
\end{enumerate}
\end{assumption}
See also \citet{andrews1994empirical,stock2000gmm} which assume similar conditions as 1-3 on the GMM estimation setup. Condition 4 requires that the weighted moment is bounded below by a quadratic function near $\theta_0$. Under these conditions, we have the following result.
\begin{theorem}
\label{thm:gmm}
Under the assumptions in, the unique solution $\hat{\theta}^{GMM}$
to 
\begin{align*}
\min_{\theta}\sqrt{\left(\frac{1}{n}\sum_{i}\psi_{i}(\theta)\right)^{T}W_{n}(\theta)\left(\frac{1}{n}\sum_{i}\psi_{i}(\theta)\right)}+\sqrt{\rho_{n}(1+\|\theta\|^{2})}
\end{align*}
converges to the solution $\theta^{GMM}$ of the population objective
\begin{align*}
\min_{\theta}\sqrt{m(\theta)^{T}W(\theta)m(\theta)}+\sqrt{\rho(1+\|\theta\|^{2})}.
\end{align*}
Moreover, whenever $\rho\leq\overline{\rho}$, $\theta^{GMM}=\theta_{0}$,
so that $\hat{\theta}^{GMM}\rightarrow_{p}\theta_{0}$. 
\end{theorem}
Therefore, the square root ridge regularized GMM estimator also satisfies the consistency property with a non-zero regularization parameter $\rho$. Next, we consider general $q$-Wasserstein distance with $q\neq 2$. 

\subsection{Generalization to $q$-Wasserstein DRIVE}
\label{sec:q-Wasserstein-DRIVE}
  The duality result in \cref{thm:duality} can be generalized to $q$-Wasserstein ambiguity sets. The resulting estimator can enjoy a similar consistency result as the square root Wasserstein DRIVE ($q=2$), but only when $q\in (1,2]$. This is because the limiting objective can be written as (assuming $\rho=1$ and $\lambda_p(\gamma^T \gamma)=1$)
  \begin{align*}
\sqrt{(\beta-\beta_{0})^{T}\gamma^T \gamma(\beta-\beta_{0})}+\sqrt[^p]{(\|\beta\|^{p}+1)},
\end{align*}
where $1/p+1/q=1$. When $q\in (1,2]$, $p\in [2,\infty)$, and so $\|x\|_2 \geq \|x\|_p$. As a result, the limiting objective is bounded below by  \begin{align*}
\|\beta-\beta_{0}\|_2+\sqrt[^p]{(\|\beta\|^{p}+1)} & \geq \|(\beta,-1)-(\beta_{0},-1)\|_p+\sqrt[^p]{(\|\beta\|^{p}+1)}\\
& = \|(\beta,-1)-(\beta_{0},-1)\|_p+\|(\beta,-1)\|_p\\
& \geq \|(\beta_0,-1)\|_p,
\end{align*}
with equality holding in both inequalities if and only if $\beta=\beta_0$, i.e., $\beta_0$ is again the unique minimizer of the limiting objective. We therefore have the following result.
\begin{corollary}
Under the same assumptions as \cref{thm:DRIVE-consistency}, the following regularized regression problem 
\begin{align}
\min_{\beta}\sqrt{\frac{1}{n}\sum_{i}(\Pi_{\mathbf{Z}}\mathbf{Y}-\Pi_{\mathbf{Z}}\mathbf{X}\beta)_{i}^{2}}+\sqrt[^p]{\rho_n(\|\beta\|^{p}+1)}
\end{align}
has a unique solution that converges in probability to $\beta_0$ whenever $q\in(1,2]$ and $\lim_{n\rightarrow\infty}\rho_n\leq \lambda_p(\gamma^T\Sigma_Z \gamma)$.
\end{corollary}

\section{Regularization Parameter Selection for Wasserstein DRIVE}
\label{sec:penalty-selection}
The selection of penalty/regularization parameters is an important consideration for all regularized regression problems. The most common approach is cross validation based on loss function minimization. However, for Wasserstein DRIVE, this standard cross validation procedure may not adequately address the challenges and goals of DRIVE. For example, from \cref{thm:DRIVE-consistency} we know that the Wasserstein DRIVE is only consistent when the penalty parameter is bounded above. We therefore need to take this result into account when selecting the penalty parameter.
% Now we are interested in selecting a good penalty $\rho$ from \emph{data} when estimating the Wasserstein DRIVE:
% \begin{align*}
% \min_\beta\sup_{\{Q:d(Q,P)\leq\rho\}}\mathbb{E}_{Q}\left[(P_Z(Y_i-X_i^{T}\beta))^{2}\right],
% \end{align*}
% where $P$ is the empirical distribution and $d$ is the Wasserstein distance between two distributions with square loss. Recall that this optimization problem is equivalent to the following square root ridge penalized version of IV estimation:
% \begin{align*}
% \min_{\beta}\sqrt{\frac{1}{n}\sum_{i}(\tilde{Y}_{i}-\tilde{X}_{i}^T\beta)^{2}}+\sqrt{\rho(\|\beta\|^{2}+1)},
% \end{align*}
% where $\tilde{Y}_i = P_ZY_i$ and $\tilde{X}_i=P_ZX_i$ are outcomes and covariates projected to the space spanned by instruments $Z$. 
In this section, we discuss two selection procedures, one based on the first stage regression coefficient, and the other based on quantiles of the score estimated using a nonparametric bootstrap procedure, which is also of independent interest. We connect our procedures to existing works in the literature on weak and invalid IVs and investigate their empirical performance in \cref{sec:numerical}.

\subsection{Selecting $\rho$ Based on Estimate of First Stage Coefficient}
\cref{thm:DRIVE-consistency} guarantees that as long as the regularization parameter converges to a value in the interval $[0,\sigma_{\min}(\gamma)]$, Wasserstein DRIVE is consistent. A natural procedure to select $\rho$ is thus to compute the minimum singular value ${\rho}_{\max}:=\sigma_{\min}(\hat \gamma)$ of the first stage regression coefficient $\hat{\gamma}$ and then select a regularization parameter $\rho=c\cdot {\rho}_{\max}$ for $c\in [0,1]$. In \cref{sec:numerical}, we verify that this procedure produces consistent DRIVE estimators whenever instruments are valid. Moreover, when the instrument is invalid or weak, Wasserstein DRIVE enjoys superior finite sample properties, outperforming the standard IV, OLS, and related estimators at estimation accuracy and prediction accuracy under distributional shift. This approach is also related to the test of \citet{cragg1993testing}, which is originally used to test for under-identification, and later used by \citet{stock2002testing} to test for weak instruments. In the Cragg-Donald test, the minimum eigenvalue of the first stage rank matrix is used to construct the $F$-statistic. 

Although selecting $\rho$ based on the first stage coefficient gives rise to Wasserstein DRIVE estimators that perform well in practice, there is one important challenge that remains to be addressed. Recall that violations of the exclusion restriction can be viewed as a form of distributional shift. We therefore expect that as the degree of invalidity increases, the distributional shift becomes stronger. From the DRO formulation of DRIVE in \cref{eq:Wasserstein-DRIVE}, we know that the regularization parameter $\rho$ is also the radius of the Wasserstein distribution set. Therefore, $\rho$ should adaptively \emph{increase} with increasingly invalid instruments. However, as the selection procedure proposed here only depends on the first stage estimate, it does not take this consideration into account. More importantly, when the instruments are weak, the smallest singular of the first stage coefficient is likely to be very close to zero, which results in a DRIVE estimate with a very small penalty parameter and may thus have similar problems as the standard IV. We next introduce another parameter selection procedure for $\rho$ based on \cref{thm:sqrt-ridge-root-n-consistency} that is able to better handle invalid and weak instruments.

\subsection{Selecting $\rho$ Based on Nonparametric Bootstrap of Quantile of Score}
Recall that the square root LASSO uses the following valid penalty:
\begin{align*}
\lambda^{\ast} & =cn\|\tilde{S}\|_{\infty},
\end{align*}
where the score function $\tilde{S}=\nabla\hat{Q}^{1/2}(\beta_{0})=\frac{E_{n}(x\epsilon)}{\sqrt{E_{n}(\epsilon^{2})}}$
with
\begin{align*}
\hat{Q}(\beta) & =\frac{1}{n}\sum_{i}(Y_{i}-X_{i}^{T}\beta)^{2},
\end{align*}
and $c=1.1$ is a constant of \citet{bickel2009simultaneous}.
The intuition for this penalty level comes from the simplest case $\beta_{0}\equiv0$, when the optimality condition requires $\lambda\geq n\|\tilde{S}\|_{\infty}$.
To estimate $\|\tilde{S}\|_{\infty}$, \citet{belloni2011square} propose to estimate
the empirical $(1-\alpha)$-quantile (conditional on $X_{i}$) of
$\frac{\|E_{n}(x\epsilon)\|_{\infty}}{\sqrt{E_{n}(\epsilon^{2})}}$
by sampling i.i.d. errors $\epsilon$ from the \emph{known }error
distribution $\Phi$ with zero mean and variance 1, resulting in 
\begin{align}
\label{eq:sqrt-LASSO-penalty}
    \lambda^\ast = c\sqrt{n}\Phi^{-1}(1-\alpha/2p),
\end{align}
where the confidence level $1-\alpha$ is usually set to 0.95. 

The consistency result in \cref{thm:sqrt-ridge-root-n-consistency} then suggests a natural choice of penalty parameter $\rho$ for the square root ridge, given by $\sqrt \rho = \frac{\sqrt{p}}{n} \lambda^\ast$, where $\lambda^\ast$ is constructed from \eqref{eq:sqrt-LASSO-penalty}. However, there are two main challenges when applying this regularization parameter selection procedure to Wasserstein DRIVE in the instrumental variables estimation setting. First, it requires prior knowledge of the type
of distribution $\Phi$, e.g., Gaussian, of the errors $\epsilon$, even if we do not
need its variance. Second, $\sqrt \rho = \frac{\sqrt{p}}{n} \lambda^\ast$
is only valid for the square root ridge problem in the standard regression setting without instruments. When applied to the IV setting, the empirical risk is now 
\begin{align*}
\hat{Q}(\beta) & =\frac{1}{n}\sum_{i}(\tilde{Y}_{i}-\tilde{X}_{i}^{T}\beta)^{2},
\end{align*}
 where $\tilde{Y}_{i}=(\Pi_{\mathbf{Z}}\mathbf{Y})_i$ and $\tilde{X}_{i}=(\Pi_{\mathbf{Z}}\mathbf{X})_i$ are variables projected
to the instrument space. This means that ``observations'' $(\tilde{Y}_{i},\tilde{X}_{i})$
are no longer independent. Therefore, the i.i.d. assumption on the
errors in the standard regression setting no longer holds.

\begin{figure}
    \centering % <-- added
\begin{subfigure}{0.4\textwidth}
  \includegraphics[width=\linewidth]{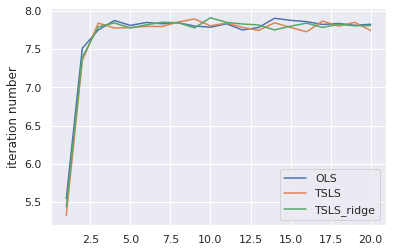}
  %\caption{$\eta=0$}
\end{subfigure}\hfil % <-- added
\begin{subfigure}{0.4\textwidth}
  \includegraphics[width=\linewidth]{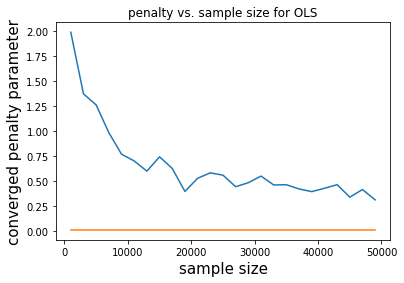}
  %\caption{$\eta=0.1$}
\end{subfigure}\hfil % <-- added

\caption{Left: Penalty parameter convergence as a function of iteration number, with $\beta^{(0)}$ starting from OLS, TSLS, and TSLS ridge estimates. Right: Converged penalty for the standard linear regression model as a function of sample size. %\textcolor{red}{Figure quality can be improved as mentioned in Figure 1}
}
\label{fig:nonparametric-bootstrap}
\end{figure}

We propose the following iterative procedure based on nonparametric
bootstrap that simultaneous addresses the two challenges above. Given a starting estimate $\beta^{(0)}$ of $\beta_{0}$
(say the IV estimator), we compute the residuals $r_{i}^{(0)}=\tilde{Y}_{i}-\tilde{X}_{i}^{T}\beta^{(0)}$.
Then we bootstrap these residuals to compute the empirical quantile
of 
\begin{align*}
\frac{\|E_{n}(\tilde{x}\epsilon)\|_{\infty}}{\sqrt{E_{n}(\epsilon^{2})}},%?\rightarrow\frac{\|\mathbb{E}(\tilde{x}\epsilon)\|_{\infty}}{\sqrt{\mathbb{E}(\epsilon^{2})}}
\end{align*}
 where $\epsilon$ is drawn uniformly with replacement from the residuals $r_{i}$.
The quantile based on bootstrap then replaces $\Phi^{-1}(1-\alpha/2p)$ in \eqref{eq:sqrt-LASSO-penalty} to give a penalty level $\rho$, which we can use to solve the square root
ridge problem to obtain a new estimate $\beta^{(1)}$. Then we use
$\beta^{(1)}$ to compute new residuals $r_{i}^{(1)}=\tilde{Y}_{i}-\tilde{X}_{i}^{T}\beta^{(1)}$,
and repeat the process. In practice, we can use the OLS or TSLS estimate as the starting point $\beta^{(0)}$. \cref{fig:nonparametric-bootstrap} shows that this procedure converges very quickly and does not depend on the initial $\beta^{(0)}$. Moreover, in \cref{sec:numerical} we demonstrate that the resulting Wasserstein DRIVE estimator has superior estimation performance in terms of $\ell^{2}$ error, as well as prediction under distributional shift. 
% Lastly, instead of estimating
% the quantile of 
% \begin{align*}
% c\sqrt{p}\frac{\|E_{n}(\tilde{x}\epsilon)\|_{\infty}}{\sqrt{E_{n}(\epsilon^{2})}},
% \end{align*}
%  via bootstrapping, we can also consider estimating the quantile of
% \begin{align*}
% c\frac{\|E_{n}(\tilde{x}\epsilon)\|_{2}}{\sqrt{E_{n}(\epsilon^{2})}}.
% \end{align*}
% In practice, we noted that the bootstrapped quantile of $c\sqrt{p}\frac{\|E_{n}(\tilde{x}\epsilon)\|_{\infty}}{\sqrt{E_{n}(\epsilon^{2})}}$
% works better. 

\subsection{Bootstrapped Score Quantile As Test Statistic for Invalid Instruments}
 When instruments are valid, one should expect the boostrapped quantiles will converge to 0. We next formalize this intuition in \cref{prop:quantiles-vanish-valid-iv} and also confirm it in numerical experiments. 
 \begin{proposition}
 \label{prop:quantiles-vanish-valid-iv}
     The bootstrapped quantiles converge to 0 when instruments are valid. 
 \end{proposition}
\begin{figure}
\begin{centering}
\includegraphics[scale=0.4]{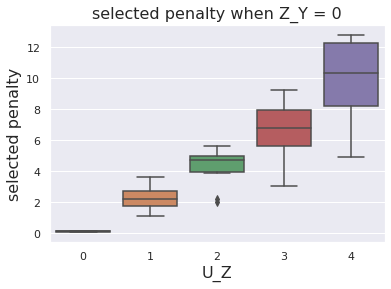}
\par\end{centering}
\caption{Penalty strength selected based on nonparametric bootstrap of score quantile vs. correlation strength between invalid instrument and unobserved confounders.}
\label{fig:quantile-increase-invalid-iv}
\end{figure}
 
 More importantly, in practice we observe that the bootstrapped quantile increases with the degree of instrument invalidity. \cref{fig:quantile-increase-invalid-iv} illustrates this phenomenon with increasing correlation between the instruments and the unobserved confounder. The intuition behind this observation is that the quantile is essentially describing the orthogonality (moment) condition for valid IVs, and so should be close to zero with valid IV. A large value therefore indicates possible violation. Thus, the bootstrapped quantile could potentially be used as a test statistic for invalid instruments, using for example permutation tests. Equivalently, in a sensitivity analysis it could be used as a sensitivity parameter, based on which we can bound the worst bias of IV/OLS models. 
 
 We provide a more detailed discussion to further justify our proposal. In a linear regression setting, the quantity 
\begin{align*}
\frac{\|\sum_{i}(x\epsilon)\|_{\infty}}{\sqrt{\sum_{i}(\epsilon^{2})}}
\end{align*}
 is a test statistic for the orthogonality condition $\mathbb{E}[X(Y-X\beta)]=0$
which holds asymptotically for $\beta$ equal to the OLS estimator.
When $\frac{\|\sum_{i}(x\epsilon)\|_{\infty}}{\sqrt{\sum_{i}(\epsilon^{2})}}$
is not zero, it indicates a violation of the orthogonality condition,
which means a non-zero penalty could be beneficial. Similarly, in
a TSLS model
\begin{align*}
\frac{\|\sum_{i}(\tilde{x}\epsilon)\|_{\infty}}{\sqrt{\sum_{i}(\epsilon^{2})}}
\end{align*}
 is a test statistic for the orthogonality condition $\mathbb{E}[\tilde{X}(\tilde{Y}-\tilde{X}\beta)]=0$
which is asymptotically correct when $\beta$ is the TSLS estimator
and $Z$ is valid instrument. A large $\frac{\|\sum_{i}(\tilde{x}\epsilon)\|_{\infty}}{\sqrt{\sum_{i}(\epsilon^{2})}}$
therefore indicates potential violations of the IV assumptions. We may also compare this quantity with the Sargan test statistic \citep{sargan1958estimation} for instrument invalidity in the over-identified setting and note similarities. 

The penalty selection proposed in \citet{belloni2011square} can therefore be seen as a test statistic
for the moment condition $E(X(Y-X\beta))=0$ which should hold asymptotically
for $\beta$ equal to the OLS estimator if the model assumption that X is
independent of the error term is correct. So if the penalty is large,
it is evidence for potential violation of $X\perp\epsilon$. Similarly,
in a TSLS model, the moment condition is $E(\tilde{X}(\tilde{Y}-\tilde{X}\beta))=0$
for beta equal to the TSLS estimator, so the penalty can be seen as
assessing potential violation of IV assumptions.

\begin{remark}
   \normalfont 
Besides the data-driven procedures discussed above, we can also consider incorporating information provided by statistical tests for IV estimation. For example, in over-identified settings, the Sargan-Hasen test \citep{sargan1958estimation,hansen1982large} can be used to test for the exclusion restriction. We can use this test to provide evidence on the validity of the instrument. For testing weak instruments, the popular test of \citet{stock2002testing} can be used. This proposal is also related to our observation that $\rho$ based on bootstrapped quantiles increase with the degree of invalidity, i.e., direct effect on the outcome or correlation with confounders, and can therefore potentially be used as a test statistic for the \emph{reliability} of the IV estimator. We leave a detailed investigation of this proposal to future work. 
\end{remark}

\section{Proofs}
\label{sec:proofs}
\subsection{Proof of Theorem \ref{thm:duality}}
\begin{proof}
The proof of Theorem \ref{thm:duality} relies on a general duality result on Wasserstein DRO, with different variations derived in \citep{gao2016distributionally,blanchet2019quantifying,sinha2017certifying}.
We start with the inner problem in the objective in \eqref{eq:Wasserstein-DRIVE}: \begin{align*}
\sup_{\{\mathbb{Q}:D(\mathbb{Q}, \tilde{\mathbb{P}}_n)\leq\rho\}}\mathbb{E}_{\mathbb{Q}}\left[(\tilde{Y}-\tilde{X}^{T}\beta)^{2}\right],
\end{align*}
 where $D$ is the 2-Wasserstein distance and $\tilde {\mathbb{P}}_n$ is the
empirical distribution on the projected data $\{\tilde Y_i, \tilde X_i\}_{i=1}^{n}\equiv \{(\Pi_Z \mathbf{Y})_i, (\Pi_Z \mathbf{X})_i\}_{i=1}^n$. Proposition 1 of \cite{sinha2017certifying} and Proposition 1 of \cite{blanchet2019robust} both imply that 
\begin{align*}
\sup_{\{\mathbb{Q}:D(\mathbb{Q},\mathbb{P})\leq\rho\}}\mathbb{E}_{\mathbb{Q}}\left[(\tilde{Y}-\tilde{X}^{T}\beta)^{2}\right] & =\inf_{\gamma\geq0}\gamma\rho+\frac{1}{n}\sum_{i=1}^{n}\phi_{\gamma}(\beta;(\tilde{X}_i,\tilde{Y}_i)),
\end{align*}
 where the ``robust'' loss function is
 \begin{align*}
\phi_{\gamma}(\beta;(\tilde{X},\tilde{Y})) & =\sup_{(X,Y)}(Y-X^{T}\beta)^{2}-\gamma\|X-\tilde{X}\|_{2}^{2}-\gamma(Y-\tilde{Y})^{2}\\
 & =\sup_{W}W^{T}\alpha\alpha^{T}W-\gamma\|W-\tilde{W}\|_{2}^{2},
\end{align*}
with $W=(X,Y)$, $\tilde W=(\tilde X,\tilde Y)$ and $\alpha=(-\beta,1)$. Note that $\gamma$ is always chosen large enough, i.e., $\gamma I-\alpha\alpha^{T} \succeq 0$, so that the objective $W^{T}\alpha\alpha^{T}W-\gamma\|W-\tilde{W}\|_{2}^{2}$ is concave in $W$. Otherwise, the supremum over $W$ in the inner problem is unbounded. Therefore, the first order condition is sufficient: 
\begin{align*}
\alpha\alpha^{T}W-\gamma(W-\tilde{W}) & =0,
\end{align*}
so that 
\begin{align*}
(\alpha\alpha^{T}-\gamma I)W & =-\gamma\tilde{W},
\end{align*}
and
\begin{align*}
W & =\gamma(\gamma I-\alpha\alpha^{T})^{-1}\tilde{W}\\
 & =(I-\alpha\alpha^{T}/\gamma)^{-1}\tilde{W},
\end{align*}
where $I-\alpha\alpha^{T}/\gamma$ is invertible if $\gamma I-\alpha\alpha^{T}$
is positive definite, which is required to make sure that the quadratic
is concave in $W$. The supremum is then given by 
\begin{align*}
&\tilde{W}^{T}(I-\alpha\alpha^{T}/\gamma)^{-1}\alpha\alpha^{T}(I-\alpha\alpha^{T}/\gamma)^{-1}\tilde{W}-\gamma(\tilde{W}^{T}((I-\alpha\alpha^{T}/\gamma)^{-1}-I)^{2}\tilde{W})\\
&=\tilde{W}^{T}((I-\alpha\alpha^{T}/\gamma)^{-1}\alpha\alpha^{T}(I-\alpha\alpha^{T}/\gamma)^{-1}-\gamma((I-\alpha\alpha^{T}/\gamma)^{-1}-I)^{2})\tilde{W} \equiv\|\tilde{W}\|_{A}^{2},
\end{align*}
where 
\begin{align*}
A & =((I-\alpha\alpha^{T}/\gamma)^{-1}\alpha\alpha^{T}(I-\alpha\alpha^{T}/\gamma)^{-1}-\gamma((I-\alpha\alpha^{T}/\gamma)^{-1}-I)^{2}).
\end{align*}
Using the Sherman-Morrison Lemma \citep{sherman1950adjustment}, whose condition is satisfied if
$\gamma I-\alpha\alpha^{T}$ is positive definite, 
\begin{align*}
(I-\alpha\alpha^{T}/\gamma)^{-1}=I+\frac{1}{\gamma-\alpha^{T}\alpha}\alpha\alpha^{T},
\end{align*}
and $A$ can be simplified as 
\begin{align*}
A=\frac{\gamma}{\gamma-\alpha^{T}\alpha}\alpha\alpha^{T}.
\end{align*}
%Compare this with $\gamma\rightarrow\infty$, which recovers the classical
%IV setting since then $A\rightarrow\alpha\alpha^{T}$.

In summary, for each projected observation (for the IV estimate)
$\tilde{W}_i=(\tilde{X}_i,\tilde{Y}_i)$, we can obtain a new ``robustified''
sample using the above operation, then minimize the following modified empirical risk
constructed from the robustified samples: 
\begin{align*}
\min_\beta \sup_{\{\mathbb{Q}:D(\mathbb{Q},\mathbb{P})\leq\rho\}}\mathbb{E}_{\mathbb{Q}}\left[(\tilde{Y}_{i}-\tilde{X}_{i}^{T}\beta)^{2}\right] &\Leftrightarrow \min_\beta \inf_{\gamma\geq0}\gamma\rho+\frac{1}{n}\sum_{i=1}^{n}(\phi_{\gamma}(\beta;(\tilde{X}_{i},\tilde{Y}_{i}))\\
&\Leftrightarrow \min_{\beta}\inf_{\gamma\geq0}\gamma\rho+\frac{1}{n}\sum_{i}\|(\tilde{X}_{i},\tilde{Y}_{i})\|_{A}^{2},
\end{align*}
where for fixed $\beta$, $\gamma\geq0$ is always chosen large enough
so that $\phi_{\gamma}(\beta;X,Y)$ is finite. 

Now, the inner minimization problem can be further solved explicitly. Recall that it is equal to
\begin{align*}
\inf_{\gamma\geq0}\gamma\rho+\frac{1}{n}\sum_{i}\tilde{W}_{i}^{T}(\frac{\gamma}{\gamma-\alpha^{T}\alpha}\alpha\alpha^{T})\tilde{W}_{i},
\end{align*}
which is convex in $\gamma$ hence minimized at the first order condition: 
\begin{align*}
\rho=\frac{1}{n}\frac{\sum_{i}\tilde{W}_{i}^{T}\alpha\alpha^{T}\tilde{W}_{i}\alpha^{T}\alpha}{(\gamma-\alpha^{T}\alpha)^{2}},
\end{align*}
or $\gamma=\sqrt{\frac{1}{n}\frac{\sum_{i}\tilde{W}_{i}^{T}\alpha\alpha^{T}\tilde{W}_{i}\alpha^{T}\alpha}{\rho}}+\alpha^{T}\alpha$, where we have chosen the larger root since only it is guaranteed to satisfy $\gamma I-\alpha\alpha^{T} \succeq 0$ for any $\alpha = (-\beta,1)$.

Plugging this expression of $\gamma$ into the objective, and using the notation $\ell_{IV}:= \frac{1}{n}\sum_{i}(\tilde{Y}_{i}-\beta^{T}\tilde{X}_{i})^{2}$,
\begin{align*}
\gamma\rho+\frac{1}{n}\sum_{i}\|(\tilde{X}_{i},\tilde{Y}_{i})\|_{A}^{2} & =\sqrt{\rho\alpha^{T}\alpha\cdot\frac{1}{n}\sum_{i}\tilde{W}_{i}^{T}\alpha\alpha^{T}\tilde{W}_{i}}+\rho\alpha^{T}\alpha\\
 & +\frac{1}{n}\sum_{i}\tilde{W}_{i}^{T}(\frac{1}{1-\frac{\alpha^{T}\alpha}{\sqrt{\frac{1}{n}\frac{\sum_{i}\tilde{W}_{i}^{T}\alpha\alpha^{T}\tilde{W}_{i}\alpha^{T}\alpha}{\rho}}+\alpha^{T}\alpha}}\alpha\alpha^{T})\tilde{W}_{i}\\
 & =\sqrt{\rho\alpha^{T}\alpha\cdot\ell_{IV}}+\rho\alpha^{T}\alpha +\frac{\ell_{IV}}{\frac{\sqrt{\frac{1}{n}\frac{\sum_{i}\tilde{W}_{i}^{T}\alpha\alpha^{T}\tilde{W}_{i}}{\rho}}}{\sqrt{\frac{1}{n}\frac{\sum_{i}\tilde{W}_{i}^{T}\alpha\alpha^{T}\tilde{W}_{i}}{\rho}}+\sqrt{\alpha^{T}\alpha}}}\\
 & =\sqrt{\rho\alpha^{T}\alpha\cdot\ell_{IV}}+\rho\alpha^{T}\alpha+\ell_{IV}+\frac{\sqrt{\alpha^{T}\alpha}}{\sqrt{\frac{1}{n}\frac{\sum_{i}\tilde{W}_{i}^{T}\alpha\alpha^{T}\tilde{W}_{i}}{\rho}}}\ell_{IV}\\
 & =2\sqrt{\rho\alpha^{T}\alpha\cdot\ell_{IV}}+\rho\alpha^{T}\alpha+\ell_{IV}\\
 & =(\sqrt{\ell_{IV}}+\sqrt{\rho\alpha^{T}\alpha})^{2}.
\end{align*}
Therefore, we have proved that the Wasserstein DRIVE objective
\begin{align*}
\min_\beta \sup_{\{\mathbb{Q}:D(\mathbb{Q}, \tilde{\mathbb{P}}_n)\leq\rho\}}\mathbb{E}_{\mathbb{Q}}\left[(\tilde{Y}-\tilde{X}^{T}\beta)^{2}\right]
\end{align*}
is equivalent to the following square root ridge regularized IV objective:
\begin{align*}
\min_{\beta}\sqrt{\frac{1}{n}\sum_{i}(\tilde{Y}_{i}-\beta^{T}\tilde{X}_{i})^{2}}+\sqrt{\rho(\|\beta\|^{2}+1)}.
\end{align*}
\end{proof}
\subsection{Proof of \cref{thm:DRIVE-consistency}}
\begin{proof}
We will show that as $n\rightarrow\infty$, $\hat \beta^{\text{DRIVE}}\rightarrow\beta_{0}$
as long as $\rho_n\rightarrow \rho \leq \sqrt{\lambda_{p}(\gamma\Sigma_Z\gamma^{T})}$. Recall the linear IV model \eqref{eq:IV-model}
\begin{align*}
        Y &= \beta^T_0 X %+ Z\eta + 
        +\epsilon,\\
    X &= \gamma^T Z + \xi.
\end{align*} 
with instrument relevance and exogeneity conditions
\begin{align*}
\text{rank}(\mathbb{E}\left[ZX^T\right]) & = p,\\
\mathbb{E}\left[Z\epsilon\right]=0, \mathbb{E}\left[Z\xi^T\right] &= \mathbf{0}.
\end{align*}

First, we compute the limit of the objective function \eqref{eq:sqrt-IV-sample}, reproduced below
\begin{align}
\label{eq:proof-sample-objective}
    \sqrt{\frac{1}{n}\|\Pi_{Z}\mathbf{Y}-\Pi_{Z}\mathbf{X}\beta\|^{2}}+\sqrt{\rho_n(\|\beta\|^{2}+1)}.
\end{align}
For the loss term, we have 
\begin{align*}
\sqrt{\frac{1}{n}\sum_{i}(\Pi_{\mathbf{Z}}\mathbf{Y}-\Pi_{\mathbf{Z}}\mathbf{X}\beta)_{i}^{2}} & =\sqrt{\frac{1}{n}(\Pi_{\mathbf{Z}}\mathbf{Y}-\Pi_{\mathbf{Z}}\mathbf{X}\beta)^{T}(\Pi_{\mathbf{Z}}\mathbf{Y}-\Pi_{\mathbf{Z}}\mathbf{X}\beta)}\\
 & =\sqrt{\frac{1}{n}(\Pi_{\mathbf{Z}}(\mathbf{X}\beta_{0}+\mathbf{\epsilon})-\Pi_{\mathbf{Z}}\mathbf{X}\beta)^{T}(\Pi_{\mathbf{Z}}(\mathbf{X}\beta_{0}+\mathbf{\epsilon})-\Pi_{\mathbf{Z}}\mathbf{X}\beta)}\\
 & =\sqrt{\frac{1}{n}(\Pi_{\mathbf{Z}}\mathbf{X}(\beta_{0}-\beta)+\mathbf{\epsilon})^{T}(\Pi_{\mathbf{Z}}\mathbf{X}(\beta_{0}-\beta)+\mathbf{\epsilon})}\\
 & =\sqrt{\frac{1}{n}(\mathbf{\epsilon}^{T}\Pi_{\mathbf{Z}}\epsilon-2\mathbf{\epsilon}^{T}\Pi_{\mathbf{Z}}\mathbf{X}(\beta-\beta_0)+(\beta-\beta_{0})^{T}\mathbf{X}^{T}\Pi_{\mathbf{Z}}\mathbf{X}(\beta-\beta_{0}))}.
\end{align*}
 Note first that $\frac{1}{n}\mathbf{\epsilon}^{T}\Pi_{\mathbf{Z}}\mathbf{X}(\beta-\beta_0)=o_p(1)$
whenever the instruments are valid, since 
\begin{align*}
\frac{1}{n}\mathbf{\epsilon}^{T}\Pi_{\mathbf{Z}}\mathbf{X}(\beta-\beta_0) & =\frac{1}{n}(\sum_{i}\epsilon_{i}Z_{i})^{T}(\mathbf{Z}^{T}\mathbf{Z})^{-1}(\sum_iZ_iX_{i}^{T}(\beta-\beta_0))\\
 & =(\frac{1}{n}\sum_{i}\epsilon_{i}Z_{i})^{T}(\frac{1}{n}\mathbf{Z}^{T}\mathbf{Z})^{-1}(\frac{1}{n}\sum_iZ_iX_{i}^{T}(\beta-\beta_0))\\
 & \rightarrow_{p}\mathbb{E}[Z\epsilon]  \cdot  \Sigma^{-1}_Z \cdot \mathbb{E}[ZX^T]\cdot(\beta-\beta_0)=0,
\end{align*}
by the continuous mapping theorem. Similarly, 
\begin{align*}
\frac{1}{n}(\beta-\beta_{0})^{T}\mathbf{X}^{T}\Pi_{\mathbf{Z}}\mathbf{X}(\beta-\beta_{0}) & =\frac{1}{n}(\sum_{i}Z_{i}X_{i}^{T}(\beta-\beta_{0}))^{T}(\mathbf{Z}^{T}\mathbf{Z})^{-1}(\sum_{i}(Z_{i}X_{i}^{T}(\beta-\beta_{0}))\\
 & =(\frac{1}{n}\sum_{i}Z_{i}X_{i}^{T}(\beta-\beta_{0}))^{T}(\frac{1}{n}\mathbf{Z}^{T}\mathbf{Z})^{-1}(\frac{1}{n}\sum_{i}Z_{i}X_{i}^{T}(\beta-\beta_{0}))\\
 & \rightarrow_{p}(\beta-\beta_{0})^{T}\mathbb{E}(X_{i}Z_{i}^{T})\Sigma^{-1}_Z \mathbb{E}(Z_{i}X_{i}^{T})(\beta-\beta_{0})\\
 & =(\beta-\beta_{0})^{T}\gamma^T\Sigma_Z \Sigma^{-1}_Z \Sigma_Z \gamma (\beta-\beta_{0})\\
 & =(\beta-\beta_{0})^{T}\gamma^T \Sigma_Z \gamma(\beta-\beta_{0}).
\end{align*}
The most important part is the ``vanishing noise'' behavior, i.e.,
\begin{align*}
\frac{1}{n}\mathbf{\epsilon}^{T}\Pi_{\mathbf{Z}}\epsilon & =\frac{1}{n}(\sum_{i}\epsilon_{i}Z_{i})^{T}(\mathbf{Z}^{T}\mathbf{Z})^{-1}(\sum_{i}\epsilon_{i}Z_{i})\\
 & =\frac{1}{n}(\sum_{i}\epsilon_{i}Z_{i})^{T}(\frac{1}{n}\mathbf{Z}^{T}\mathbf{Z})^{-1}(\frac{1}{n}\sum_{i}\epsilon_{i}Z_{i})\\
 & \rightarrow_{p}(\mathbb{E}(\epsilon_{i}Z_{i}))^{T}\Sigma_Z^{-1}(\mathbb{E}(\epsilon_{i}Z_{i}))=0.
\end{align*}
 It then follows that the regularized regression objective \eqref{eq:sqrt-IV-sample} of the Wasserstein DRIVE estimator converges in probability to \eqref{eq:limiting-objective}, reproduced below
\begin{align}
\label{eq:proof-population-limit}
\sqrt{(\beta-\beta_{0})^{T}\gamma^T \Sigma_Z \gamma(\beta-\beta_{0})}+\sqrt{\rho(\|\beta\|^{2}+1)}.
\end{align}
For $\rho>0$, the population objective \eqref{eq:proof-population-limit} is continuous and strictly convex in $\beta$, and so has a unique minimizer $\beta^{\text{DRIVE}}$. Applying the convexity lemma of \citet{pollard1991asymptotics}, since \eqref{eq:proof-sample-objective}
is also strictly convex in $\beta$, the convergence to \eqref{eq:proof-population-limit} is uniform on compact sets $B\subseteq \mathbb{R}^p$ that contain $\beta^{\text{DRIVE}}$. Applying Corollary 3.2.3 of \citet{van_der_vaart1996}, we can therefore conclude that the minimizers of the empirical objectives converge in probability to the minimizer of the population objective, i.e., 
\[\hat \beta^{\text{DRIVE}} \rightarrow_p \beta^{\text{DRIVE}}.\]
% \begin{align*}
%     \arg \min_\beta \sqrt{\frac{1}{n}\|\Pi_{Z}\mathbf{Y}-\Pi_{Z}\mathbf{X}\beta\|^{2}}+\sqrt{\rho(\|\beta\|^{2}+1)} \rightarrow_p \\ \arg \min_\beta \sqrt{(\beta-\beta_{0})^{T}\gamma^T \Sigma_Z \gamma(\beta-\beta_{0})}+\sqrt{\rho(\|\beta\|^{2}+1)}. 
% \end{align*}
 
 Next, we consider minimizing the population objective \eqref{eq:proof-population-limit}. If $\rho$ is bounded above by the smallest singular value of $\gamma^T\Sigma_Z\gamma$, i.e., 
\begin{align*}
\rho & \leq \lambda_{p}(\gamma^T\Sigma_Z\gamma^{T}),
\end{align*}
the population objective is lower bounded by
\begin{align*}
\sqrt{(\beta-\beta_{0})^{T}\gamma^T \Sigma_Z \gamma(\beta-\beta_{0})}+\sqrt{\rho(\|\beta\|^{2}+1)} & \geq \sqrt{\rho}\|\beta-\beta_{0}\|_{2}+\sqrt{\rho}\sqrt{\|\beta\|^{2}+1}\\
 & =\sqrt{\rho}\|(\beta,1)-(\beta_{0},1)\|_{2}+\sqrt{\rho}\|(\beta,1)\|_{2}\\
 & \geq\sqrt{\rho}\|(\beta_{0},1)\|_{2},
\end{align*}
where in the second line we augment the vectors $\beta,\beta_0$ with an extra coordinate equal to 1. The last line follows from the triangle inequality, with equality if and only if $\beta\equiv \beta_0$. We can verify that the lower bound $\sqrt{\rho}\|(\beta_{0},1)\|_{2}$ of the population objective is therefore achieved uniquely at $\beta\equiv\beta_{0}$ due to strict convexity. We
have thus proved that when $0<\rho\leq\sqrt{\lambda_{p}(\gamma^T\Sigma_Z\gamma)}$,
the population objective has a unique minimizer at $\beta_{0}$. When $\rho=0$, the consistency of $\hat \beta^{\text{DRIVE}}$ can be similarly proved as long as $\lambda_{p}(\gamma^T\Sigma_Z\gamma)>0$, which guarantees that $\beta_0$ is the unique minimizer of \eqref{eq:proof-population-limit}. Therefore, whenever $\rho \leq \lambda_{p}(\gamma^T\Sigma_Z\gamma^{T})$, we have $\hat \beta^{\text{DRIVE}}\rightarrow_{p}\beta_{0}$.
% Note that this result requires that $\lambda_{p}(\gamma\gamma^{T})>0$,
% which exactly requires that the IV model is \emph{identified}. In
% practice, we can set the penalty parameter $\lambda$ to be equal
% to the square root of the smallest eigenvalue of the OLS estimator
% $\hat{\gamma}$ of the first stage. 
% We need to check
% that, for the objective functions $M_{n}$ and limiting objective
% function $M$,
% \begin{align*}
% \sup_{\beta\in B}|M_{n}(\beta)-M(\beta)| & \rightarrow_{p}0\\
% \sup\{M(\beta):\|\beta-\beta_{0}\|\geq\epsilon\} & >M(\beta_{0})\\
% M_{n}(\beta_{n}^{\text{DRIVE}}) & \leq M_{n}(\beta_{0})+o_{P}(1).
% \end{align*}
% The first condition is satisfied whenever $B$ compact. The second
% condition is satisfied since $M$ has a unique minimizer at $\beta_{0}$.
% The third condition trivially holds in our case since $M_{n}$ has
% a unique minimizer at $\beta_{n}^{\text{DRIVE}}$.
\end{proof}
\subsection{Proof of \cref{thm:asymptotic-distribution}}
\begin{proof}
Define the objective function $H_n(\delta)$ of a local parameter $\delta\in\mathbb{R}^p$ as follows:
\begin{align*}
\phi_{n}(\beta) & :=\sqrt{\frac{1}{n}\|\Pi_{Z}\mathbf{Y}-\Pi_{Z}\mathbf{X}\beta\|^{2}}+\sqrt{\rho_n(\|\beta\|^{2}+1)}\\
H_{n}(\delta) & :=\sqrt{n}\left[\phi_{n}(\beta_{0}+\delta/\sqrt{n})-\phi_{n}(\beta_{0})\right].
\end{align*}
Note that $H_{n}(\delta)$ is minimized at $\delta=\sqrt{n}(\hat{\beta}^{\text{DRIVE}}_{n}-\beta_{0})$.
The key components of the proof are to compute the uniform limit $H(\delta)$ of $H_{n}(\delta)$ on compact sets in the weak topology, and to verify that their minimizers are uniformly tight, i.e., $\sqrt{n}(\hat{\beta}^{\text{DRIVE}}_{n}-\beta_{0})=O_p(1)$. We can then apply Theorem 3.2.2 of \citet{van_der_vaart1996} to conclude that the sequence of minimizers $\sqrt{n}(\hat{\beta}_{n}-\beta_{0})$ of $H_{n}(\delta)$ converges in
distribution to the minimizer of the limit $H(\delta)$. We have 
\begin{align*}
H_{n}(\delta)=\sqrt{n}\cdot(\phi_{n}(\beta_{0}+\delta/\sqrt{n})-\phi_{n}(\beta_{0})) & =\underset{\textbf{I}}{\underbrace{\sqrt{\|\Pi_{Z}\mathbf{Y}-\Pi_{Z}\mathbf{X}(\beta_{0}+\delta/\sqrt{n})\|^{2}}-\sqrt{\|\Pi_{Z}\mathbf{Y}-\Pi_{Z}\mathbf{X}\beta_{0}\|^{2}}}}\\
 & +\underset{\textbf{II}}{\underbrace{\sqrt{n\rho_n(1+\|\beta_{0}+\delta/\sqrt{n}\|^{2})}-\sqrt{n\rho_n(1+\|\beta_{0}\|^{2})}}}.
\end{align*}
We first focus on \textbf{I}:
\begin{align*} 
    \textbf{I} & =\sqrt{\|\Pi_{Z}\mathbf{Y}-\Pi_{Z}\mathbf{X}(\beta_{0}+\delta/\sqrt{n})\|^{2}}-\sqrt{\|\Pi_{Z}\mathbf{Y}-\Pi_{Z}\mathbf{X}\beta_{0}\|^{2}}\\
 & =\sqrt{F_{n}(\beta_{0}+\delta/\sqrt{n})}-\sqrt{F_{n}(\beta_{0})},
\end{align*}
 where 
\begin{align*}
F_{n}(\beta) & =\|\Pi_{Z}\mathbf{Y}-\Pi_{Z}\mathbf{X}\beta\|^{2}.
\end{align*}
We have, with $\psi_{i}(\beta)\equiv Z_{i}(Y_{i}-\beta^{T}X_{i})$, 
\begin{align*}
F_{n}(\beta_{0}+\delta/\sqrt{n}) & =\|\Pi_{Z}\mathbf{Y}-\Pi_{Z}\mathbf{X}(\beta_{0}+\delta/\sqrt{n})\|^{2}\\
 & =(\frac{1}{\sqrt{n}}\sum_{i}\psi_{i}(\beta_{0}+\delta/\sqrt{n}))^{T}(\frac{1}{n}\mathbf{Z}^{T}\mathbf{Z})^{-1}(\frac{1}{\sqrt{n}}\sum_{i}\psi_{i}(\beta_{0}+\delta/\sqrt{n})),\\
F_{n}(\beta_{0}) & =\|\Pi_{Z}\mathbf{Y}-\Pi_{Z}\mathbf{X}\beta_{0}\|^{2}\\
 & =(\frac{1}{\sqrt{n}}\sum_{i}\psi_{i}(\beta_{0}))^{T}(\frac{1}{n}\mathbf{Z}^{T}\mathbf{Z})^{-1}(\frac{1}{\sqrt{n}}\sum_{i}\psi_{i}(\beta_{0})).
\end{align*}
We compute the limits of $F_{n}(\beta_{0}+\delta/\sqrt{n}) 
$ and $F_{n}(\beta_{0})$. 
We have
\begin{align*}
\|\psi_{i}(\beta_{1})-\psi_{i}(\beta_{2})\| & \leq\|Z_{i}X_{i}^{T}(\beta_{1}-\beta_{2})\|\\
 & =\|Z(Z^{T}\gamma+\xi^{T})(\beta_{1}-\beta_{2})\|\\
 & \leq(\|ZZ^{T}\|\|\gamma\|+\|Z\xi^{T}\|)\cdot\|(\beta_{1}-\beta_{2})\|\\
 & \leq(\|Z\|^{2}\|\gamma\|+\|Z\|\|\xi\|)\cdot\|(\beta_{1}-\beta_{2})\|
\end{align*}
 where $\|\cdot\|$ denotes operator norm for matrices and Euclidean
norm for vectors. We have, for some constant $c$ that depends on
$k$,
\begin{align*}
\mathbb{E}(\|Z\|^{2}\|\gamma\|+\|Z\|\|\xi\|)^{k} & \leq c(\mathbb{E}\|Z\|^{2k}\|\gamma\|^{k}+\mathbb{E}\|Z\|^{k}\|\xi\|^{k})\\
 & \leq c(\mathbb{E}\|Z\|^{2k}\|\gamma\|^{k}+\sqrt{\mathbb{E}\|Z\|^{2k}}\cdot\sqrt{\mathbb{E}\|\xi\|^{2k}})\\
 & <\infty
\end{align*}
 using the assumptions that $\mathbb{E}\|Z\|^{2k}<\infty$ and $\mathbb{E}\|\xi\|^{2k}<\infty$.
Moreover, we have 
\begin{align*}
\mathbb{E}\|\psi(\beta)\|^{k} & =\mathbb{E}\|Z(Y-\beta^{T}X)\|^{k}\\
 & =\mathbb{E}\|Z(X^{T}(\beta_{0}-\beta)+\epsilon)\|^{k}\\
 & =\mathbb{E}\|Z((Z^{T}\gamma+\xi^{T})(\beta_{0}-\beta)+\epsilon)\|^{k}\\
 & \leq c\left(\mathbb{E}\|ZZ^{T}\gamma(\beta_{0}-\beta)\|^{k}+\mathbb{E}\|Z\xi^{T}(\beta_{0}-\beta)\|^{k}+\mathbb{E}\|Z\epsilon\|^{k}\right)\\
 & \leq c\left(\mathbb{E}\|Z\|^{2k}\|\gamma(\beta_{0}-\beta)\|^{k}+\sqrt{\mathbb{E}\|Z\|^{2k}}\sqrt{\mathbb{E}\|\xi\|^{2k}}\|(\beta_{0}-\beta)\|^{k}+\sqrt{\mathbb{E}\|Z\|^{2k}}\sqrt{\mathbb{E}\|\epsilon\|^{2k}}\right)
\end{align*}
 which is uniformly bounded on compact subsets. The consistency result in \cref{thm:DRIVE-consistency} combined with the above bounds guarantee stochastic equicontinuity \citep{andrews1994empirical}, so that as $n\rightarrow\infty$, \emph{uniformly} in $\delta$ on compact sets that contain $\delta=\sqrt{n}(\hat{\beta}^{\text{DRIVE}}_{n}-\beta_{0})$,
\begin{align*}
\frac{1}{\sqrt{n}}\left (\sum_{i}\psi_{i}(\beta_{0}+\delta/\sqrt{n})-\mathbb{E}\psi_{i}(\beta_{0}+\delta/\sqrt{n})\right) & \rightarrow_{d}\mathcal{N}(0,\Omega(\beta_0))\equiv \mathcal{Z},
\end{align*}
where $\Omega(\beta)=\frac{1}{\sqrt{n}}\mathbb{E}\sum_{i}(\psi_{i}(\beta)\psi_{i}^{T}(\beta))$, so that
\begin{align*}
\Omega(\beta_{0})=\frac{1}{\sqrt{n}}\mathbb{E}\sum_{i}(\psi_{i}(\beta_{0})\psi_{i}^{T}(\beta_{0})) & =\frac{1}{\sqrt{n}}\mathbb{E}\sum_{i}(Y_{i}-X_{i}^{T}\beta)^{2}Z_{i}Z_{i}^{T}\\
 & =\frac{1}{\sqrt{n}}\mathbb{E}\sum_{i}\epsilon_{i}^{2}Z_{i}Z_{i}^{T}=\sigma^{2}\Sigma_{Z},
 \end{align*}
 using independence and homoskedasticity. Moreover,
\begin{align*}
\frac{1}{\sqrt{n}}\sum_{i}\mathbb{E}\psi_{i}(\beta_{0}+\delta/\sqrt{n}) & =\sqrt{n}\mathbb{E}\left[X^{T}(\beta_{0}+\delta/\sqrt{n})-Y\right]Z\\
 & =\sqrt{n}\mathbb{E}\left[X^{T}(\beta_{0}+\delta/\sqrt{n})-(X^{T}\beta_{0}+\epsilon)\right]Z\\
 & =\mathbb{E}ZX^{T}\delta=\mathbb{E}Z(Z^{T}\gamma+\xi)\delta\\
 & =\Sigma_{Z}\gamma\delta.
\end{align*}
Combining these, we have 
\begin{align*}
\frac{1}{\sqrt{n}}\sum_{i}\psi_{i}(\beta_{0}+\delta/\sqrt{n}) & \rightarrow_{d}\mathcal{Z}+\Sigma_Z\gamma\delta,
\end{align*}
 uniformly in $\delta$ on compact sets, so that 
\begin{align*}
F_{n}(\beta_{0}+\delta/\sqrt{n}) & \rightarrow_{d}(\mathcal{Z}+\Sigma_Z\gamma\delta)^{T}\Sigma_Z^{-1}(\mathcal{Z}+\Sigma_Z\gamma\delta)\\
F_{n}(\beta_{0}) & \rightarrow_{d}\mathcal{Z}^{T}\Sigma_Z^{-1}\mathcal{Z},
\end{align*}
and applying the continuous mapping theorem to the square root function,
\begin{align*}
     \textbf{I} & =\sqrt{F_{n}(\beta_{0}+\delta/\sqrt{n})}-\sqrt{F_{n}(\beta_{0})}\rightarrow_d
    \sqrt{(\mathcal{Z}+\Sigma_Z\gamma\delta)^{T}\Sigma_Z^{-1}(\mathcal{Z}+\Sigma_Z\gamma\delta)} - \sqrt{\mathcal{Z}^{T}\Sigma_Z^{-1}\mathcal{Z}}.
\end{align*}
 Next we have 
\begin{align*}
\mathbf{II} & =\sqrt{n\rho_n(1+\|\beta_{0}+\delta/\sqrt{n}\|^{2})}-\sqrt{n\rho_n(1+\|\beta_{0}\|^{2})}\\
 & =\frac{n\rho_n\beta_{0}^{T}}{\sqrt{n\rho_n(1+\|\beta_{0}\|^{2})}}\cdot\delta/\sqrt{n}+o(\delta/\sqrt{n})\\
 & \rightarrow\frac{\sqrt{\rho_n}\beta_{0}^{T}}{\sqrt{(1+\|\beta_{0}\|^{2})}}\cdot\delta.
\end{align*}

Combining the analyses of $\mathbf{I}$ and $\mathbf{II}$, we have
\begin{align*}
H_{n}(\delta) & \rightarrow_d
\sqrt{(\mathcal{Z}+\Sigma_Z\gamma\delta)^{T}\Sigma_Z^{-1}(\mathcal{Z}+\Sigma_Z\gamma\delta)} - \sqrt{\mathcal{Z}^{T}\Sigma_Z^{-1}\mathcal{Z}}+\frac{\sqrt{\rho}\beta_{0}^{T}}{\sqrt{(1+\|\beta_{0}\|^{2})}}\cdot\delta
\end{align*}
uniformly. Because $H_{n}(\delta)$ is convex
and $H(\delta)$ has a unique minimum, $\arg\min_{\delta}H_{n}(\delta)=\sqrt{n}(\hat{\beta}^{\text{DRIVE}}_{n}-\beta_{0})=O_{p}(1)$.
Applying Theorem 3.2.2 of \citet{van_der_vaart1996} allows
us to conclude that 
\begin{align*}
\sqrt{n}(\hat{\beta}^{\text{DRIVE}}_{n}-\beta_{0}) & \rightarrow_{d}\arg\min_{\delta}\sqrt{(\mathcal{Z}+\Sigma_Z\gamma\delta)^{T}\Sigma_Z^{-1}(\mathcal{Z}+\Sigma_Z\gamma\delta)} - \sqrt{\mathcal{Z}^{T}\Sigma_Z^{-1}\mathcal{Z}}+\frac{\sqrt{\rho}\beta_{0}^{T}}{\sqrt{(1+\|\beta_{0}\|^{2})}}\cdot\delta.
\end{align*}
 In fact, we may drop the term $\sqrt{\mathcal{Z}^{T}\Sigma_Z^{-1}\mathcal{Z}}$ since it does not depend on $\delta$. Therefore,
\begin{align*}
\sqrt{n}(\hat{\beta}^{\text{DRIVE}}_{n}-\beta_{0})=\arg\min_{\delta}H_{n}(\delta) & \rightarrow_{d}\arg\min_{\delta}\sqrt{(\mathcal{Z}+\Sigma_Z\gamma\delta)^{T}\Sigma_Z^{-1}(\mathcal{Z}+\Sigma_Z\gamma\delta)}+\frac{\sqrt{\rho}\beta_{0}^{T}}{\sqrt{(1+\|\beta_{0}\|^{2})}}\cdot\delta.
\end{align*}

Now when $\rho=0$, the objective above
reduces to 
\begin{align*}
\sqrt{(\mathcal{Z}+\Sigma_Z\gamma\delta)^{T}\Sigma_Z^{-1}(\mathcal{Z}+\Sigma_Z\gamma\delta)},
\end{align*}
which recovers the same minimizer as the  TSLS objective
\begin{align*}
(\mathcal{Z}+\Sigma_Z\gamma\delta)^{T}\Sigma_Z^{-1}(\mathcal{Z}+\Sigma_Z\gamma\delta)-\mathcal{Z}^{T}\Sigma_Z^{-1}\mathcal{Z} & =2\delta^{T}\gamma^{T}\mathcal{Z}+\delta^{T}\gamma^{T}\Sigma_Z\gamma\delta,
\end{align*}
 since the first order condition of the former is 
\begin{align*}
\frac{\gamma^{T}\mathcal{Z}+\gamma^{T}\Sigma_Z\gamma\delta}{\sqrt{(\mathcal{Z}+\Sigma_Z\gamma\delta)^{T}(\mathcal{Z}+\Sigma_Z\gamma\delta)}} & =0,
\end{align*}
 and of the latter is 
\begin{align*}
\gamma^{T}\mathcal{Z}+\gamma^{T}\Sigma_Z\gamma\delta & =0.
\end{align*}

We can therefore conclude that with vanishing $\rho_n\rightarrow \rho=0$, regardless
of the rate, the asymptotic distribution of Wasserstein DRIVE coincides with that
of the standard TSLS estimator.
\end{proof}
\subsection{Proof of \cref{cor:asymptotic-distribution-coincide}}

\begin{proof}
When $\rho_n\rightarrow0$, the limiting objective is $\sqrt{(\mathcal{Z}+\gamma\delta)^{T}(\mathcal{Z}+\gamma\delta)}$
which is minimized at the same $\delta$ that minimizes the standard
limit $(\mathcal{Z}+\gamma\delta)^{T}(\mathcal{Z}+\gamma\delta)$. 

If $0<\rho\leq|\gamma|$, then FOC gives 
\begin{align*}
\frac{\gamma^{T}\mathcal{Z}+\delta^{T}\gamma^{T}\gamma}{\sqrt{(\mathcal{Z}+\gamma\delta)^{T}(\mathcal{Z}+\gamma\delta)}}+\frac{\sqrt{\rho}\beta_{0}}{\sqrt{(1+\|\beta_{0}\|^{2})}} & =0.
\end{align*}
If $\beta_{0}$ is one-dimensional (but $\gamma$ can be a vector,
i.e., multiple instruments), then FOC reduces to 
\begin{align*}
\gamma^{T}\mathcal{Z}+\delta^{T}\gamma^{T}\gamma & =0,
\end{align*}
%(\zq{TODO: Elaborate on this derivation.})
 which is the same FOC as the standard IV limiting objective.

If both $\gamma$ and $\beta_{0}$ are one-dimensional, but $\beta_{0}$
is not necessarily 0, we have that
\begin{align*}
\sqrt{(\mathcal{Z}+\gamma\delta)^{T}(\mathcal{Z}+\gamma\delta)}+\frac{\sqrt{\rho}\beta_{0}^{T}}{\sqrt{(1+\|\beta_{0}\|^{2})}}\cdot\delta & =|\mathcal{Z}+\gamma\delta|+\frac{\sqrt{\rho}\beta_{0}}{\sqrt{(1+\|\beta_{0}\|^{2})}}\cdot\delta
\end{align*}
The objective is $\mathcal{Z}+\gamma\delta+\frac{\sqrt{\rho}\beta_{0}}{\sqrt{(1+\|\beta_{0}\|^{2})}}\cdot\delta$
when $\gamma\delta+\mathcal{Z}\geq0$ and $-\mathcal{Z}-\gamma\delta+\frac{\sqrt{\rho}\beta_{0}}{\sqrt{(1+\|\beta_{0}\|^{2})}}\cdot\delta$
when $\gamma\delta+\mathcal{Z}\leq0$. Recall that by assumption $\sqrt{\rho}\leq|\gamma|$.

%\yc{Can we polish the statements? It's hard to follow.} 
If $\beta_{0}>0$ and $\gamma>0$, then $\gamma\delta+\frac{\sqrt{\rho}\beta_{0}}{\sqrt{(1+\|\beta_{0}\|^{2})}}\cdot\delta$
when $\gamma\delta+\mathcal{Z}\geq0$ is minimized at $\delta=-\gamma^{-1}\mathcal{Z}$,
and $-\gamma\delta+\frac{\sqrt{\rho}\beta_{0}}{\sqrt{(1+\|\beta_{0}\|^{2})}}\cdot\delta$
when $\gamma\delta+\mathcal{Z}\leq0$ is minimized at $-\gamma^{-1}\mathcal{Z}$.

If $\beta_{0}>0$ and $\gamma<0$, then $\gamma\delta+\frac{\sqrt{\rho}\beta_{0}}{\sqrt{(1+\|\beta_{0}\|^{2})}}\cdot\delta$
when $\gamma\delta+\mathcal{Z}\geq0$ is again minimized at $\delta=-\gamma^{-1}\mathcal{Z}$
(since $\delta\leq-\gamma^{-1}\mathcal{Z}$), and $-\gamma\delta+\frac{\sqrt{\rho}\beta_{0}}{\sqrt{(1+\|\beta_{0}\|^{2})}}\cdot\delta$
when $\gamma\delta+\mathcal{Z}\leq0$ is again minimized at $-\gamma^{-1}\mathcal{Z}$. 

If $\beta_{0}<0$ and $\gamma>0$, then $\gamma\delta+\frac{\sqrt{\rho}\beta_{0}}{\sqrt{(1+\|\beta_{0}\|^{2})}}\cdot\delta$
when $\gamma\delta+\mathcal{Z}\geq0$ is minimized at $\delta=-\gamma^{-1}\mathcal{Z}$,
and $-\gamma\delta+\frac{\sqrt{\rho}\beta_{0}}{\sqrt{(1+\|\beta_{0}\|^{2})}}\cdot\delta$
when $\gamma\delta+\mathcal{Z}\leq0$ is minimized at $-\gamma^{-1}\mathcal{Z}$.

If $\beta_{0}<0$ and $\gamma>0$, then $\gamma\delta+\frac{\sqrt{\rho}\beta_{0}}{\sqrt{(1+\|\beta_{0}\|^{2})}}\cdot\delta$
when $\gamma\delta+\mathcal{Z}\geq0$ is minimized at $\delta=-\gamma^{-1}\mathcal{Z}$,
and $-\gamma\delta+\frac{\sqrt{\rho}\beta_{0}}{\sqrt{(1+\|\beta_{0}\|^{2})}}\cdot\delta$
when $\gamma\delta+\mathcal{Z}\leq0$ is minimized at $-\gamma^{-1}\mathcal{Z}$.

We can therefore conclude that the objective is always minimized at
$\delta=-\gamma^{-1}\mathcal{Z}$, which is the limiting distribution
of TSLS. 
\end{proof}

\subsection{Proof of \cref{thm:gmm}}
\begin{proof}
We can write 
\begin{align*}
\frac{1}{n}\sum_{i}\psi_{i}(\theta) & =\frac{1}{n}\sum_{i}[\psi_{i}(\theta)-\mathbb{E}\psi_{i}(\theta)]+\frac{1}{n}\sum_{i}\mathbb{E}\psi_{i}(\theta).
\end{align*} 
\cref{ass:gmm}.1 guarantees that 
\begin{align*}
\frac{1}{n}\sum_{i}[\psi_{i}(\theta)-\mathbb{E}\psi_{i}(\theta)] & =o_{p}(1),
\end{align*}
 using for example Andrews.

Next, \cref{ass:gmm}.2 guarantees that $\frac{1}{n}\sum_{i}\mathbb{E}\psi_{i}(\theta)\rightarrow m(\theta)$
uniformly in $\theta$, and \cref{ass:gmm}.3 further guarantees that
\begin{align*}
\sqrt{\left(\frac{1}{n}\sum_{i}\psi_{i}(\theta)\right)^{T}W_{n}(\theta)\left(\frac{1}{n}\sum_{i}\psi_{i}(\theta)\right)}+\sqrt{\rho_{n}(1+\|\theta\|^{2})} & \rightarrow_{p}\\
\sqrt{m(\theta)^{T}W(\theta)m(\theta)}+\sqrt{\rho(1+\|\theta\|^{2})}
\end{align*}
 uniformly in $\theta$. Applying Corollary 3.2.3 of \citet{van_der_vaart1996}, we can conclude
$\hat{\theta}^{GMM}\rightarrow_{p}\theta^{GMM}$. 

Next, we consider the minimizer of the population objective. Applying
\cref{ass:gmm}.4, when $\rho\leq\overline{\rho}$, it is lower bounded
by 
\begin{align*}
\sqrt{m(\theta)^{T}W(\theta)m(\theta)}+\sqrt{\rho(1+\|\theta\|^{2})} & \geq\sqrt{\overline{\rho}\|\theta-\theta_{0}\|^{2}}+\sqrt{\rho(1+\|\theta\|^{2})}\\
 & \geq\sqrt{\overline{\rho}}\cdot(\sqrt{\|\theta-\theta_{0}\|^{2}}+\sqrt{(1+\|\theta\|^{2})})\\
 & =\sqrt{\overline{\rho}}\cdot(\|(\theta,1)-(\theta_{0},1)\|_{2}+\sqrt{\rho}\|(\theta,1)\|_{2})\\
 & \geq\sqrt{\rho}\|(\theta_{0},1)\|_{2},
\end{align*}
where again the last inequality follows from the triangle inequality.
We can verify that equalities are achieved if and only if $\theta=\theta_{0}$,
which guarantees that $\hat{\theta}^{GMM}\rightarrow_{p}\theta_{0}$.
The condition $m(\theta)^{T}W(\theta)m(\theta)\geq\overline{\rho}\|\theta-\theta_{0}\|^{2}$
is satisfies by many GMM estimators, including the TSLS, so this proof applies to \cref{thm:DRIVE-consistency} as well. 
\end{proof}
\subsection{Proof of \cref{thm:sqrt-ridge-root-n-consistency}}
\begin{proof}
First, we use optimality condition of $\hat{\beta}$ to bound 
\begin{align*}
\sqrt{\hat{Q}(\hat{\beta})}-\sqrt{\hat{Q}(\beta_{0})} & \leq\frac{\lambda}{n}\sqrt{\|\beta_{0}\|^{2}+1}-\frac{\lambda}{n}\sqrt{\|\hat{\beta}\|^{2}+1}
\end{align*}

On the other hand, by convexity of $\sqrt{\hat{Q}(\beta)}$,
\begin{align*}
\sqrt{\hat{Q}(\hat{\beta})}-\sqrt{\hat{Q}(\beta_{0})} & \geq\tilde{S}^{T}(\hat{\beta}-\beta_{0})\geq-\|\tilde{S}\|_{2}\|\hat{\beta}-\beta_{0}\|_{2}\geq-\frac{\lambda}{cn}\|\hat{\beta}-\beta_{0}\|_{2}
\end{align*}

Now the estimation error in terms of the ``prediction norm'' (which
is just the norm defined using the Gram matrix)
\begin{align*}
\|\hat{\beta}-\beta_{0}\|_{2,n}^{2} & :=\frac{1}{n}\sum_{i}(X_{i}^{T}(\hat{\beta}-\beta_{0}))^{2}\\
 & =(\hat{\beta}-\beta_{0})^{T}\frac{1}{n}\sum_{i}X_{i}X_{i}^{T}(\hat{\beta}-\beta_{0})
\end{align*}
 is related to the difference $\hat{Q}(\hat{\beta})-\hat{Q}(\beta_{0})$
as follows:
\begin{align*}
\hat{Q}(\hat{\beta})-\hat{Q}(\beta_{0}) & =\frac{1}{n}\sum_{i}(Y_{i}-X_{i}^{T}\hat{\beta})^{2}-\frac{1}{n}\sum_{i}(Y_{i}-X_{i}^{T}\beta_{0})^{2}\\
 & =\frac{1}{n}\sum_{i}(Y_{i}-X_{i}^{T}\beta_{0}+X_{i}^{T}\beta_{0}-X_{i}^{T}\hat{\beta})^{2}-\frac{1}{n}\sum_{i}(Y_{i}-X_{i}^{T}\beta_{0})^{2}\\
 & =\|\hat{\beta}-\beta_{0}\|_{2,n}^{2}+2\frac{1}{n}\sum_{i}(Y_{i}-X_{i}^{T}\beta_{0})(X_{i}^{T}\beta_{0}-X_{i}^{T}\hat{\beta})\\
 & =\|\hat{\beta}-\beta_{0}\|_{2,n}^{2}+2\frac{1}{n}\sum_{i}(\sigma\epsilon_{i})X_{i}^{T}(\beta_{0}-\hat{\beta})\\
 & =\|\hat{\beta}-\beta_{0}\|_{2,n}^{2}-2E_{n}(\sigma\epsilon X^{T}(\hat{\beta}-\beta_{0}))
\end{align*}
 On the other hand, 
\begin{align*}
\hat{Q}(\hat{\beta})-\hat{Q}(\beta_{0}) & =\left[\sqrt{\hat{Q}(\hat{\beta})}+\sqrt{\hat{Q}(\beta_{0})}\right]\cdot\left[\sqrt{\hat{Q}(\hat{\beta})}-\sqrt{\hat{Q}(\beta_{0})}\right]
\end{align*}
 and using Holder's inequality, 
\begin{align*}
2E_{n}(\sigma\epsilon X^{T}(\hat{\beta}-\beta_{0})) & =2\frac{1}{n}\sum_{i}(\sigma\epsilon_{i})X_{i}^{T}(\hat{\beta}-\beta_{0})\\
 & =2\sqrt{\frac{1}{n}\sum_{i}(\sigma\epsilon_{i})^{2}}\frac{\frac{1}{n}\sum_{i}(\sigma\epsilon X_{i}^{T})}{\sqrt{\frac{1}{n}\sum_{i}(\sigma^{2}\epsilon_{i}^{2})}}(\hat{\beta}-\beta_{0})\\
 & =2\sqrt{\hat{Q}(\beta_{0})}\cdot\tilde{S}^{T}(\hat{\beta}-\beta_{0})\\
 & \leq2\sqrt{\hat{Q}(\beta_{0})}\|\tilde{S}\|_{2}\|\hat{\beta}-\beta_{0}\|_{2}
\end{align*}

Combining these, we can bound the estimation error $\|\hat{\beta}-\beta_{0}\|_{2,n}^{2}$
as 
{\small
\begin{align*}
&\|\hat{\beta}-\beta_{0}\|_{2,n}^{2}\\
 & =2E_{n}(\sigma\epsilon X^{T}(\hat{\beta}-\beta_{0}))+\hat{Q}(\hat{\beta})-\hat{Q}(\beta_{0})\\
 & \leq2\sqrt{\hat{Q}(\beta_{0})}\|\tilde{S}\|_{2}\|\hat{\beta}-\beta_{0}\|_{2}+(\frac{\lambda}{n}\sqrt{\|\beta_{0}\|^{2}+1}-\frac{\lambda}{n}\sqrt{\|\hat{\beta}\|^{2}+1})\cdot(\sqrt{\hat{Q}(\hat{\beta})}+\sqrt{\hat{Q}(\beta_{0})})\\
 & \leq2\sqrt{\hat{Q}(\beta_{0})}\|\tilde{S}\|_{2}\|\hat{\beta}-\beta_{0}\|_{2}+(\frac{\lambda}{n}\sqrt{\|\beta_{0}\|^{2}+1}-\frac{\lambda}{n}\sqrt{\|\hat{\beta}\|^{2}+1})\cdot(2\sqrt{\hat{Q}(\beta_{0})}+\frac{\lambda}{n}\sqrt{\|\beta_{0}\|^{2}+1}-\frac{\lambda}{n}\sqrt{\|\hat{\beta}\|^{2}+1})\\
 & =2\sqrt{\hat{Q}(\beta_{0})}\|\tilde{S}\|_{2}\|\hat{\beta}-\beta_{0}\|_{2}+(\frac{\lambda}{n}\sqrt{\|\beta_{0}\|^{2}+1}-\frac{\lambda}{n}\sqrt{\|\hat{\beta}\|^{2}+1})^{2}+2\sqrt{\hat{Q}(\beta_{0})}(\frac{\lambda}{n}\sqrt{\|\beta_{0}\|^{2}+1}-\frac{\lambda}{n}\sqrt{\|\hat{\beta}\|^{2}+1})\\
 & \leq2\sqrt{\hat{Q}(\beta_{0})}\|\tilde{S}\|_{2}\|\hat{\beta}-\beta_{0}\|_{2}+(\frac{\lambda}{n})^{2}\|\hat{\beta}-\beta_{0}\|_{2}^{2}+2\sqrt{\hat{Q}(\beta_{0})}\frac{\lambda}{n}\|\hat{\beta}-\beta_{0}\|_{2}\\
 & \leq2\sqrt{\hat{Q}(\beta_{0})}\frac{\lambda}{n}(\frac{1}{c}+1)\|\hat{\beta}-\beta_{0}\|_{2}+(\frac{\lambda}{n})^{2}\|\hat{\beta}-\beta_{0}\|_{2}^{2}
\end{align*}
}
 %
\begin{comment}
The key is to cancel the first term with the last term using $\lambda\geq cn\|\tilde{S}\|_{2}$,
which yields 
\begin{align*}
2\sqrt{\hat{Q}(\beta_{0})}\|\tilde{S}\|_{2}\|\hat{\beta}-\beta_{0}\|_{2}-2\sqrt{\hat{Q}(\beta_{0})}\frac{\lambda}{n}\sqrt{\|\hat{\beta}\|^{2}+1} & \leq2\sqrt{\hat{Q}(\beta_{0})}\|\tilde{S}\|_{2}\|\hat{\beta}-\beta_{0}\|_{2}-2\sqrt{\hat{Q}(\beta_{0})}\|\tilde{S}\|_{2}\sqrt{\|\hat{\beta}\|^{2}+1}\\
 & =2\sqrt{\hat{Q}(\beta_{0})}\|\tilde{S}\|_{2}(\|(\hat{\beta},1)-(\beta_{0},1)\|_{2}-\|(\hat{\beta},1)\|_{2})\\
 & \leq2\sqrt{\hat{Q}(\beta_{0})}\|\tilde{S}\|_{2}\|(\beta_{0},1)\|_{2}\\
 & \leq2\sqrt{\hat{Q}(\beta_{0})}\frac{\lambda}{cn}\|(\beta_{0},1)\|_{2}
\end{align*}
 using again $\|\tilde{S}\|_{2}\leq\frac{\lambda}{cn}$. Plugging
in this inequality back, we get 
\begin{align*}
\|\hat{\beta}-\beta_{0}\|_{2,n}^{2} & \leq(\frac{\lambda}{n}\sqrt{\|\beta_{0}\|^{2}+1}-\frac{\lambda}{n}\sqrt{\|\hat{\beta}\|^{2}+1})^{2}+2\frac{\lambda}{n}(1+\frac{1}{c})\sqrt{\hat{Q}(\beta_{0})}\|(\beta_{0},1)\|_{2}\\
 & \leq(\frac{\lambda}{n})^{2}\|\hat{\beta}-\beta_{0}\|_{2}^{2}+
\end{align*}
\end{comment}
Now the norms $\|\hat{\beta}-\beta_{0}\|_{2,n}^{2}$ and $\|\hat{\beta}-\beta_{0}\|_{2}$
differ by the Gram matrix $\frac{1}{n}\sum_{i}X_{i}X_{i}^{T}$, which
by the assumption $\frac{1}{n}\sum_{i}X_{ij}^{2}=1$ has diagonal
entries equal to 1. Recall that $\kappa$ is the tight constant such
that 
\begin{align*}
\kappa\|\hat{\beta}-\beta_{0}\|_{2} & \leq\|\hat{\beta}-\beta_{0}\|_{2,n}
\end{align*}
 for any $\hat{\beta}-\beta_{0}$, so we get 
\begin{align*}
\|\hat{\beta}-\beta_{0}\|_{2}^{2}\leq\frac{1}{\kappa^{2}}\|\hat{\beta}-\beta_{0}\|_{2,n}^{2} & \leq2\frac{1}{\kappa^{2}}\sqrt{\hat{Q}(\beta_{0})}\frac{\lambda}{n}(\frac{1}{c}+1)\|\hat{\beta}-\beta_{0}\|_{2}+\frac{1}{\kappa^{2}}(\frac{\lambda}{n})^{2}\|\hat{\beta}-\beta_{0}\|_{2}^{2}
\end{align*}
 which yields 
\begin{align*}
\|\hat{\beta}-\beta_{0}\|_{2} & \leq\frac{1}{1-\frac{1}{\kappa^{2}}(\frac{\lambda}{n})^{2}}2\frac{1}{\kappa^{2}}\sqrt{\hat{Q}(\beta_{0})}\frac{\lambda}{n}(\frac{1}{c}+1)\\
 & =\frac{2\sqrt{\hat{Q}(\beta_{0})}\frac{\lambda}{n}(\frac{1}{c}+1)}{\kappa^{2}-(\frac{\lambda}{n})^{2}}
\end{align*}
provided that 
\begin{align*}
(\frac{\lambda}{n})^{2} & \leq\kappa^{2}.
\end{align*}

As $\lambda/n\rightarrow0$ and $\kappa$ is a universal constant
linking the two norms, this condition will be satisfied for all $n$
large enough if Assumption 2 holds, so that the rate of convergence
of $\|\hat{\beta}-\beta_{0}\|_{2,n}\rightarrow0$ is governed by that
of $\frac{\lambda}{n}\rightarrow0$: 
\begin{align*}
\|\hat{\beta}-\beta_{0}\|_{2}\leq\frac{2\frac{\lambda}{n}(\frac{1}{c}+1)}{\kappa^{2}-(\frac{\lambda}{n})^{2}}\cdot\sqrt{\hat{Q}(\beta_{0})}\lesssim\sigma\sqrt{p\log(2p/\alpha)/n}.
\end{align*}
\end{proof}

\end{appendices}
% \putbib[ref_DRIVE]
% \end{bibunit}

\end{document}